\documentclass[twocolumn,amsfont,amssymb,amsmath, showpacs,balancelastpage, nofootinbib]{revtex4-1}
\pdfoutput=1

\usepackage{graphicx}
\usepackage{dcolumn}
\usepackage{bm}
\usepackage{amssymb,amsmath,bm}  
\usepackage{color}
\usepackage[colorlinks,linkcolor=red,citecolor=blue,urlcolor=blue ]{hyperref}
\usepackage{multirow}
\usepackage[utf8]{inputenc}
\usepackage{balance}
\usepackage{enumitem}
\usepackage{lipsum}

\newcommand{\nv}{\hat{\bf n}}
\newcommand{\jcap}{JCAP}
\newcommand{\mnras}{MNRAS}
\newcommand{\aap}{A\&A}

\newcommand{\apjl}{ApJL}
\newcommand{\aj}{Astron. Journal}

\newcommand{\procspie}{Proceedings of the SPIE}

\newcommand{\zph}{z_{\rm ph}}
\newcommand{\fsh}{\hat{\sf F}}
\newcommand{\cov}{\hat{\sf C}}
\newcommand{\dhi}{\delta_{\rm HI}}
\newcommand{\mv}{{\bf m}}
\newcommand{\dmat}{\delta_{\rm M}}
\newcommand{\dgal}{\delta_{\rm g}}

\begin{document}
\title{Calibrating photometric redshifts with intensity mapping observations}
\author{David Alonso$^1$, Pedro G. Ferreira$^1$, Matt J. Jarvis$^{1,2}$, Kavilan Moodley$^3$}
\affiliation{$^{1}$Oxford Astrophysics, Department of Physics, Keble Road, Oxford, OX1 3RH, UK\\
             $^{2}$Department of Physics, University of the Western Cape, Bellville 7535, South Africa\\
             $^{3}$Astrophysics and Cosmology Research Unit, School of Mathematics.\\ Statistics and Computer Science, University of KwaZulu-Natal, Durban, 4041, South Africa}

\begin{abstract}
  Imaging surveys of galaxies will have a high number density and angular resolution yet a poor
  redshift precision. Intensity maps of neutral hydrogen (HI) will have accurate redshift
  resolution yet will not resolve individual sources. Using this complementarity, we show how
  the clustering redshifts approach, proposed for spectroscopic surveys can also be used in
  combination with intensity mapping observations to calibrate the redshift distribution of
  galaxies in an imaging survey and, as a result, reduce uncertainties in photometric redshift
  measurements. We show how the intensity
  mapping surveys to be carried out with the MeerKAT, HIRAX and SKA instruments can improve
  photometric redshift uncertainties to well below the requirements of DES and LSST. The
  effectiveness of this method as a function of instrumental parameters, foreground subtraction
  and other potential systematic errors is discussed in detail.
\end{abstract}

  \date{\today}
  \maketitle

\section{Introduction}\label{sec:intro}
  Photometric redshift surveys are an economic way of building up a detailed map of the large
  scale structure of the Universe. By imaging large swathes of the sky, it is possible to
  construct catalogues of individually resolved galaxies with high number density
  (and therefore a low ``shot'' noise). The trade-off for such a large number of objects is the
  inability to obtain accurate redshift measurements for individual objects. Thus, photometric redshift
  surveys are orders of magnitude less resolved in the radial direction than the sparser
  spectroscopic redshift surveys. The uncertainty in the individual redshifts and in the
  overall galaxy redshift distribution can severely degrade the constraining power of such
  datasets for cosmology.

  Galaxies cluster to form the cosmic web, and one expects structures in the galaxy
  distribution to be spatially correlated with structures in any other tracer of the dark
  matter density. For example, if one has an imaging survey of galaxies (where redshifts
  are poorly resolved) and a spectroscopic catalog (where redshifts are well resolved), they
  should have non-trivial cross-correlations; in particular, structures in the imaging survey
  should be mirrored in the spectroscopic survey. A natural step is to use these
  cross-correlations so that the precise redshift measurements of the spectroscopic survey
  can be used to sharpen the photometric redshifts in the imaging survey, or at least calibrate
  its redshift distribution. These types of methods have been advocated in
  \cite{2008ApJ...684...88N,2010MNRAS.408.1168B,2010ApJ...721..456M,
  2013MNRAS.431.3307S,2013arXiv1303.4722M,2017arXiv170305326V}, and employed in the analysis
  of several datasets (e.g. \cite{2016MNRAS.460..163R,2016MNRAS.463.3737C,2016MNRAS.462.1683S,
  2016MNRAS.457.3912R,2017MNRAS.465.4118J}).

  One does not necessarily have to use a catalogue of resolved sources to follow this rationale.
  In particular, if one can accurately map out, in redshift, any tracer of the dark matter, it
  can in principle be used to improve redshift measurements in a sister imaging survey. A
  notable example is that of an unresolved map of neutral hydrogen, HI, through a technique known
  as intensity mapping \citep{Battye:2004re,2006ApJ...653..815M,Chang:2007xk,2008MNRAS.383..606W,
  2008PhRvL.100p1301L,2009astro2010S.234P,Bagla:2009jy,2013MNRAS.434.1239B,2013ApJ...763L..20M,
  2013MNRAS.434L..46S,2015ApJ...803...21B}. Radio observations at a GHz or below will map out
  the distribution of neutral hydrogen out to redshifts, $z\sim2$ or higher. The neutral
  hydrogen traces the large scale structure of the dark matter and thus, inevitably, will be
  correlated with any other tracer. Maps of HI will be exquisitely resolved in the frequency
  domain and therefore will map out the density distribution, in detail, in redshift. Although
  intensity mapping observations will not resolve individual objects, they will be able to
  achieve sufficient angular resolution for cosmological studies (although this statement
  depends on the observing mode).

  In this paper we explore the use of HI intensity mapping to calibrate photometric redshift surveys.
  In particular we show that forthcoming intensity mapping experiments such as those undertaken
  by MeerKAT \cite{2017MNRAS.466.2780F}, HIRAX \cite{2016SPIE.9906E..5XN} and the SKA
  \cite{2015aska.confE..19S} can be used to reduce the uncertainties related to photo-$z$
  systematics well below the requirements currently posited by the DES and LSST surveys, thus
  improving final constraints on cosmological parameters. We structure this paper as follows:
  in Section \ref{sec:method} we describe the formalism and discuss, in detail, various
  aspects of the instrumental and observational models we are assuming. Section
  \ref{sec:results} presents our results as a function of experimental configuration and
  foreground uncertainties. In Section \ref{sec:discuss} we discuss the prospects of using
  such a method and compare with other proposals currently being developed. The appendices
  present a number of calculations which are essential for the models considered here.
  
    \begin{figure*}
      \centering
      \includegraphics[width=0.49\textwidth]{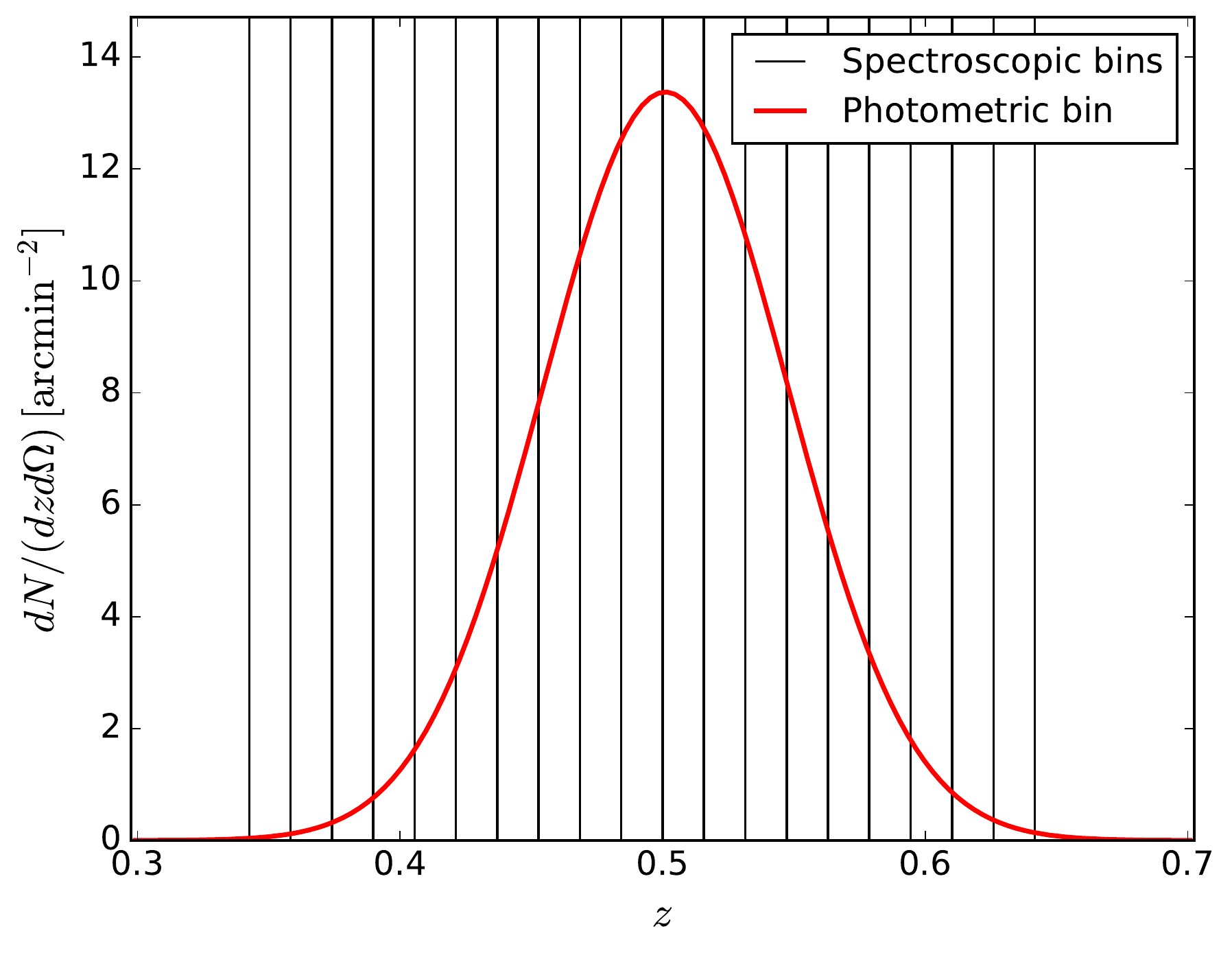}
      \includegraphics[width=0.49\textwidth]{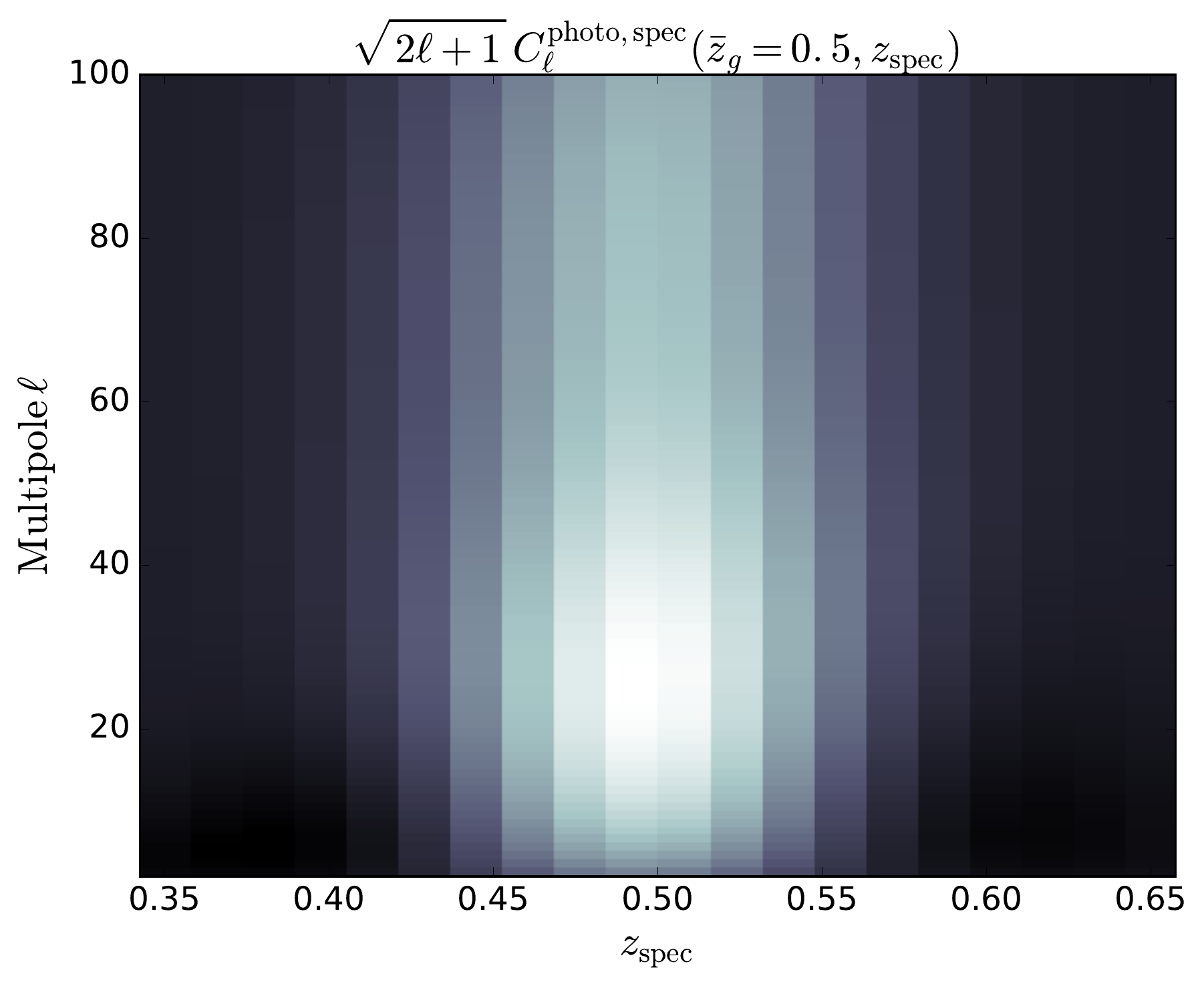}
      \caption{{\sl Left panel:} example of a redshift bin for a photometric survey and the
               redshift bins chosen for an overlapping spectroscopic survey.
               {\sl Right panel:} amplitude of the cross-correlation with an overlapping
               spectroscopic survey as a function of spectroscopic redshift bin ($x$ axis)
               and angular scale ($y$ axis). The amplitude of the cross-correlation traces
               the redshift distribution, and can therefore be used to constrain it.}
      \label{fig:clustred_example}
    \end{figure*}
\section{Formalism}\label{sec:method}
  \subsection{Clustering-based photo-$z$ calibration}\label{ssec:method.clustred}
    Consider two galaxy samples with redshift distributions $\phi_i(z)$ ($i=\{1,2\}$),
    and let $a^i_{\ell m}$ be the harmonic coefficients of their projected overdensity
    of counts on the sky. Their cross-correlation is given by:
    \begin{align}
      &\langle a^i_{\ell m}a^{j*}_{\ell m}\rangle=N^{ij}_\ell+S^{ij}_\ell\\\label{eq:cl1}
      &S^{ij}_\ell=\frac{2}{\pi}\int dz \int dz'\,\phi_i(z)\phi_j(z')\times\\\nonumber
      &\times\int dk\,k^2\,b_i(z)b_j(z')P_m(k,z,z')\,j_\ell(k\chi(z))\,j_\ell(k\chi(z')),
    \end{align}
    where $P_m$ is the matter power spectrum, $\chi$ is the radial comoving distance,
    $j_\ell(x)$ is a spherical Bessel function,
    $N^{ij}_\ell$ is the cross-noise power spectrum between samples $i$ and $j$, $b_i$
    is the linear bias of the $i$-th sample and we have neglected redshift-space distortions
    and all other sub-dominant contributions to the observed power spectrum. In the Limber
    approximation - where $j_\ell(x)\rightarrow\sqrt{\pi/(2\ell+1)}\delta^\mathcal{D}(\ell+1/2-x)$ -
    this expression simplifies to:
    \begin{equation}\label{eq:cl2}
      S^{ij}_\ell=\int dk\,P_m(k,z_\ell)\,\frac{H^2(z_\ell)b^i(z_\ell)b^j(z_\ell)}{\ell+1/2}
      \phi_i(z_\ell)\phi_j(z_\ell),
    \end{equation}
    where $\chi(z_\ell)\equiv(\ell+1/2)/k$.

    For the purposes of this discussion, the most important feature of Equation \ref{eq:cl2}
    is the fact that the amplitude of the cross-correlation is proportional to the overlap
    between the redshift distributions of those samples. This is especially relevant if one
    of the samples has good radial resolution, in which case it can be split into narrow bins
    of redshift. The cross-correlations of all narrow bins with the other sample will
    therefore trace the amplitude of its redshift distribution, and can effectively be used
    to constrain it. This is illustrated in Fig. \ref{fig:clustred_example}, which shows the
    cross-power spectrum between a Gaussian photo-$z$ bin of width $\sigma=0.05$ and a set
    of narrow redshift bins ($\delta z\sim0.002$).
    
    Note also that Eq. \ref{eq:cl2} implies that the redshift distribution and the
    redshift-dependent galaxy bias of the photometric sample are completely degenerate in
    this method, and therefore additional information is needed in order to separate both
    quantities (e.g. including prior information or lensing data). Since this is an inherent
    problem of the method, and not specific to the case of intensity mapping, we will simply
    assume that $b(z)$ is a sufficiently smoothly-varying function of $z$, thus treating IM
    and spectroscopic surveys on an equal footing. The more complicated biasing
    scheme that arises on small scales also prevents the use of those modes to constrain
    $\phi(z)$ \cite{2013MNRAS.431.3307S}, and therefore one must be conservative when deciding
    the range of scales to include in the analysis.

    Different recipes have been formulated to carry out this kind of analysis, such as the
    optimal quadratic estimator method of \cite{2013MNRAS.433.2857M}. The forecasts
    presented here will interpret the redshift distribution (in a parametric or
    non-parametric form) as a set of extra nuisance parameters, on which we will carry out
    the Fisher matrix analysis described in Section \ref{ssec:method.fisher}. Thus, even
    though our results will be optimistic in as much as the Fisher matrix saturates the
    Rao-Cramer bound, they will account for all correlations between redshift distribution
    parameters and with the cosmological parameters, as well as the presence of
    redshift-space distortions and magnification bias (effects that have been overseen in
    previous works).

    For the purposes of estimating the ability of future surveys to calibrate photometric
    redshift distributions through cross-correlations, we will always consider an individual
    redshift bin for a photometric sample with unknown distribution, together with a set of
    overlapping narrow redshift bins of spectroscopic galaxies or intensity mapping
    observations. Let $N^p(z)$ be the overall true redshift distribution of the photometric
    sample, and let $p(\zph|z)$ be the conditional distribution for a photo-$z$ $\zph$
    given the true redshift $z$. Then, the redshift distribution in a photo-$z$ redshift bin
    $b$ with bounds $z_b^i<\zph<z_b^f$ is given by
    \begin{equation}\label{eq:phz_dist}
      \phi_b(z)\propto N^p(z)
      \int_{z_b^i}^{z_b^f}d\zph\,p(\zph|z).
    \end{equation}
    In what follows we will consider two degrees of complexity in terms of describing the
    unknown redshift distribution:
    \begin{enumerate}
      \item We will assume Gaussian photo-$z$s with a given variance ($\sigma_z^2$)
        and bias $\Delta z$:
        \begin{align}\nonumber
          p(\zph|z)&\equiv \mathcal{N}(\zph-\Delta z;z,\sigma_z)\\\label{eq:photoz_gaussian}
          &\equiv\frac{\exp\left[-\frac{1}{2}\frac{(\zph-z-\Delta z)^2}
          {\sigma_z^2}\right]}{\sqrt{2\pi\sigma_z}},
        \end{align}
        and we will assume that the uncertainty in the redshift distribution is fully
        described by $\Delta z$ and $\sigma_z$.
      \item We will use a non-parametric form for $\phi_b(z)$, given as a piecewise function with
        a free amplitude for each spectroscopic redshift bin.
    \end{enumerate}
    Our assumed fiducial value for $\Delta z$ and $\sigma_z$, as well as the binning scheme
    used are described in Section \ref{ssec:method.phz}.
   
    We finish this section by noting that the use of cross-correlations with spectroscopic surveys or
    intensity mapping observations for photo-$z$ calibration is not limited to the measurement of
    the redshift distribution of a given galaxy sample, but that they can also be used to improve the
    precision of photometric redshift estimates for individual galaxies (e.g. \cite{2012MNRAS.425.1042J}). 
    Although we leave the discussion of this possibility for future work, we describe a Bayesian
    formalism for this task in Appendix \ref{app:ind_phz}.
    
  \subsection{Photometric redshift surveys}\label{ssec:method.phz}
    \begin{figure}
      \centering
      \includegraphics[width=0.49\textwidth]{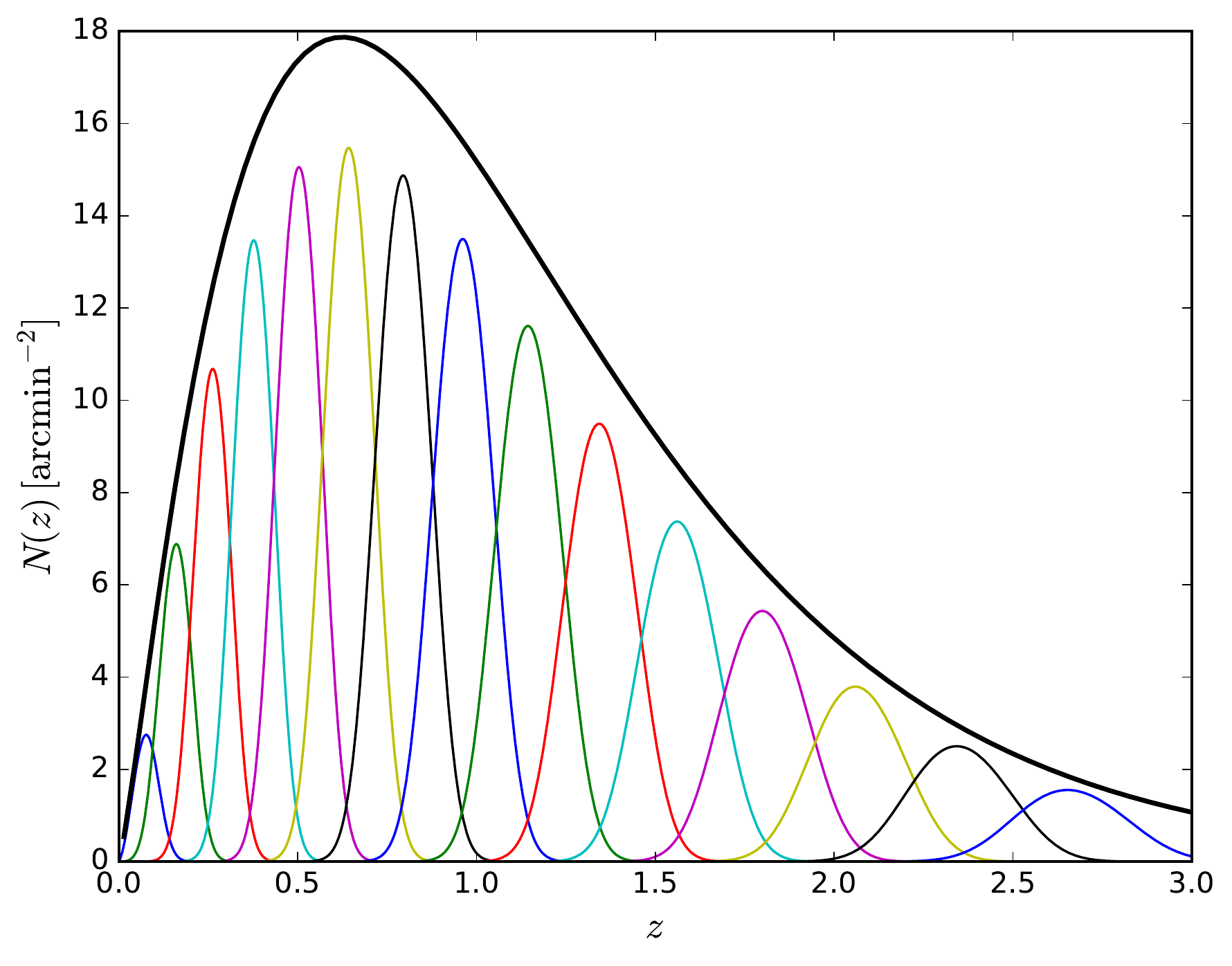}
      \caption{Angular number density of galaxies as a function of redshift for the LSST
               gold sample (solid black line). The colored lines show the window functions
               of the 15 redshift bins considered here.}
      \label{fig:nz_lsst}
    \end{figure}
    This section describes the model used here for a LSST-like photometric redshift survey.
    As in \cite{2015PhRvD..92f3525A}, we base our description of the number density of sources
    and their magnification bias on the measurements of the luminosity function of
    \cite{2006A&A...448..101G}, with $k$-corrections computed with {\tt kcorrect}
    \cite{2007AJ....133..734B}. We assume a magnitude cut of $25.3$ in the $i$ band, corresponding
    to the so-called ``gold'' sample \cite{2009arXiv0912.0201L}. Unlike \cite{2015PhRvD..92f3525A},
    and for simplicity, we will consider a single galaxy population,
    instead of splitting it into ``red'' and ``blue'' sources. The resulting redshift
    distribution is shown by the solid black line in Figure \ref{fig:nz_lsst}.

    We model the linear galaxy bias as a function of redshift as $b(z)=1+0.84z$, based
    on the simulations of \cite{2004ApJ...601....1W}, and quoted in the LSST science book
    \cite{2009arXiv0912.0201L}.
    
    The photometric redshift requirement for the gold sample as stated in the LSST science
    book are $\sigma_z/(1+z)<0.05$, with a goal of $0.02$. Here we have taken a conservative
    estimate, assuming a standard deviation $\sigma_z=0.03(1+z)$. We then split the full
    sample into redshift bins with a width given by $3\times\bar{\sigma}_z$, where
    $\bar{\sigma}_z$ is the photo-$z$ variance at the bin centre. This binning scheme is
    chosen to reduce the correlation between bins induced by the tails of the photo-$z$
    distribution, and results in the 15 redshift bins shown in Fig. \ref{fig:nz_lsst} (where
    the redshift distributions are computed with Eq. \ref{eq:phz_dist}). Our fiducial
    photo-$z$ model will assume biased Gaussian distributions, fully determined by
    $\sigma_z$ and $\Delta z$.

  \subsection{Intensity mapping}\label{ssec:method.imap}
    \begin{table}
      \centering{
      \renewcommand*{\arraystretch}{1.2}
      \begin{tabular}{|c|c|c|c|}
        \hline
        Experiment     & SKA          & MeeKAT       & HIRAX              \\
        \hline
        $T_{\rm inst}$ & 25K          & 25 K         & 50 K               \\
        $t_{\rm tot}$  & 10000 h      & 4000 h       & $2.8\times10^4$ h  \\
        $N_{\rm dish}$ & 197          & 64           & 1024 ($32\times32$)\\
        $D_{\rm dish}$ & 15 m         & 13.5 m       & 6 m                \\
        $\nu$ range    & 350-1050 MHz & 600-1050 MHz & 400-800 MHz        \\
        $f_{\rm sky}$  & 0.4          & 0.1          & 0.4                \\
        \hline
      \end{tabular}}
      \caption{Experimental specifications assumed for SKA, MeerKAT and HIRAX. The baseline
               distributions for each experiment are described in Section \ref{ssec:method.imap}.
               Note that the frequency ranges above correspond to the UHF band of SKA and 
               MeerKAT.}
      \label{tab:im_exp}
    \end{table}
    Intensity mapping (IM) is a novel observational technique that circumvents the
    long integration times needed to obtain reliable spectroscopic redshifts for individual
    objects through an approach that is transverse to that used by photometric surveys.
    The idea \citep{2006ApJ...653..815M,2008MNRAS.383..606W,2009astro2010S.234P,
    2013MNRAS.434L..46S} is to observe the unresolved combined emission of many line-emitting
    sources in a relatively wide pixel at different frequencies. The signal-to-noise ratio
    of the corresponding line emission is much stronger than that of the individual sources,
    and thus, combining the intensity measured across the sky and relating the intensity
    observed at a given frequency to the rest-frame wavelength of the emission line it is
    possible to produce three-dimensional maps of the density of the line-emitting species.
    This technique is particularly appealing for isolated spectral lines, as is the case
    of the 21cm line caused by the spin-flip transition in neutral hydrogen atoms (HI),
    and thus HI intensity mapping has been proposed as an ideal method to cover vast
    volumes at relatively low cost.
    
    A number of experiments have been proposed to carry out IM measurements of the baryon
    acoustic oscillation scale, such as BINGO \cite{2012arXiv1209.1041B}, CHIME
    \cite{2014SPIE.9145E..4VN}, FAST \cite{2016ASPC..502...41B},
    HIRAX \cite{2016SPIE.9906E..5XN}, SKA \cite{2015aska.confE..19S} and Tianlai
    \cite{2011SSPMA..41.1358C}. The different instrumental approaches to IM can
    be broadly classified into two camps:
    \begin{itemize}
      \item {\sl Interferometers:} the sky emission is measured by a set of antennas, and
      the measurements of pairs of antennas separated by a given baseline ${\bf d}$ are
      cross-correlated to produce the measurement of an angular Fourier mode with scale
      ${\bf l}\sim2\pi{\bf d}/\lambda$ (where $\lambda$ is the observed wavelength).
      The intensity map is then reconstructed by combining pairs with different baselines.
      \item {\sl Single-dish:} in this case the sky emission is measured and auto-correlated
      by individual antennas. A band-limited intensity map with a resolution
      $\delta\theta\sim\lambda/D_{\rm dish}$ is then produced by varying the antenna pointing,
      where $D_{\rm dish}$ is the antenna diameter.
    \end{itemize}
    The expressions for the noise power spectrum for both cases are derived in Appendix
    \ref{app:noise_im}, and can be summarized as:
    \begin{equation}\label{eq:nl_im}
      N^\nu_{\bf l}=\frac{T_{\rm sys}^24\pi f_{\rm sky}}{\eta^2\Delta\nu t_{\rm tot}}
      \left\{\begin{array}{ll}
              \frac{1}{N_{\rm dish}B^2({\bf l})}, & \text{single dish}\\
              \frac{\Omega_p}{N_d({\bf d}={\bf l}\lambda/(2\pi))\lambda^2}, & \text{interferometer}.
             \end{array}\right.
    \end{equation}
    Here $T_{\rm sys}$ is the system temperature, given as a combination of instrumental and sky
    temperature (see Appendix \ref{app:noise_im}), $f_{\rm sky}$ is the sky fraction covered
    by the observations, $\eta^2$ is the antenna efficiency\footnote{$\eta$ is defined as the
    ratio of the effective to real antenna area.}, $\Delta\nu$ is the bandwidth in that channel,
    $t_{\rm tot}$ is the total observation time for the survey, $N_{\rm dish}$ is the number of
    dishes, $B({\bf l})$ is the harmonic transform of the antenna beam, $N_d({\bf d})$ is the
    distribution of baselines and $\Omega_p$ is the solid angle covered per pointing. For all
    experiments discussed here we will assume $\eta=1$, Gaussian beams so that
    $B({\bf l})=\exp[-\ell(\ell+1)\theta_{\rm FWHM}^2/(16\log2)]$, and $\Omega_p=\theta_{\rm FWHM}^2$,
    where $\theta_{\rm FWHM}$ is the beam full-width at half maximum, which will approximate
    as $\theta_{\rm FWHM}=1.22\lambda/D_{\rm dish}$. Note that the baseline distribution $N_d$
    is normalized such that:
    \begin{equation}
      \frac{N_{\rm dish}(N_{\rm dish}-1)}{2}=\int d{\bf d}^2\,N_d({\bf d}),
    \end{equation}
    where $N_{\rm dish}(N_{\rm dish}-1)/2$ is the total number of independent baselines.
    
    Given their expected full overlap with LSST, we will consider here the two main currently
    envisaged southern-hemisphere intensity mapping experiments: SKA (and its pathfinder,
    MeerKAT) and HIRAX.

    \subsubsection{MeerKAT and the SKA}\label{sssec:method.imap.ska}

MeerKAT is the 64-dish precursor to the mid-frequency component of the
SKA. MeerKAT is comprised of 13.5 metre dishes and will operate
between $\sim 550-3$~GHz using three separate receivers. Although, it
will predominantly used as an interferometer, and as such only be
sensitive to relatively small spatial scales, there is a proposed
project to use MeerKAT in single-dish mode \cite{2017MNRAS.466.2780F}.
If such a mode of operation is viable, then MeerKAT will become an
extremely efficient intensity-mapping facility operating at frequencies
that allows the detection of H{\sc i} to $z\sim 1.5$. Indeed, a
proposed open-time survey would provide a several thousand square
degree sky survey over the Dark Energy Survey and/or Kilo-degree
Survey areas, which will provide excellent visible wavelength coverage.

In the 2020s, MeerKAT will be enhanced by the addition of 130, 15
metre dishes to form the mid-frequency SKA. Operating at similar
frequencies to MeerKAT, the additional 130 dishes will provide much
more sensitivity for all science aims, and is capable of carrying out a
$\sim 10,000$ deg$^2$ intensity mapping survey \cite{2015aska.confE..19S}.

As such, both MeerKAT and the SKA will provide a unique view on the
H{\sc i} Universe, and as we will show, can enhance the cosmological
science with the LSST with cross-correlations. 
      
    \subsubsection{HIRAX}\label{sssec:method.imap.hirax}
      The Hydrogen Intensity mapping and Real-time Analysis eXperiment (HIRAX) is a proposed
      close-packed radio array comprising 1024 six metre dishes disposed in a $32\times32$
      grid and operating at 400-800 MHz. The telescope will be located on the South African
      Karoo site, which has very low levels of RFI in this band, and provides an ideal
      location to overlap in sky coverage with other planned southern sky cosmological surveys.
      The large collecting area and field-of-view provide excellent sensitivity and mapping
      speed, with the high density of short baselines allowing for sensitive measurements of
      the baryon acoustic oscillation (BAO) scale in the cosmic HI distribution from redshift
      $\sim$0.8 to 2.5, which in turn will provide competitive constraints on dark energy
      \citep{2016SPIE.9906E..5XN}. HIRAX will make high signal-to-noise maps of 21cm intensity
      fluctuations over 15,000 sq degrees (taken to overlap fully with LSST) on cosmological
      scales of interest, with the relatively high frequency resolution (1024 channels over
      the 400 MHz bandwidth) allowing for accurate redshift calibration of 21cm intensity
      maps. This makes it ideal for calibration of LSST photometric redshifts through the
      cross-correlation technique.
      
    \subsubsection{Generic IM experiment}\label{sssec:method.imap.generic}
      Besides SKA and HIRAX we will also explore the capabilities of a generic intensity mapping
      experiment in terms of photo-$z$ calibration. The performance of a given experiment is
      roughly determined by three quantities:
      \begin{itemize}
        \item The range of angular scales over which the noise power spectrum is low enough
              to probe the cosmological HI emission. This range can be characterized by the
              minimum and maximum baselines $d_{\rm min}$ and $d_{\rm max}$.
        \item The noise level (normalized by the bandwidth $\Delta\nu$) $\sigma_{T}$ on
              this range of scales. For a fixed integration time, this is determined by the
              system temperature $T_{\rm sys}$ and the observed sky area $f_{\rm sky}$.
      \end{itemize}
      Here we will model the effects of the minimum and maximum baselines as a sharp and an
      inverse-Gaussian cutoff respectively. Thus, our model for the angular noise power spectrum
      is:
      \begin{equation}\label{eq:exp_generic}
        N^\nu_\ell=\frac{\sigma^2_{T}}{\Delta\nu}
        \left[\Theta\left(\frac{\ell\lambda}{2\pi},d_{\rm min}\right)\right]^{-1}
        \exp\left[\ell(\ell+1)\frac{\theta_{\rm beam}^2}{8\log2}\right],
      \end{equation}
      where $\theta_{\rm beam}\equiv1.22\lambda/d_{\rm max}$ and $\Theta(x,x_i)$ is 1 if
      $x_i<x$ and 0 otherwise. Note that by definition $\sigma_{T}$ has units of
      $[{\rm mK}\,{\rm rad}\,{\rm MHz}^{1/2}]$. For comparison, the equivalent
      values of these parameters that roughly reproduce the noise curves for HIRAX are:
      \begin{align}\nonumber
        &d^{\rm HIRAX}_{\rm min}=6\,{\rm m},\hspace{12pt}d^{\rm HIRAX}_{\rm max}\sim300\,{\rm m},\\\nonumber
        &\sigma^{\rm HIRAX}_{T}\sim10^{-3}\,{\rm mK}\,{\rm rad}\,{\rm MHz}^{1/2}
      \end{align}
      
    \subsubsection{Foregrounds}\label{sssec:method.imap.foregrounds}
      One of the main obstacles that HI intensity mapping must overcome to become a useful
      cosmological tool is the presence of galactic and extragalactic foregrounds several orders
      of magnitude larger than the 21cm cosmological signal \cite{2005ApJ...625..575S,
      2014MNRAS.441.3271W}. Under the assumption that foregrounds are coherent in frequency
      (as opposed to the cosmic signal tracing the density inhomogeneities along the line of
      sight), these foreground sources can be in principle efficiently removed using
      component-separation methods \cite{2014MNRAS.441.3271W,2015PhRvD..91h3514S,
      2015MNRAS.447..400A}. However,
      instrumental imperfections, such as frequency-dependent beams or polarisation leakage,
      can generate foreground residuals with a non-trivial frequency structure that could 
      strongly bias cosmological constraints from 21cm data alone. In any case, the removal of
      frequency-smooth components will introduce large uncertainties on the large-scale radial
      modes of the 21cm fluctuations.
      
      Here we have introduced the effect of foregrounds by including an extra component, $f$,
      in the sky model for HI accounting for foreground residuals. Thus we will assume that the
      measured harmonic coefficients at a given frequency $\nu$ are given by:
      \begin{equation}
        a^\nu_{\ell m}=s^\nu_{\ell m}+f^\nu_{\ell m}+n^\nu_{\ell m},
      \end{equation}
      where $s^\nu_{\ell m}$ and $n_{\ell m}$ are the true cosmological signal and the instrumental
      noise contribution. We will model $f$ as an almost-correlated  component with a
      cross-frequency power spectrum given by
      \begin{align}\nonumber
        C^{\nu\nu'}_{f,\ell m}&\equiv\langle f^\nu_{\ell m}f^{\nu'*}_{\ell m}\rangle\\\label{eq:fgcl}
        &=A_{\rm FG}\left(\frac{\ell}{\ell_*}\right)^\beta\left(\frac{\nu\nu'}{\nu_*^2}\right)^\alpha
          \exp\left[-\frac{{\rm log}^2(\nu/\nu')}{2\xi^2}\right].
      \end{align}
      Here $A_{\rm FG}$ and $\beta$ parametrise the amplitude of the foreground residuals and their
      distribution on different angular scales, and $\alpha$ describes their mean frequency
      dependence. Finally, $\xi$ parametrises the characteristic frequency scale over which
      foregrounds are decorrelated. When including the effects of foregrounds (Section
      \ref{ssec:results.foregrounds}) we will also marginalize over $(A_{\rm FG},\alpha,\beta,\xi)$.
      For $\alpha$ and $\beta$ we will use the fiducial values $\alpha=-2.7$ and $\beta=-2.4$,
      corresponding to galactic synchrotron emission \cite{2005ApJ...625..575S,
      2016A&A...594A..10P}, and we will set $A_{\rm FG}=1\,{\rm mK}^2$, large enough for the
      residuals to dominate the equal-$\nu$ power spectrum\footnote{We use a pivot scale
      $\ell_*=1000$ an a privot frequency $\nu_*=130\,{\rm MHz}$, as in
      \cite{2005ApJ...625..575S}.}. We will study the final constraints as a function of $\xi$.
      
      Effectively, this extra component cancels the constraining power of all radial modes of the
      21cm fluctuations with comoving radial wavenumbers $k_\parallel$ below a scale
      $k_\parallel^{\rm FG}$, related to $\xi$ through
      \begin{equation}\label{eq:fgkmin}
        k_\parallel^{\rm FG}\sim\frac{\pi\,H(z)}{c\,(1+z)\,\xi}
      \end{equation}
      
    The effect of foregrounds on the ability to constrain redshift distributions can be readily
    understood as loss of information in the $k_\parallel-k_\perp$ space. In the flat-sky
    approximation, and on linear scales, the angular power spectrum between two tracers $i$ and
    $j$ of the matter density can be computed as:
    \begin{equation}
      C^{ij}_{k_\perp}=\int\frac{dk_\parallel}{2\pi}P(k_\parallel,k_\perp)W^i(k_\parallel)
      W^{j*}(k_\parallel),
    \end{equation}
    where we have again ignored the effect of redshift-space distortions and:
    \begin{equation}
      W^k(k_\parallel)\equiv\int dx_\parallel\,\phi^k(x_\parallel)b^k(x_\parallel)D(x_\parallel)
      e^{ix_\parallel k_\parallel}.
    \end{equation}
    Here $D$ is the linear growth factor and $b^k$ and $\phi^k$ are the linear bias and selection
    function for the $k$-th tracer. Let us assume that $i$ is a photometric redshift bin
    and $j$ is a narrow intensity mapping frequency shell with comoving width $\delta\chi$
    centered at $\chi_*$. Assuming $D$ and the bias $b^j$ to be slowly-varying functions of
    $\chi$ we obtain:
    \begin{equation}
      C^{ij}_{k_\perp}=\int\frac{dk_\parallel}{2\pi}P(k_\parallel,k_\perp)
      W^i(k_\parallel)D(\chi_*)b^j(\chi_*)j_0(k_\parallel\delta\chi/2).
    \end{equation}
    Now, assuming that $\phi^i$ has support over a wide range of redshifts, corresponding to a
    comoving width $\Delta\chi$, its Fourier transform ($\sim W^i$) will only have support over
    wavenumbers $k_\parallel\lesssim1/\Delta\chi$. Since the Bessel function $j_0$ provides
    support over all values of $k_\parallel\lesssim1/\delta\chi$, and under the assumption
    that $\delta\chi<\Delta\chi$, the total number of modes that contribute to $C^{ij}$ is
    bound by $\sim1/\Delta\chi$. Since foreground contamination will mostly affect large radial
    modes, eventually a large fraction of this $k_\parallel$-range becomes dominated by
    foreground uncertainties and stops contributing efficiently to the overall signal-to-noise
    ratio, thus degrading the final constraints on any model parameter.

      We finish this Section by noting that, as described in Section \ref{ssec:method.clustred},
      the main constraining power for photo-$z$ calibration comes from the cross-correlation of the
      photometric and spectroscopic samples. Since the photometric sample would not suffer from
      foreground contamination, these cross-correlations are very robust against foreground biasing,
      which makes photo-$z$ calibration an ideal application of IM.
 
  \subsection{Spectroscopic surveys}\label{ssec:method.spec}
    \begin{figure}
      \centering
      \includegraphics[width=0.49\textwidth]{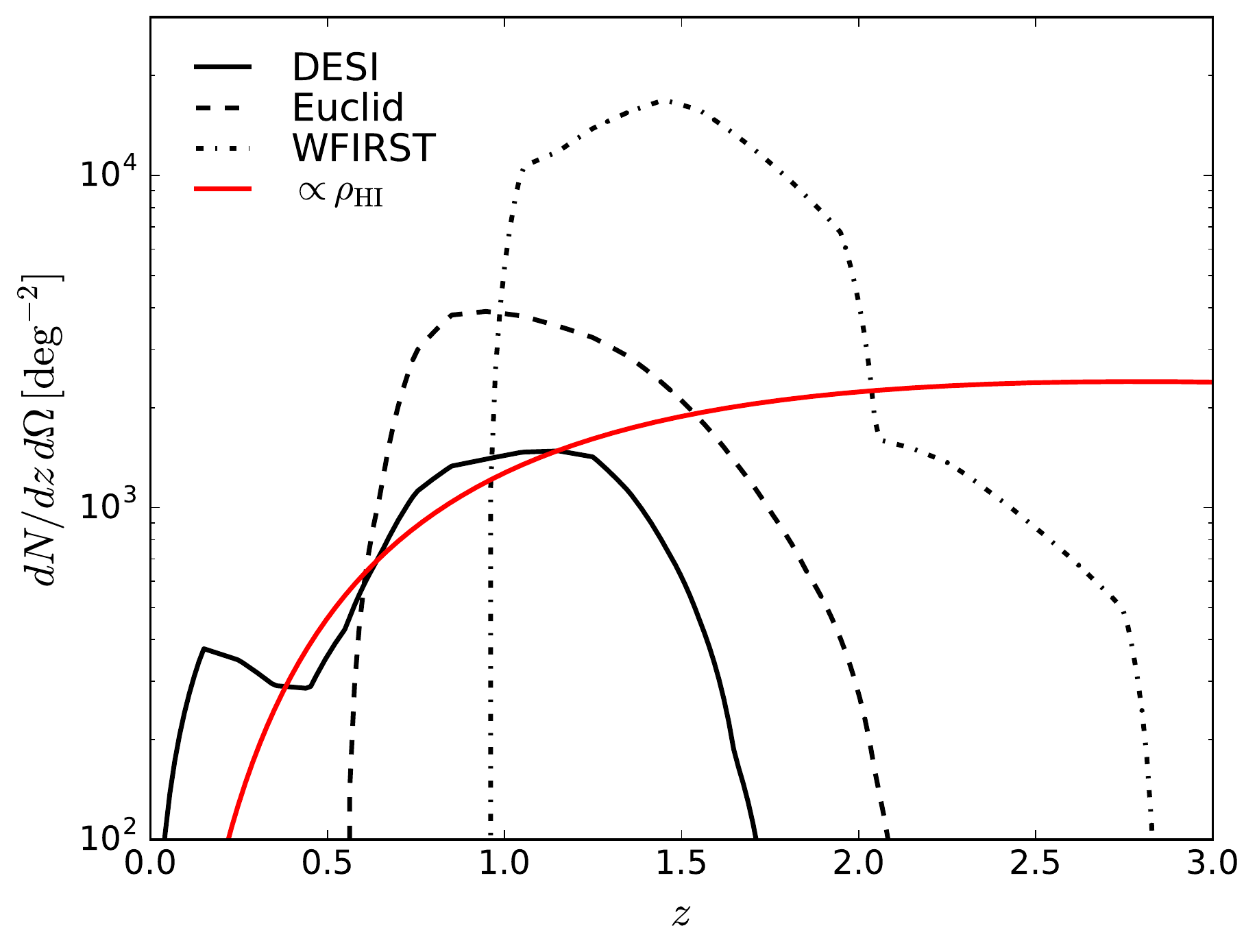}
      \caption{Angular number density of galaxies as a function of redshift for the three
               spectroscopic surveys considered here: DESI (solid), Euclid (dashed) and
               WFIRST (dot-dashed). The red solid line shows the redshift dependence of the
               mean HI density (arbitrarily normalized) for comparison.}
      \label{fig:nz_spec}
    \end{figure}
    In order to showcase the possibility of calibrating redshift distributions through
    cross-correlation with future intensity mapping experiments we will compare their
    forecast performance against that of the most relevant future spectroscopic suveys:
    \begin{itemize}
      \item The Dark Energy Spectroscopic Instrument (DESI) \cite{2013arXiv1308.0847L} is a
            spectroscopic galaxy survey planned to cover $\sim14000\,{\rm deg}^2$ from its
            northern-hemisphere location at Kitt Peak National Observatory. We assume an area
            overlap of $f_{\rm sky}=0.2$ with LSST, and we model the number density and
            clustering bias of the two galaxy samples considered here (Luminous Red Galaxies
            and Emission Line Galaxies) as done in \cite{2014JCAP...05..023F}.
      \item The Euclid galaxy survey \cite{2011arXiv1110.3193L} is a space-borne infrared
            spectrograph that will aim to detect $\sim5\times10^7$ H$\alpha$-emitting
            galaxies in the redshift range $0.65<z<2$ over $\sim15000\,{\rm deg}^2$. We
            assume full overlap with LSST, and we model the number density and bias as in
            \cite{2014JCAP...05..023F}.
      \item The Wide Field Infrared Survey Telescope (WFIRST) \cite{2013arXiv1305.5422S} is a
            future space observatory in the infrared that will measure redshifts for
            $\sim2.6\times10^7$ objects over $\sim2000\,{\rm deg}^2$. The deep nature of
            WFIRST will make it ideal to calibrate the LSST redshift distribution at high
            redshifts. We model the number density and bias of the WFIRST sample as in
            \cite{2014JCAP...05..023F}, and we assume a full overlap with LSST
            ($f_{\rm sky}=0.05$).
    \end{itemize}
    Figure \ref{fig:nz_spec} shows the redshift distributions for galaxies detected by these
    three experiments.
 
  \subsection{Forecasting formalism}\label{ssec:method.fisher}
    Our formalism will distinguish between two types of tracers of the density field:
    \begin{itemize}
      \item Spectroscopic: tracers whose redshift distribution is well known. This would
        correspond to tracers with good radial resolution such as a narrow redshift bin
        of spectroscopic sources or an intensity map in a narrow frequency band, as well
        as other tracers with a well-known window function, such as a CMB lensing map.
      \item Photometric: tracers whose redshift distribution is unknown or uncertain.
        This would correspond to e.g. a photometric-redshift bin, a radio continuum
        survey or a map of the Cosmic Infrared Background.
    \end{itemize}

    Let us start by considering a set of sky maps corresponding to a number of tracers,
    and let ${\sf a}$ be the corresponding vector of maps expressed in a given basis.
    In the following sections we will assume that ${\sf a}$ is stored in terms of
    spherical harmonic coefficients and that it takes the form
    ${\sf a}_{\ell m}=(p_{\ell m},s^1_{\ell m},...s^{N_s}_{\ell m})$,
    where $p_{\ell m}$ is a photometric tracer and $s^i_{\ell m}$ is a set of
    spectroscopic tracers. For the moment, however, we will keep the discussion general.
    
    Assuming that ${\sf a}$ is Gaussianly distributed with zero mean and covariance
    $\cov\equiv\langle {\sf a}{\sf a}^\dag\rangle$, its log-likelihood is given by:
    \begin{equation}\label{eq:like}
      \mathcal{L}\equiv-2\log p({\sf a})={\sf a}^\dag\cov^{-1}{\sf a}+\log(\det(2\pi\cov)).
    \end{equation}
    Now let $q_i$ be a set of parameters modelling $\cov$, including (but not limited to)
    the parameters describing the photometric redshift distribution. A maximum-likelihood
    estimator for $q_i$ can be defined by using an iterative Newton-Raphson method to
    minimize Eq. \ref{eq:like}. This is described in \cite{1998ApJ...499..555T,
    1998PhRvD..57.2117B,2013MNRAS.433.2857M}, and yields the iterative algorithm:
    \begin{align}\label{eq:nr_ml}
      &q_i^{n}=q_i^{n-1}+[\fsh^{-1}]_{ij}
      \left[{\sf a}^\dag\cov^{-1}\cov_{,j}\cov^{-1}{\sf a}-
        {\rm Tr}(\cov_{,j}\cov^{-1})\right],\\\nonumber
      &\fsh_{ij}\equiv\left\langle\frac{\partial^2\mathcal{L}}{\partial q_i\partial q_j}\right\rangle=
      {\rm Tr}\left(\cov^{-1}\cov_{,i}\cov^{-1}\cov_{,j}\right),
    \end{align}
    where, in Eq. \ref{eq:nr_ml} there is an implicit summation over $j$, the sub-index $_{,i}$
    implies differentiation with respect to $q_i$, $\fsh$ is the Fisher matrix, $q_i^{n}$
    is the $n-$th iteration of the solution for $q_i$ and the previous iteration $q_i^{n-1}$ is
    used to compute $\cov$ and $\cov_{,i}$ in the second term. Note that we have
    simplified a pure Newton-Raphson iteration by taking the ensemble average of the likelihood
    Hessian (i.e. the Fisher matrix). Furthermore, in the case where the likelihood is
    well-approximated by a Gaussian, $\fsh^{-1}$ is the covariance matrix of the $q_i$.
    Eq. \ref{eq:nr_ml} is the basis of the method proposed in \cite{2013MNRAS.433.2857M} (with a
    number of simplifications) and used in \cite{2017MNRAS.465.4118J} to constrain the redshift
    distribution of galaxies in the KiDS survey.

    In our case, we mainly care about the uncertainty in the redshift distribution parameters
    included in the $q_i$, and therefore we will simply estimate the Fisher matrix $\fsh$. In
    the case where ${\sf a}$ is a set of spherical harmonic coefficients with power spectrum
    $\langle{\sf a}_{\ell m}{\sf a}^\dag_{\ell' m'}\rangle=
    \delta_{\ell\ell'}\delta_{mm'}\cov_\ell$, $\fsh$ is given by
    \begin{equation}\label{eq:fisher}
      \fsh_{ij}=\sum_{\ell=2}^{\ell_{\rm max}}
      f_{\rm sky}(\ell+1/2)\,{\rm Tr}\left(\cov^{-1}_\ell\cov_{\ell,i}\cov^{-1}_\ell\cov_{\ell,j}\right),
    \end{equation}
    where we have approximated the effects of a partial sky coverage by scaling the number of
    independent modes per $\ell$ by the sky fraction $f_{\rm sky}$. The form of the power
    spectra $\cov_\ell$ for the different tracers considered in this work is given in Appendix
    \ref{app:cls}.
    
    As explicitly shown in Eq. \ref{eq:fisher}, smaller-scale modes carry a higher statistical
    weight (proportional to $\sim\ell$), and would in principle dominate the redshift
    distribution constraints. The smallest scales are, however, dominated by theoretical
    uncertainties from non-linearities in the evolution of the density field and the galaxy-halo
    connection, and therefore a multipole cutoff $\ell_{\rm max}$ must be used to contain
    the constraining power of systematics-dominated modes. In this paper we use a
    redshift-dependent cutoff defined as follows. Let $z$ be the mean redshift of a given
    redshift bin, and let $\sigma^2(k_*)$ be the variance of the linear density field
    at that redshift on modes with wavenumber $k<k_*$:
    \begin{equation}\label{eq:sthr}
      \sigma^2(k_*,z)\equiv\frac{1}{2\pi^2}\int_0^{k_*}dk\,k^2\,P_m(k,z).
    \end{equation}
    We then define the cutoff scale as $\ell_{\rm max}(z)=\chi(z)\,k_{\rm max}(z)$,
    where $k_{\rm max}(z)$ satisfies $\sigma(k_{\rm max},z)=\sigma_{\rm thr}$ for some
    choice of $\sigma_{\rm thr}$. In what follows we will use a fiducial threshold
    $\sigma_{\rm thr}=1$, corresponding to $k_{\rm max}(z=0)\simeq0.3\,h\,{\rm Mpc}^{-1}$,
    and we will study the dependence of our results on this choice. Besides this choice of
    $\ell_{\rm max}$, we will also impose a hard cutoff for all galaxy-survey and
    intensity-mapping tracers of $\ell<2000$ (thus, in reality,
    $\ell_{\rm max}={\rm min}(\chi k_{\rm max},2000)$).

\section{Results} \label{sec:results}
  \begin{figure*}
    \centering
    \includegraphics[width=0.49\textwidth]{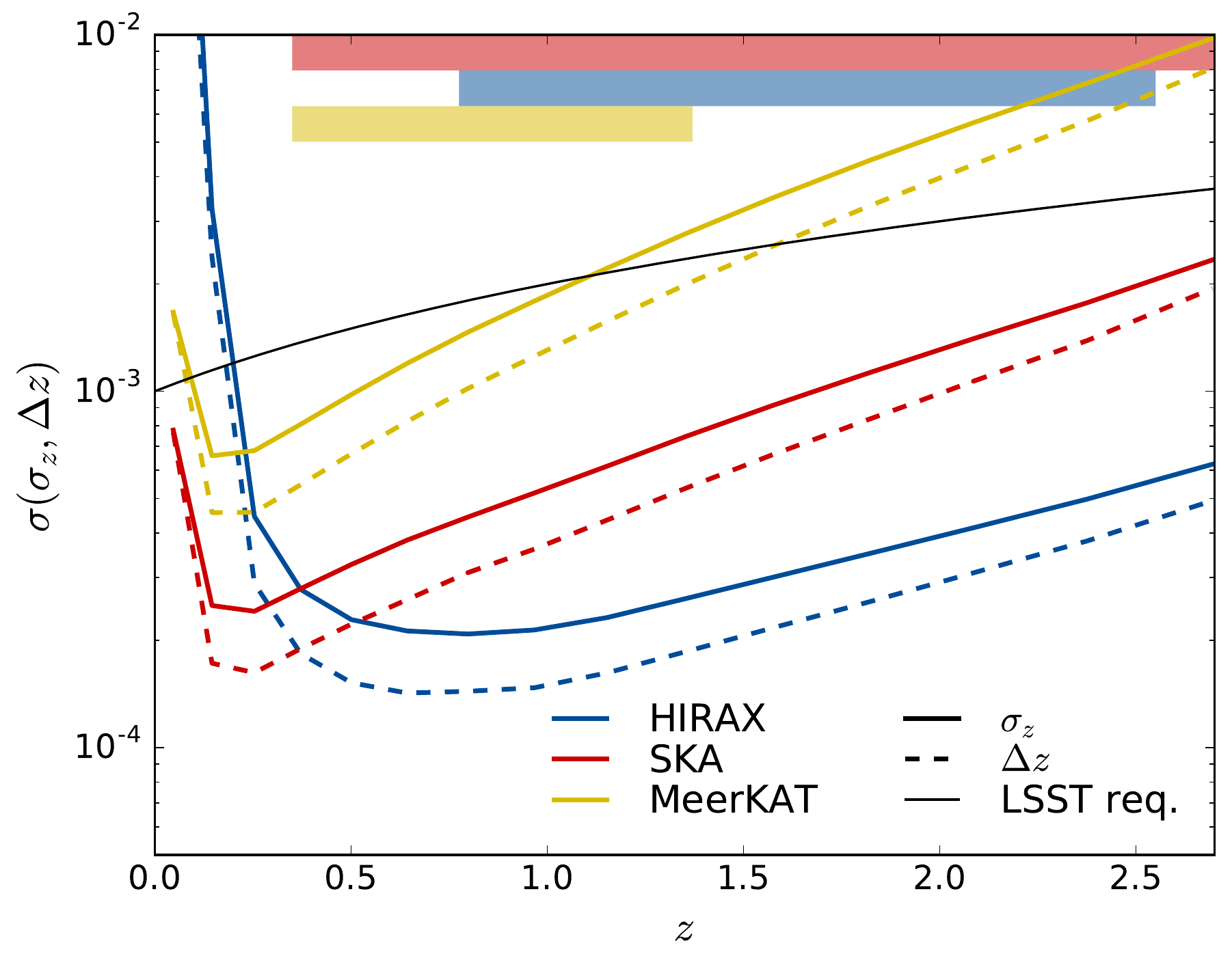}
    \includegraphics[width=0.49\textwidth]{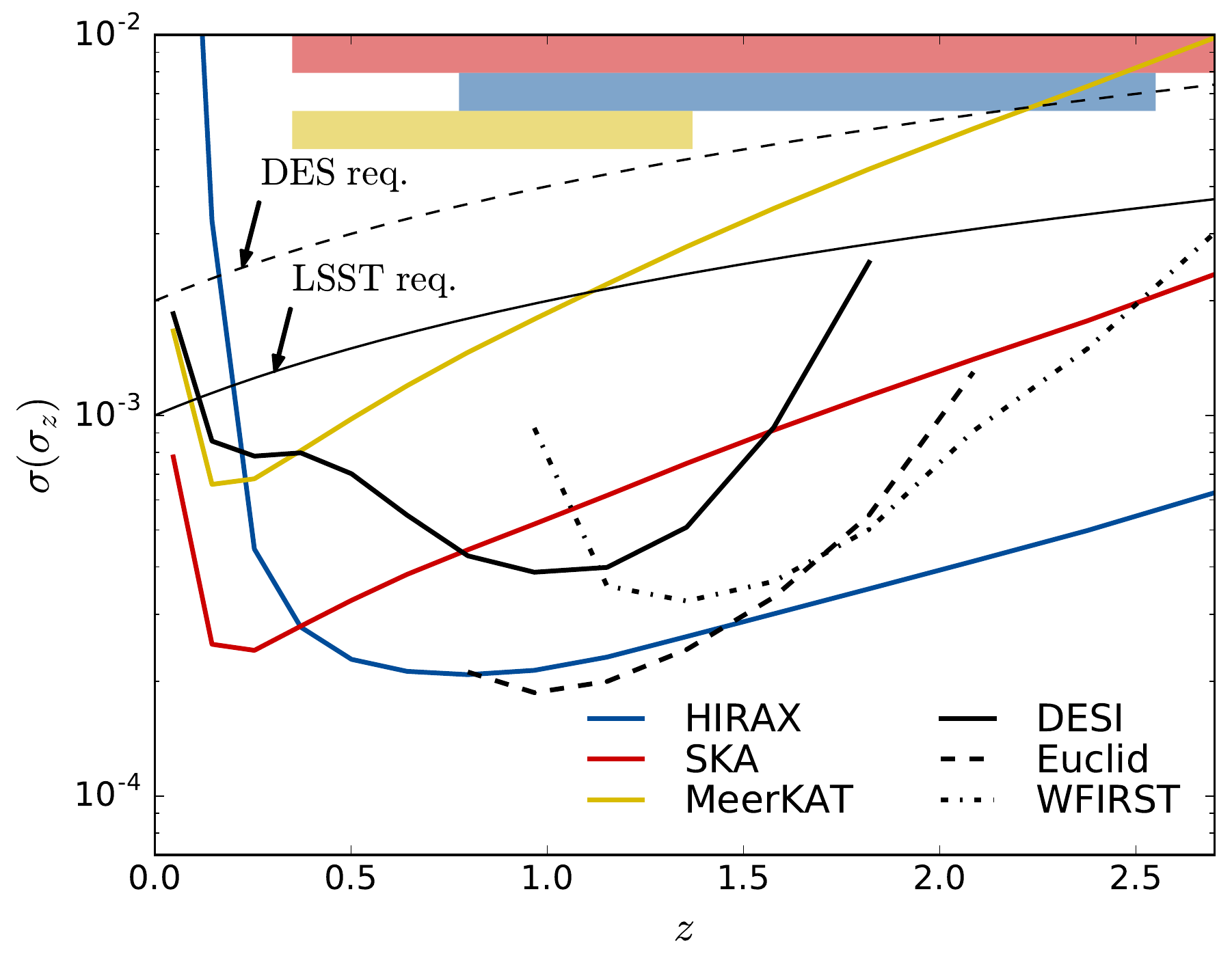}
    \caption{{\sl Left panel:} forecast 1$\sigma$ constraints on the photo-$z$ scatter
             $\sigma_z$ (solid lines) and bias $\Delta z$ (dashed lines) for the three
             IM experiments under consideration: HIRAX (blue), SKA (red) and
             MeerKAT (yellow). {\sl Right panel:} comparison of the previous three
             intensity mapping experiments, in terms of the forecast constraints on
             $\sigma_z$, with three future spectroscopic surveys: DESI (black solid),
             Euclid (black dashed) and WFIRST (black dot-dashed). In both panels, the
             thin solid line shows the photo-$z$ calibration requirement on both
             $\sigma_z$ and $\Delta z$ for LSST, with the corresponding requirement
             for DES shown as a thin dashed line in the right panel. The coloured bands
             in the upper part of all plots show the proposed frequency ranges
             for the three 21cm experiments (same color code). We have assumed using
             the UHF band for SKA and MeerKAT. The L-band would be able to cover all
             redshifts below $z\sim0.35$.}
    \label{fig:compare_spec}
  \end{figure*}
  In order to forecast for the ability of future experiments to constrain photometric
  redshift distributions, in the following sections we will use the formalism 
  described in Section \ref{ssec:method.fisher} with a data vector given by
  ${\sf a}_{\ell m}=(p_{\ell m},s^1_{\ell m},...,s^{N_s}_{\ell m})$, where $p$ is
  a photometric redshift bin and $s^i$ are a set of overlapping redshift bins for
  a spectroscopic tracer (either an intensity mapping experiment or a spectroscopic
  galaxy survey). The number $N_s$, width and redshift range of the spectroscopic
  redshift bins is chosen in order to adequately sample the changes in the photometric
  redshift distribution. We choose the redshift bin width to be $33\%$ of the
  photo-$z$ uncertainty $\sigma_z$, which governs the variability of the redshift
  distribution (i.e. each redshift interval of $\sigma_z$ is sampled in 3 points).
  In order to sample the tails of the distribution we then define the redshift 
  range of the set of spectroscopic bins as $[z_b^i-3\sigma_z,z_b^f+3\sigma_z]$,
  where $z_b^i$ and $z_b^f$ are the edges of the photometric redshift bin. The
  number of spectroscopic redshift bins $N_s$ is then defined in terms of these
  numbers.
  
  The model parameters $q_i$ in the following sections will be given by:
  \begin{itemize}
    \item All of the parameters needed to determine the redshift distribution
          ($\sigma_z$, $\Delta z$ or the amplitude $N(z)$ in different
          spectroscopic bins, depending on the case).
    \item Two overall clustering bias parameters, $b_{\rm p}$ and $b_{\rm s}$,
          corresponding to the bias of the photometric and spectroscopic tracers.
    \item We will also include two cosmological parameters: the fractional matter
          density $\Omega_M$ and the amplitude of scalar perturbations
          $A_{\rm s}$, in $q_i$ in order to account for the possible cosmology
          dependence of the results.
  \end{itemize}
  
  We will change this setup in Section \ref{ssec:results.cosmo}, where we will
  explore the impact of the achieved constraints on the photo-$z$ parameters on
  the final cosmological constraints. In this section
  ${\sf a}$ will correspond to the 15 photometric redshift bins for LSST, for
  both galaxy clustering and weak lensing (i.e. 30 sets of spherical harmonics).
  Likewise $q_i$ will contain the cosmological parameters
  $(\omega_c,\omega_b,h,w_0,w_a,A_s,n_s,\tau_{\rm reio})$ as well as all the
  baseline photo-$z$ parameters ($\Delta z$ and $\sigma_z$ for all redshift
  bins), with priors corresponding to the constraints found in the preceding
  sections.

  \subsection{Baseline forecasts} \label{ssec:results.baseline}
    \begin{figure}
      \centering
      \includegraphics[width=0.49\textwidth]{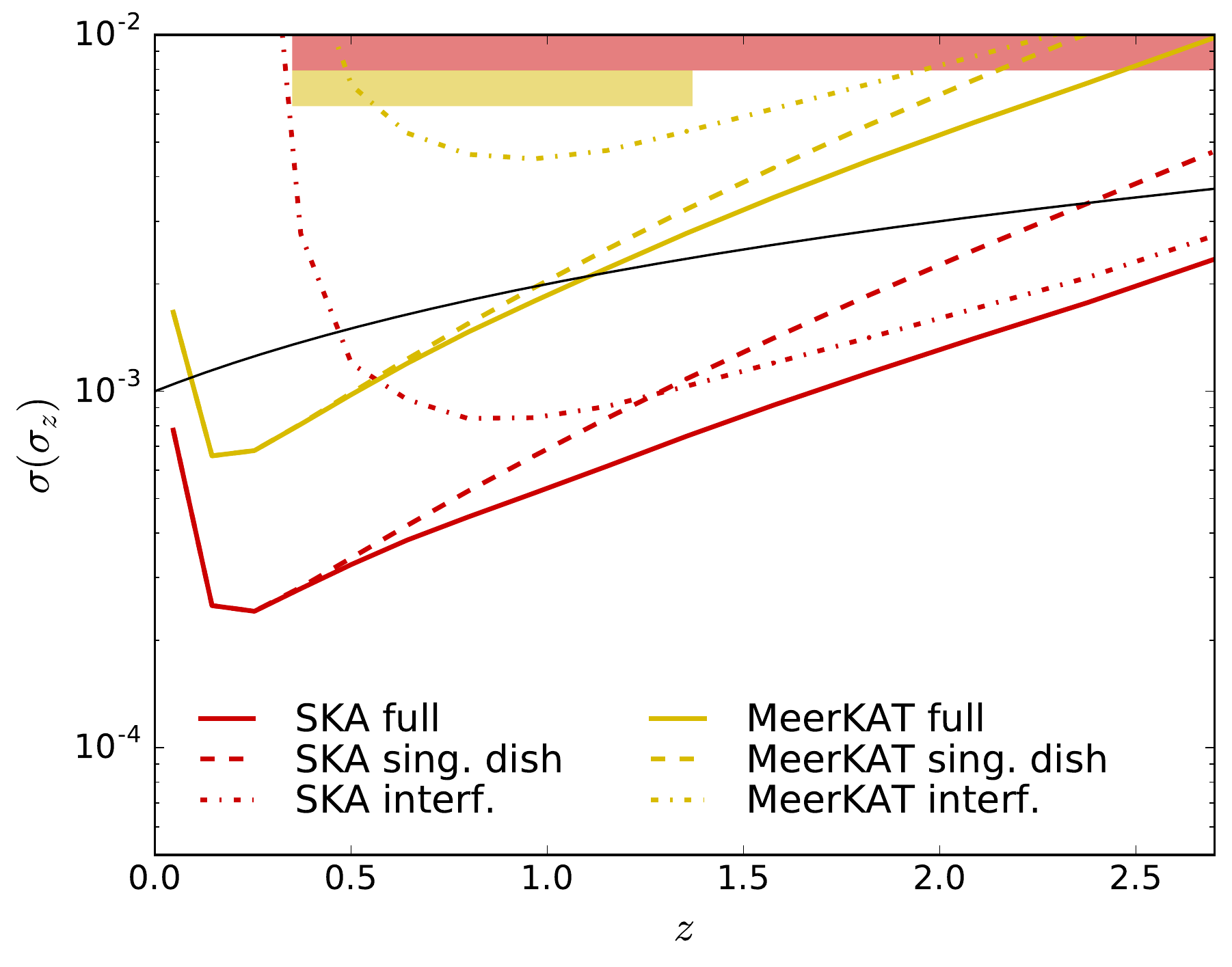}
      \caption{Forecast constraints on the LSST photo-$z$ scatter $\sigma_z$ for SKA (red)
               and MeerKAT (yellow) assuming only interferometric observations
               (dot-dashed lines), single-dish observations (dashed lines) and both
               simultaneously (solid lines).}
      \label{fig:compare_if_sd}
    \end{figure}
    Using the formalism described above, and in the simplified scenario of Gaussian photo-$z$s,
    we present, in the left panel of Figure \ref{fig:compare_spec}, the forecast constraints on the 
    photo-$z$ bias ($\Delta z$) and variance ($\sigma_z$) for the key intensity mapping experiments
    introduced in Section \ref{ssec:method.imap}. In this and all subsequent plots, the thin black
    solid line shows the LSST requirement of $\sigma(\Delta z,\sigma_z)\lesssim10^{-3}(1+z)$
    \cite{2015APh....63...81N,2014ApJ...780..185D}, while the thin dashed line corresponds to the
    nominal requirement for the Dark Energy Survey ($2\times10^{-3}$) \cite{2005astro.ph.10346T}.
    The coloured bars in these and all subsequent plots show the redshift ranges corresponding to
    the proposed frequency bands of the three IM experiments explored here.
    
    Two key features must be noted in this figure: first, the uncertainties grow steeply at low
    redshifts. This is due to the reduced number of modes available in that regime, associated
    with the smaller comoving volume and the impact of non-linearities on lower values of $k$.
    The latter effect is especially severe for HIRAX, given its inability to measure angular modes
    larger than its beam size. Note, however, that this regime lies outside the proposed frequency
    ranges for both HIRAX ($0.8\lesssim z\lesssim2.5$) and SKA ($0.35\lesssim z\lesssim3$).
    \begin{figure}
      \centering
      \includegraphics[width=0.49\textwidth]{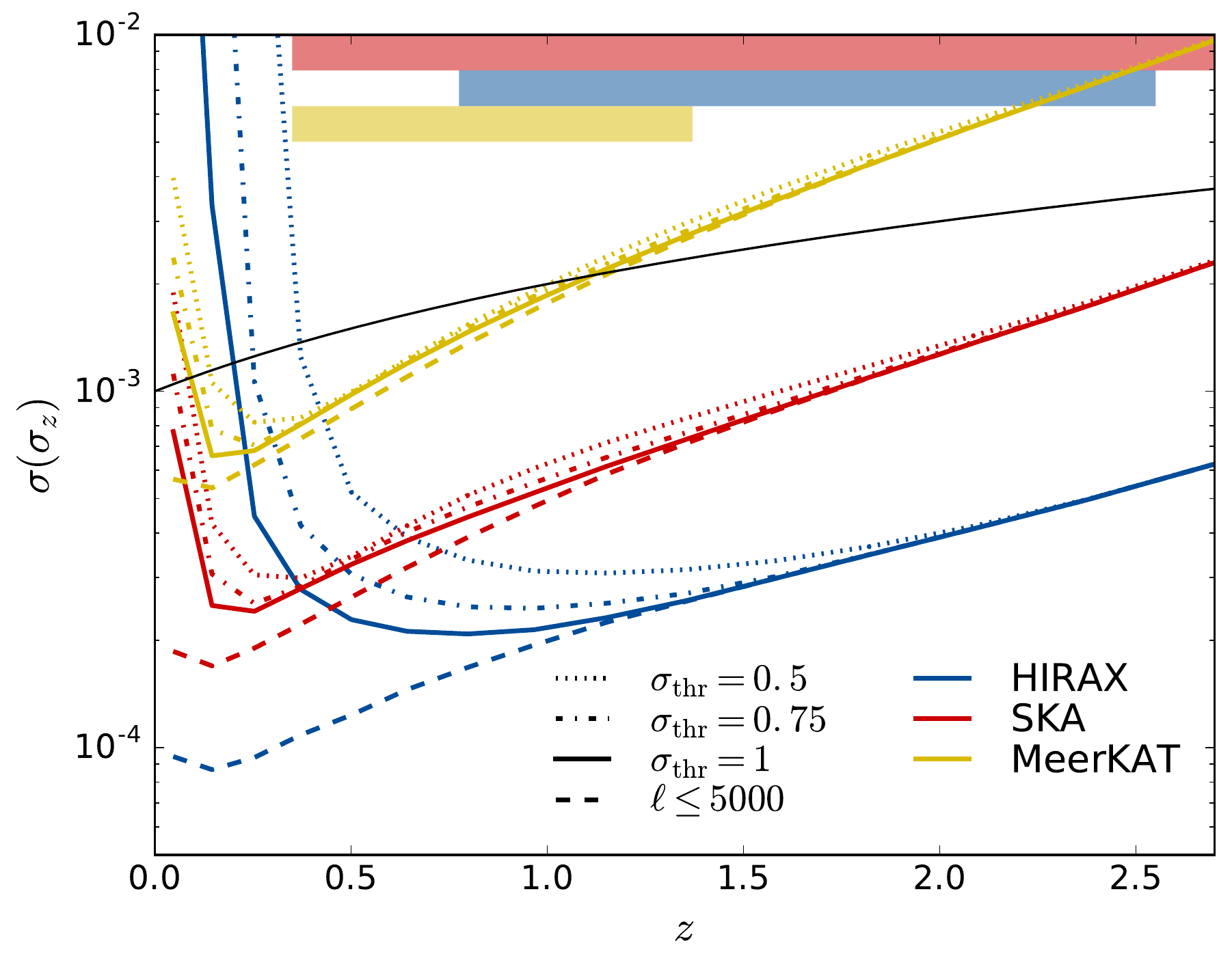}
      \caption{Dependence of the constraints on the LSST photo-$z$ scatter $\sigma_z$ on the
               overdensity variance threshold used to filter out non-linear scales (see 
               Equation \ref{eq:sthr}). The results are shown for HIRAX (blue), SKA (red)
               and MeerKAT (yellow). The fiducial value used in this analysis is shown as
               solid lines, while a more optimistic scenario where all multipoles up to
               $\ell=5000$ are included at all redshifts is shown as a dashed line. This case
               would mostly benefit interferometric observations, given their higher
               sensitivity on small angular scales.}
      \label{fig:compare_nlin}
    \end{figure}

    Secondly, the ratio between $\sigma(\sigma_z)$ and $\sigma(\Delta z)$ stays roughly constant
    ($\sim1.4$). This is compatible with the expected ratio between the uncertainties associated
    to the maximum-likelihood estimates of the mean and standard deviation of a Gaussian
    distribution from a finite number of samples ($\sigma(\sigma)/\sigma(\mu)=\sqrt{2}$).
    This result holds for most of the cases explored here (see Section
    \ref{ssec:results.foregrounds} for an exception), and thus we have omitted the curves for
    $\sigma(\Delta z)$ in most of the subsequent figures.
    
    The right panel of Fig. \ref{fig:compare_spec} compares the constraints achievable by IM
    experiments with those forecast for the spectroscopic surveys described in Section
    \ref{ssec:method.spec}. We see that both SKA and HIRAX would be able to satisfy the LSST
    requirements over the redshift range of interest. The SKA precursor MeerKAT would fall short
    except at low redshifts. However, the shorter-term timeline of MeerKAT (2018 onward)
    would make it an ideal experiment to prove the viability of this technique in cross-correlation
    with the Dark Energy Survey (DES) \cite{2005astro.ph.10346T}, particularly in the light of the
    proposed intensity mapping surveys \cite{2017MNRAS.466.2780F} targeting a full overlap with
    DES\footnote{Note that the photo-$z$ calibration requriements, defined in terms of the
    degradation of the final constraints, should be less stringent for DES}.
    
    As discussed in \cite{2015ApJ...803...21B}, the dish size of SKA is not ideal for cosmological
    observations, since it is not large/small enough to resolve the angular BAO scale sufficiently
    well in either single-dish or interferometric modes, although single-dish observations are able
    to address important science cases such as primordial non-Gaussianity
    \cite{2013PhRvL.111q1302C,2015PhRvD..92f3525A}. Small scales carry a larger statistical weight,
    however, and it is not clear that a single-dish strategy would also be ideal for the purposes
    of photo-$z$ calibration. This is explored in Figure \ref{fig:compare_if_sd}, which shows the
    constraints on $\sigma_z$ achievable with single-dish (dashed lines) and interferometric
    (dash-dotted lines) observations for SKA (red) and MeerKAT (yellow). The constraints from a
    joint auto- and cross-correlation analysis are shown as solid lines, and correspond to the
    results reported here. We see that, in the case of SKA, the single-dish mode outperforms the
    interferometer up to $z\sim1.4$, when a sufficiently large number of usable modes enter the
    regime probed by the latter. This suggests that, if simultaneous single-dish and
    interferometric observations proved to be unfeasible, the photo-$z$ calibration
    requirements could still be met by using either mode in different redshift ranges.
    
    The performance of this method at low redshift depends crucially on the prescription used to
    isolate the effect of non-linearities. Here we have done this in terms of the threshold
    rms variance $\sigma_{\rm thr}$ defined in Eq. \ref{eq:sthr} for a fiducial value of
    $\sigma_{\rm thr}=1$, corresponding to $k_{\rm max}\sim0.3\,{\rm Mpc}^{-1}\,h$ at $z=0$.
    Figure \ref{fig:compare_nlin} shows the result of relaxing or tightening this criterion.
    The effect on SKA and MeerKAT is only moderate, since these experiments gather most of their
    sensitivity from the large, linear scales in auto-correlation mode. HIRAX, on the other hand,
    loses sensitivity more rapidly as the scale of non-linearities removes a larger fraction 
    of the available modes. Nevertheless, even for $\sigma_{\rm thr}=0.5$ (corresponding to
    $k_{\rm max}=0.1\,{\rm Mpc}^{-1}\,h$ at $z=0$) the LSST calibration requirements are satisfied
    in the redshift range corresponding to the HIRAX frequency band.
    
  \subsection{Dependence on experimental parameters} \label{ssec:results.params}
    \begin{figure}
      \centering
      \includegraphics[width=0.49\textwidth]{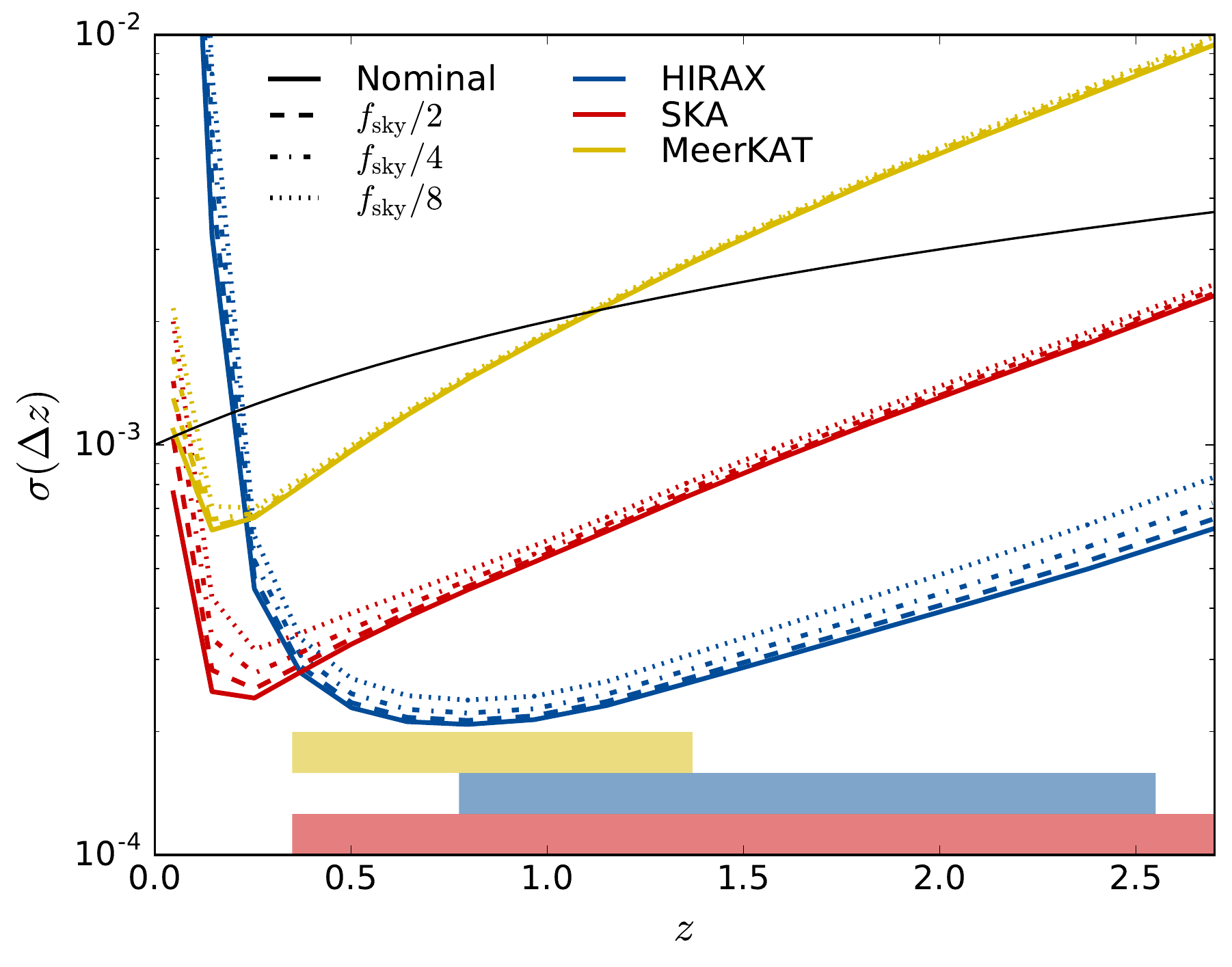}
      \caption{Dependence of the forecast constraints on the LSST photo-$z$ scatter $\sigma_z$
               on the overlap sky fraction for constant total observation time. In all cases
               the constraints are mostly insensitive to the trade-off between sky area and noise
               level, although larger areas are marginally preferred, which reflects the fact that
               these measurements are mostly dominated by cosmic variance and not noise.}
      \label{fig:compare_fsky}
    \end{figure}
    We have so far quantified the potential of currently-proposed experimental configurations
    for photo-$z$ calibration. The aim of this section is to identify the optimal instrumental
    specifications for this task.
    \begin{figure}
      \centering
      \includegraphics[width=0.49\textwidth]{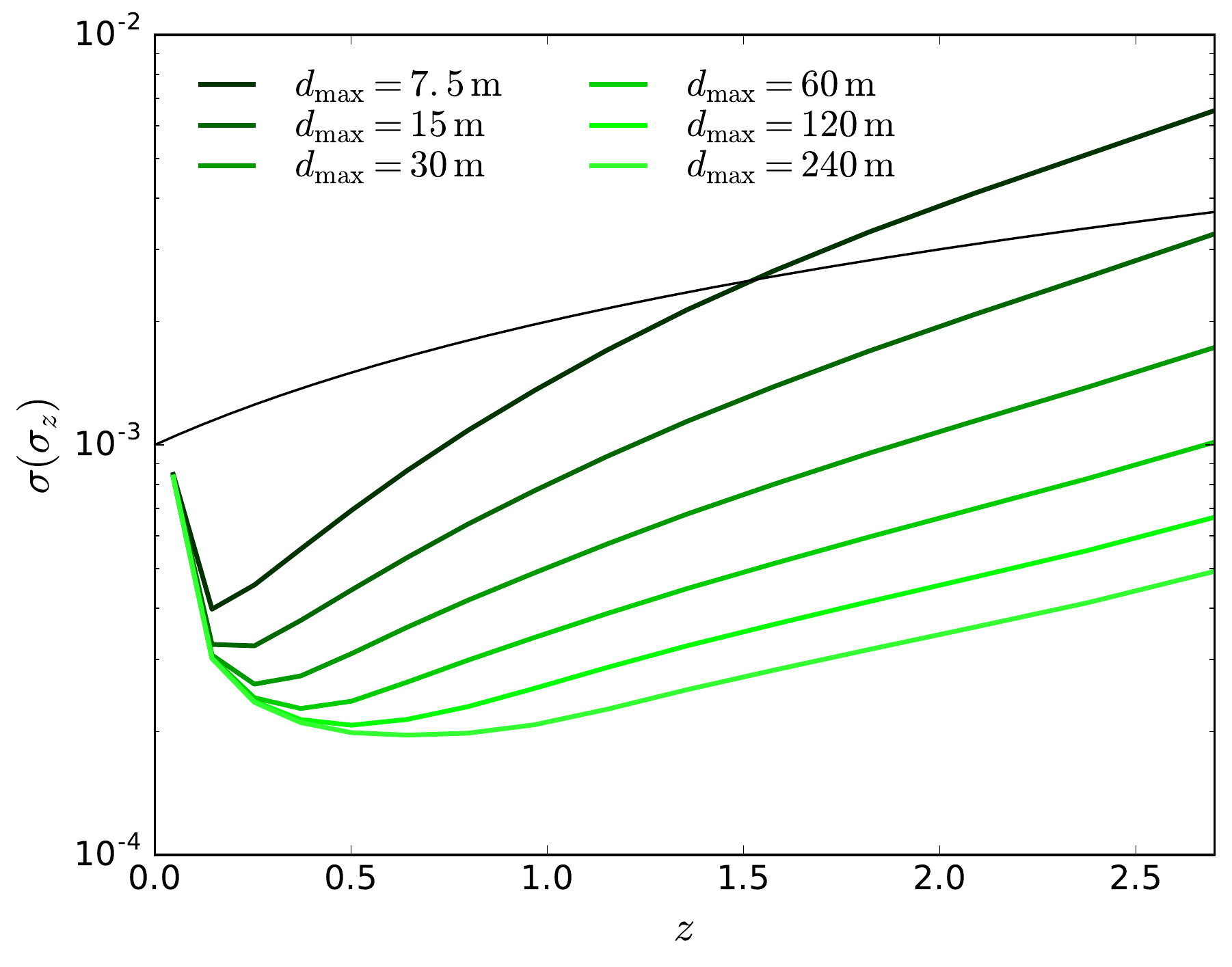}
      \includegraphics[width=0.49\textwidth]{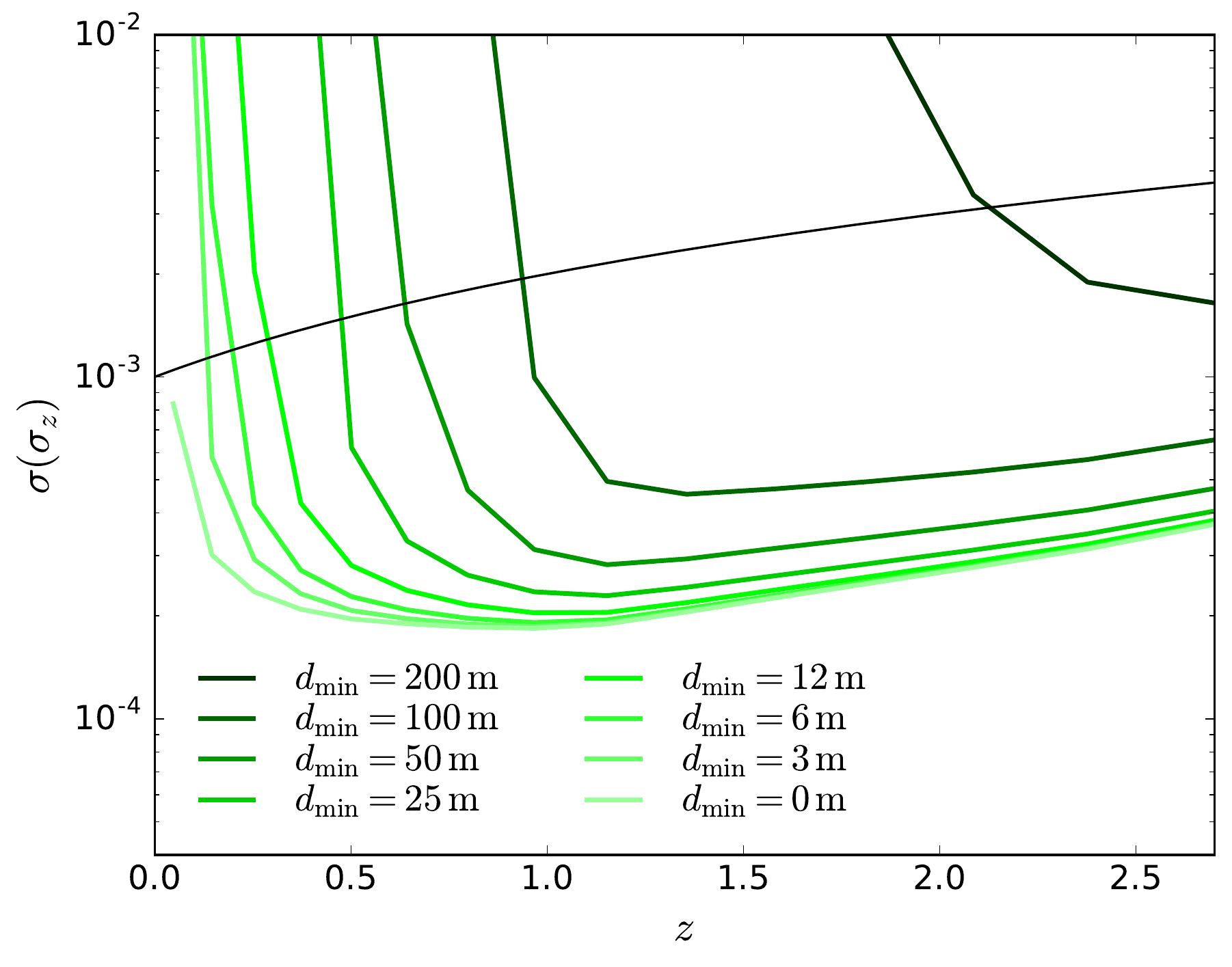}
      \caption{{\sl Upper panel:} dependence of the forecast constraints on the LSST photo-$z$
               scatter $\sigma_z$ on the dish size for single-dish IM observations. A dish size
               of at least $\sim15{\rm m}$ is needed to match the LSST requirements throughout
               the whole redshift range. {\sl Lower panel:} dependence on the minimum baseline
               for interferometers. Baselines of at most $\sim12{\rm m}$ are needed to
               successfully constrain photo-$z$ systematics below $z\sim0.5$.}
      \label{fig:compare_baselines}
    \end{figure}
    
    We start by exploring the balance between noise level and sky fraction, varying the total
    overlapping area of the three experiments explored in the previous section (SKA, MeerKAT and
    HIRAX) keeping the total observation time fixed. The result is presented in Figure
    \ref{fig:compare_fsky}, which shows the achievable constraints on $\sigma_z$ when reducing
    the sky area by successive factors of 2. We find that, although the results are almost
    insensitive to the reduction in $f_{\rm sky}$, larger sky areas are always preferred, a reflection
    of the fact that the constraints are dominated by cosmic variance rather than noise.
    
    As we have discussed in the previous section, the key drawback of single-dish experiments
    is their inability to access angular scales smaller than the beam size (with their higher
    statistical weight). Conversely, interferometers are unable to cover scales larger than
    that probed by their smallest baseline, and therefore they have access to a limited number
    of reliable (i.e. mildly non-linear) modes. Using the generic instrument parametrization
    given by Eq. \ref{eq:exp_generic}, and the fiducial parameters corresponding to HIRAX,
    Figure \ref{fig:compare_baselines} explores these issues.
    
    The upper panel shows the constraints achievable by a single-dish experiment for different dish
    sizes. Dishes of at least 15m (corresponding to the SKA case) are needed in order to achieve
    the LSST requirements at all redshifts, while a $\sim100{\rm m}$ dish (e.g. such as the Green
    Bank Telescope \cite{2013ApJ...763L..20M}) would be able to achieve constraints similar to those
    of HIRAX. The largest currently planned experiment is FAST \cite{2016ASPC..502...41B}, with a
    dish size of 500m.
    
    The lower panel of the figure shows the performance of an interferometer as a function of the
    minimum baseline $d_{\rm min}$. In this case the main effect is the loss, at lower redshifts,
    of the large, mildly non-linear scales. A maximum baseline of at most 12m is needed in order
    to calibrate redshift distributions below $z\sim0.5$ with sufficient accuracy.

  \subsection{Foregrounds} \label{ssec:results.foregrounds}
    \begin{figure}
      \centering
      \includegraphics[width=0.49\textwidth]{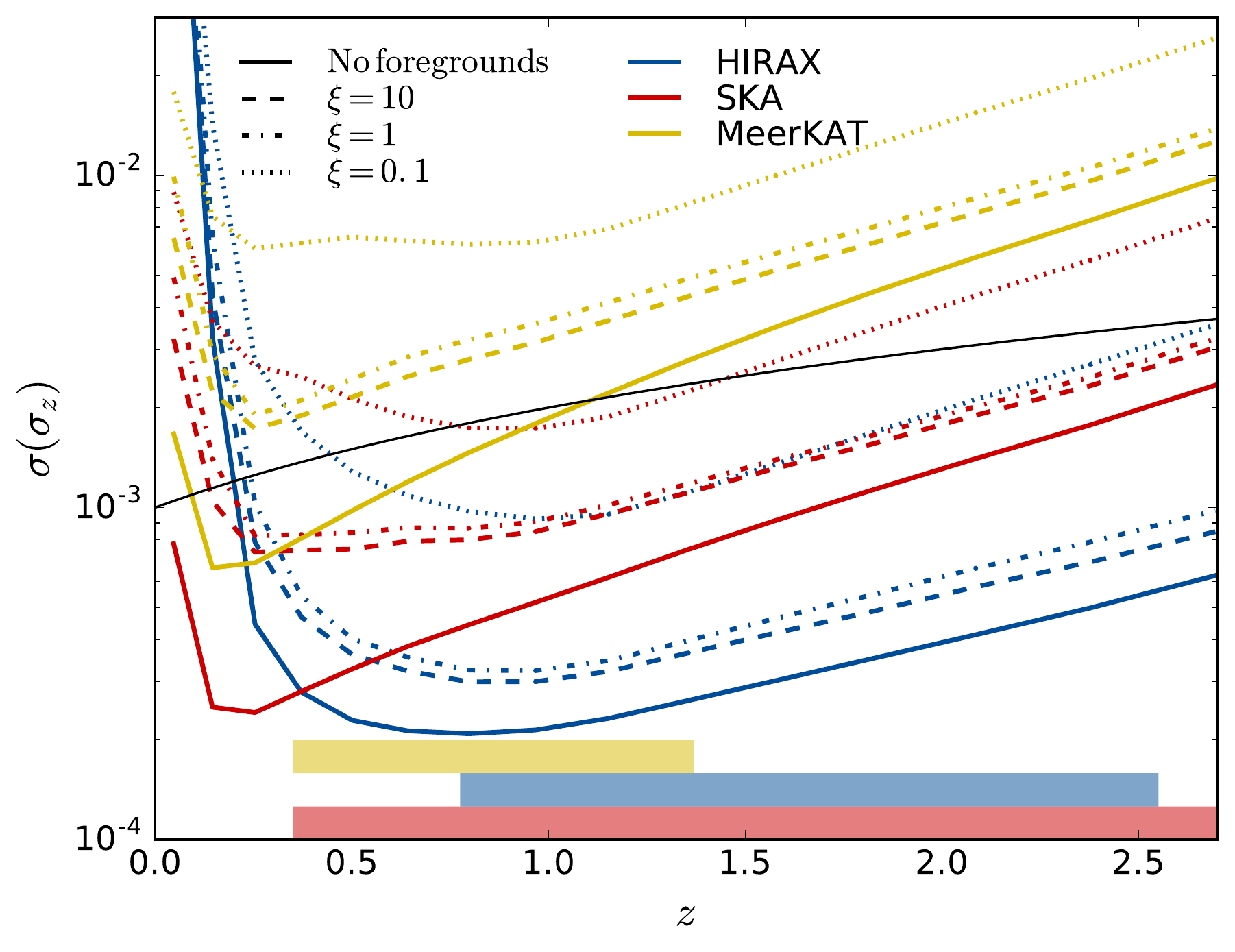}
      \includegraphics[width=0.49\textwidth]{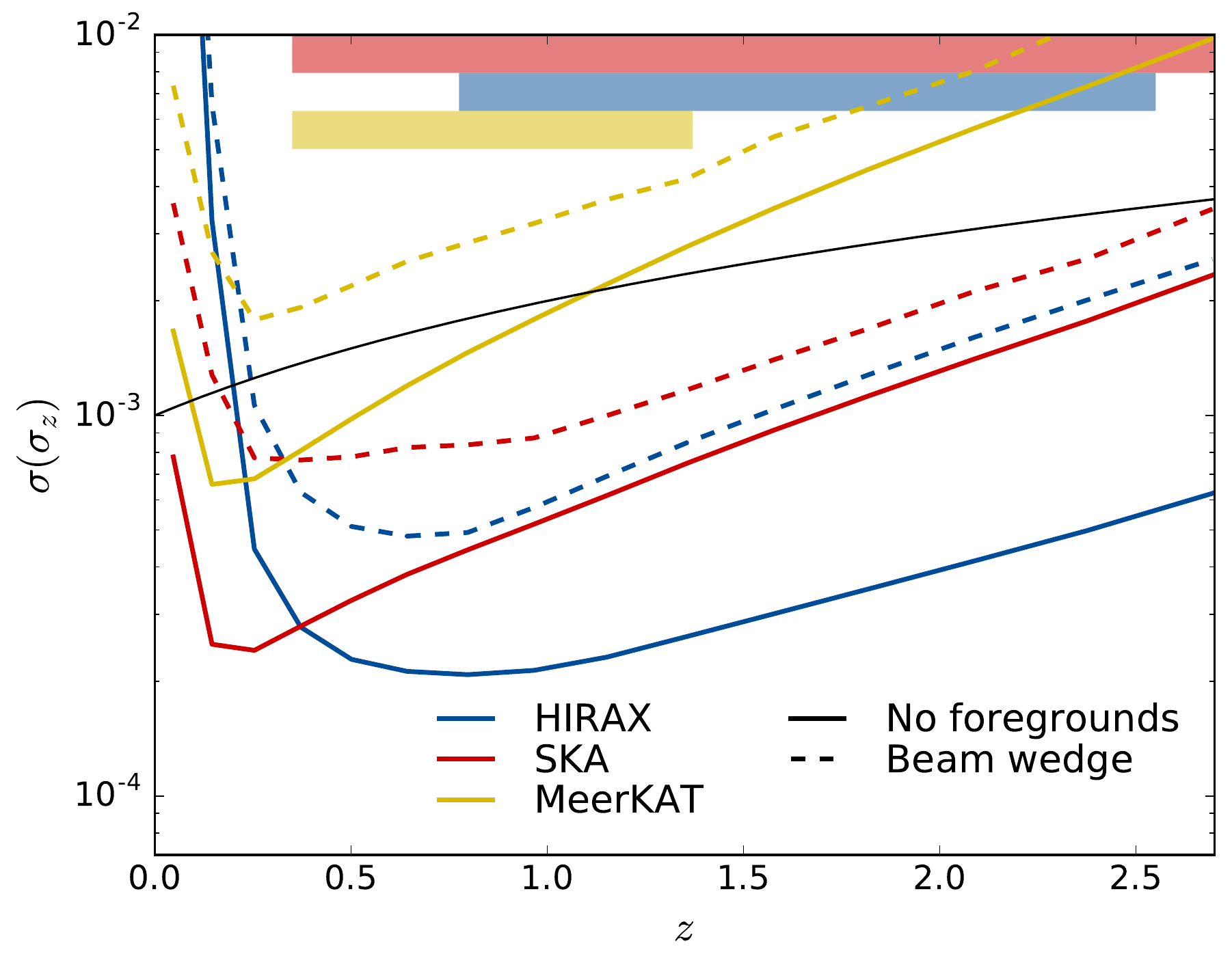}
      \caption{{\sl Upper panel:} dependence of the forecast constraints on the LSST photo-$z$
               scatter on the frequency correlation length of foreground residuals (a measure
               of the coherence of these residuals across frequencies). Correlation lengths
               above $\xi\sim0.1$ (corresponding to radial scales
               $k_\parallel\sim0.01\,h\,{\rm Mpc}^{-1}$) are necessary to limit the
               degradation of the photo-$z$ calibration.
               {\sl Lower panel:} impact of the foregrounds wedge on the final constraints on
               the photo-$z$ scatter.}
      \label{fig:fg_xi}
    \end{figure}
    \begin{figure*}
      \centering
      \includegraphics[width=0.98\textwidth]{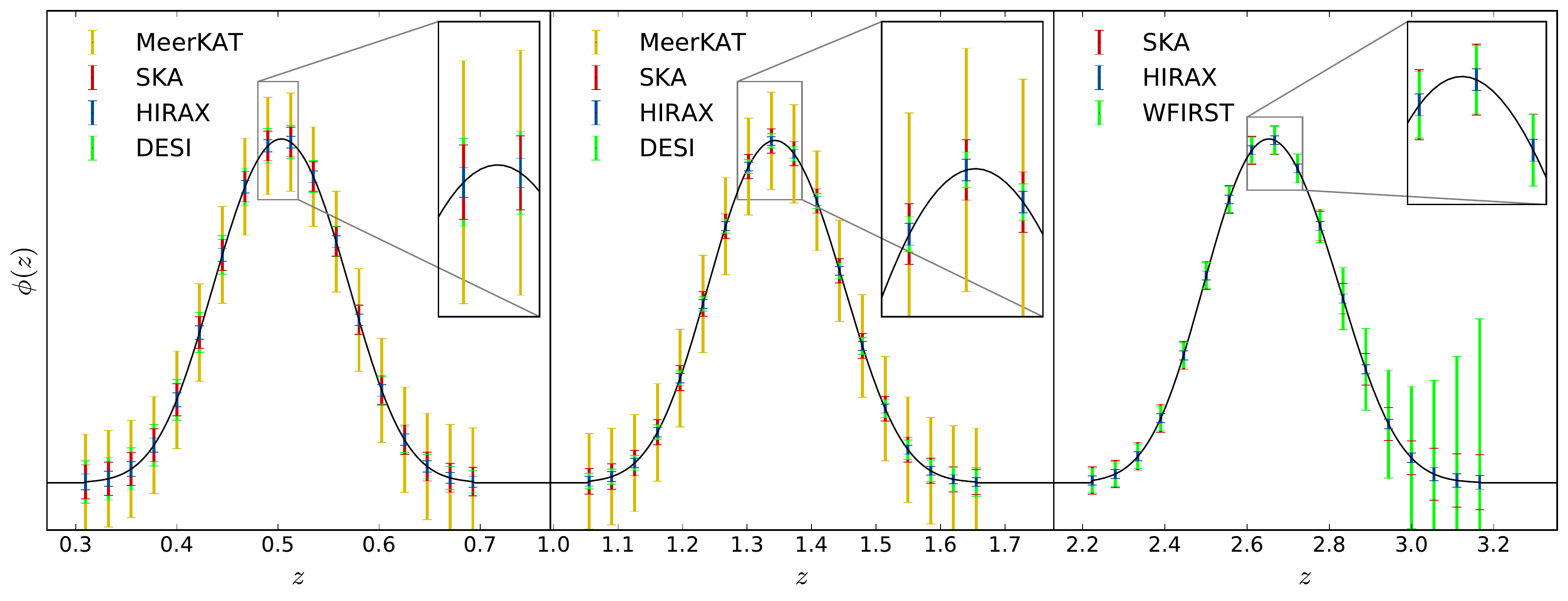}
      \caption{Non-parametric reconstruction of the redshift distribution for four different
               redshift bins. The points with error bars show the $1\sigma$ constraints on
               $\phi(z)$ achievable with HIRAX (blue), SKA (red) and MeerKAT (yellow), as well
               as two spectroscopic surveys (green), DESI and WFIRST, at low and high redshifts
               respectively. Note that, given the proposed frequency band for HIRAX, it would
               not be able to calibrate the low and high-redshift bins shown here.}
      \label{fig:compare_nonpar}
    \end{figure*}
    As we discussed in Section \ref{sssec:method.imap.foregrounds}, the main effect of radio
    foregrounds for 21cm observations is to make large radial scales (i.e. modes with
    $k_\parallel$ smaller than some $k_\parallel^{\rm FG}$) useless. We have parametrised
    this by introducing an extra component corresponding to foreground residuals characterised
    by an amplitude $A_{\rm FG}$ and a frequency correlation length $\xi$ (see Section
    \ref{sssec:method.imap.foregrounds} and Eq. \ref{eq:fgcl} for a full description). We set
    $A_{\rm FG}$ to a value large enough to dominate the emission on radial scales larger
    than the comoving length corresponding to $\xi$ (see Eq. \ref{eq:fgkmin}), and study
    the final constraints as a function of $\xi$.
    
    Figure \ref{fig:fg_xi} shows the result of this analysis: while sufficiently coherent
    foregrounds ($\xi\gtrsim1$) do not qualitatively modify the final results for the three 
    experiments under consideration, correlation lengths smaller than $\xi\sim0.1$ would
    result in a fast degradation of the performance for photo-$z$ calibration. In particular,
    the associated loss of $k$-space volume would prevent MeerKAT and SKA from reaching the
    calibration requirements for DES and LSST. The peformance of HIRAX would also be severely
    compromised by foreground contamination, although it would still be able to reach the
    required constraints within its proposed frequency range.
    
    These results pose the question of how uncorrelated we can expect foreground residual to be.
    For reference, raw foreground components are constrained to have correlation lengths of
    $\xi\sim1-10$ \cite{2005ApJ...625..575S}. On the other hand, more complicated residuals
    arising from leaked polarised synchrotron would exhibit a much richer frequency structure
    caused by Faraday rotation, with correlation lengths $\xi\sim0.1$ at high frequencies 
    ($\sim800\,{\rm MHz}$) decreasing for longer wavelengths
    \cite{2010MNRAS.409.1647J,2015PhRvD..91h3514S,2015MNRAS.447..400A}. An exquisite
    instrumental calibration will therefore be necessary in order to optimise the scientific
    output of 21cm experiments. It is worth noting that the comoving scale 
    corresponding to $\xi\sim0.1$ ($k_\parallel\sim0.01\,h\,{\rm Mpc}^{-1}$ at $z\sim1$)
    is similar to the cut suggested by \cite{2015PhRvD..91h3514S}, although the expontial form
    assumed here for the power spectrum of the foreground residuals (Eq. \ref{eq:fgcl}) extends
    the degrading effect into larger wavenumbers.
    
    A further complication comes in the form of the so-called ``foreground wedge'', produced
    by the long time-delay contribution of foregrounds from antennas with far side-lobe
    responses \cite{2010ApJ...724..526D,2012ApJ...745..176V,2014PhRvD..90b3019L,
    2016MNRAS.456.3142S}. As proven in \cite{2012ApJ...756..165P,2014ApJ...782...66P}, this
    effect makes the region of $k$-space defined by:
    \begin{equation}
      k^{\rm hor}_\parallel<\frac{\chi(z)H(z)}{c(1+z)}k_\perp
    \end{equation}
    liable to foreground contamination. This is the so-called ``horizon'' wedge, and
    corresponds to the case where foreground contamination can be caused by the
    sources in the horizon picked up by very far sidelobes. Under optimistic assumptions,
    however, we can consider the case where this effect extends only up to foreground
    sources located in the outer fringes of the primary beam, in which case the size
    of the wedge is reduced to the so-called primary-beam wedge \cite{2014ApJ...782...66P},
    given by $k_\parallel^{\rm pb}=\sin(\theta_{\rm FWHM}/2)k^{\rm hor}_\parallel$.
    The lower panel of Figure \ref{fig:fg_xi} shows that the LSST photo-$z$ calibration
    requirements are still met after accounting for the loss of $k$-space to the primary
    beam wedge.
    
  \subsection{Generalized redshift distributions} \label{ssec:results.outliers}
    Even though the simple parametrization of photo-$z$ systematic uncertainties in terms of
    $\Delta z$ and $\sigma_z$ allows us to easily compare the performance of different
    experiments in terms of photo-$z$ calibration, in a realistic scenario we would like
    to calibrate generic redshift distributions without assuming a particular parametrization.
    
    This is typically done by promoting the amplitude of the redshift distribution of the
    photometric sample in each narrow redshift interval to a free parameter that can be constrained
    from the cross-correlation with the spectroscopic survey. In this section we explore this
    scenario for the same redshift bins considered in the previous sections.
    
    For this we use the method proposed by \cite{2013MNRAS.433.2857M}. In essence, this method
    is equivalent to the formalism outlined in Section \ref{ssec:method.fisher}, where the free
    parameters $q_i$ considered are the amplitudes of the photometric redshift distribution.
    The method is further simplified in \cite{2013MNRAS.433.2857M} to make it applicable to the
    analysis of real data using the following approximations:
    \begin{enumerate}
      \item All power spectra are computed using the Limber approximation. This implies (among
            other things) that all cross-correlations between non-overlapping redshift bins are
            neglected.
      \item The contributions from RSDs and magnification bias are not included in the
            model for the angular power spectra.
      \item No marginalization over cosmological or other nuisance parameters is carried out.
      \item The auto-correlation of photometric sources does not contain information about 
            their redshift distribution.
    \end{enumerate}
    We have adopted these same assumptions here to simplify the discussion.
    
    Figure \ref{fig:compare_nonpar} shows the constraints achievable by different IM experiments
    and spectroscopic surveys on the generalized form of the redshift distribution for three
    photometric redshift bins centered around $z\sim0.5,\,1.35$ and 2.6\footnote{Note that
    the first and last bins would lie outside the proposed frequency range of HIRAX}. The
    constraining power displayed in this figure matches the results shown in the right panel
    of Figure \ref{fig:compare_spec}.
    
    Note that the uncertainties on the amplitude of the redshift distribution found this way
    can be translated into uncertainties on the two parameters $\Delta z$ and $\sigma_z$ used
    to characterize this distribution in the previous section by performing a simple 2-parameter
    likelihood analysis for the model in Eqs. \ref{eq:phz_dist} and \ref{eq:photoz_gaussian}.
    We find that, using this procedure, we recover constraints on $\Delta z$ and $\sigma_z$ that
    are a factor $\sim2-3$ worse than in the optimal case. This can be understood in terms of
    the simplifying assumptions adopted here, such as neglecting the information encoded in
    cross-bin correlations and in the auto-correlation of the photometric sample. In all cases,
    however, we recover the same relative performance between different experiments in terms
    of $\sigma(\sigma_z,\Delta z)$.
    
  \subsection{Impact on cosmological constraints} \label{ssec:results.cosmo}
    In order to study the impact of photo-$z$ calibration on the final cosmological constraints
    we have run a Fisher matrix analysis using the formalism described in Section
    \ref{ssec:method.fisher} with the specifications for LSST described in Section
    \ref{ssec:method.phz}. In this case we consider a set of 54 free parameters:
    \begin{itemize}
      \item 3 nuisance parameters for each of the 15 redshift bins: the balaxy bias $b$, the
            photo-$z$ bias $\Delta z$ and the photo-$z$ scatter $\sigma_z$.
      \item 9 cosmological parameters: the fractional density of cold dark matter $\Omega_c\,h^2$,
            the fractional density in baryons $\Omega_b\,h^2$, the normalized local expansion rate
            $h$, the amplitude and tilt of primordial scalar perturbations $A_s$ and $n_s$, the
            optical depth to reionization $\tau$, the equation of state of dark energy in the
            $w_0$-$w_a$ parametrization and the sum of neutrino masses $\sum m_\nu$.
    \end{itemize}
    In order to pin down the early-universe parameters, we also include information from a
    hypothetical ground-based Stage-4 CMB experiment \cite{2016arXiv161002743A} using 
    the specifications assumed in \cite{2016arXiv161110269C} and complemented by Planck at
    low multipoles.
    
    \begin{figure}
      \centering
      \includegraphics[width=0.49\textwidth]{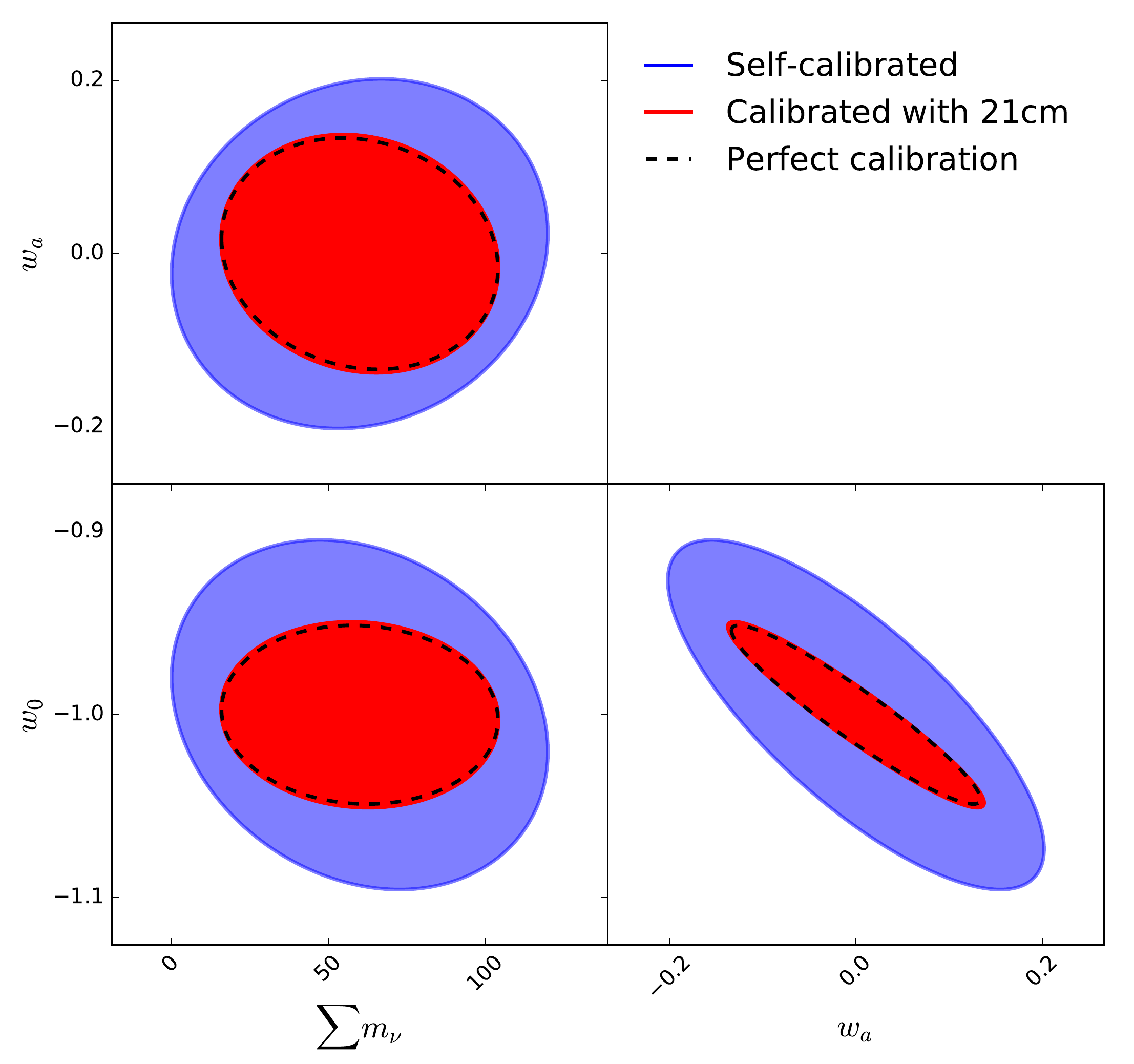}
      \caption{Constraints on the equation of state of dark energy and the sum of neutrino
               masses in the absence of external photo-$z$ calibration (blue ellipses), with
               redshift distributions calibrated through cross-correlation with a HIRAX-like
               21cm experiment (red ellipses) and in the case of perfect calibration (dashed
               lines). The constraints also include early-universe information from a Stage-4
               CMB experiment.}
      \label{fig:constraints}
    \end{figure}
    For the photo-$z$ nuisance parameters we then add Gaussian priors corresponding to the
    $1\sigma$ uncertainties on $\sigma_z$ and $\Delta z$ found using the procedure described
    in the previous sections for the different experiments considered in this paper.    
    The results of this exercise are displayed in Figure \ref{fig:constraints}, which shows the
    constraints on the most relevant late-universe parameters: the dark energy equation of state
    parameters, $w_0$ and $w_a$, and the sum of neutrino masses. The results shown correspond to
    the $1\sigma$ contours in the absence of photo-$z$ uncertainties (dashed black line), with
    photo-$z$ systematics constrained through cross-correlation with a 21cm experiment, in this
    case HIRAX (red ellipse) and in the absence of external data for photo-$z$ calibration
    (blue ellipse).
    
    It is important to stress that the overall constraints on these parameters forecast here
    depend heavily on the survey specifications assumed (e.g. photo-$z$ model and uncertainties),
    as well as the scales included in the analysis, a point where we have tried to be very
    conservative. Thus, the results shown in Figure \ref{fig:constraints} should not be
    taken to represent the final constraints achievable by LSST. The main result shown in this
    figure is the relative improvement on the final constraints after photo-$z$ calibration,
    which should be more robust to these considerations.
    
    Photo-$z$ calibration improves the constraints on each of these parametes by a factor
    $\sim1.5-2$, and the dark energy figure of merit by a factor $\gtrsim5$. Furthermore, we find
    that the level of calibration achievable through cross-correlations with intensity mapping
    experiments is nearly equivalent to the case of perfect calibration. Equivalent results are
    found for SKA and HIRAX, as well as for the combination of DESI, Euclid and WFIRST, as could
    be expected from the results displayed in the right panel of Figure \ref{fig:compare_spec}.

\section{Discussion}\label{sec:discuss}
  In this paper we have shown that intensity maps of the HI emission can be used to improve the
  scientific output of photometric redshift surveys. By exploiting the cross-correlations
  between imaging surveys of galaxies and HI maps, we find that it is possible to optimally
  calibrate the photometric redshift distributions. This is made clear when assessing
  improvements in constraints on cosmological parameters: in Figure \ref{fig:constraints} we see
  that the FoM using this method is effectively as good as having perfect calibration. This also
  means that, with the aid of future IM experiments it should be possible to, for example,
  improve the LSST equation of state figure of merit by approximately a factor of 5.

  \begin{figure}
    \centering
    \includegraphics[width=0.49\textwidth]{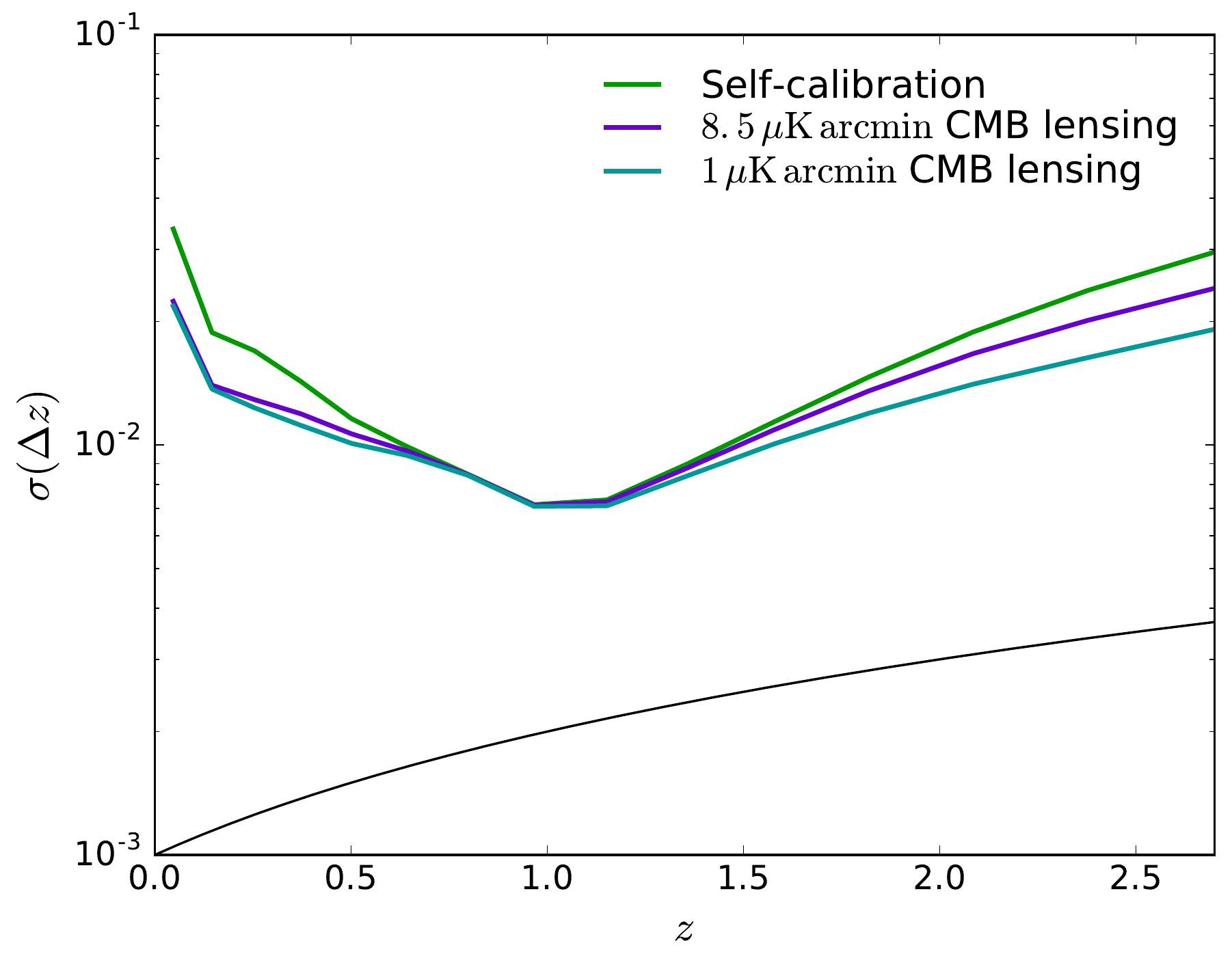}
    \caption{Forecast constraints on the LSST photo-$z$ bias $\Delta z$ in the absence of
             external datasets (green), and adding CMB lensing data from Stage-3 (purple)
             and Stage-4 (cyan) CMB experiments.}
    \label{fig:cmblens}
  \end{figure}
  This approach is promising, but it is important to highlight some of the limitations that have
  to be overcome if IM is to be used successfully in this context. For a start, it will be
  important to be able to deal with foreground contamination. In essence, as we have discussed,
  one can model the effect of foregrounds as a source of noise that cancels the information
  contained in long-wavelength radial modes. This effect is therefore greatly dependent on how
  coherent foreground residuals are in frequency. We can expect foregrounds to be highly
  correlated in intensity, with correlation lengths of order $\xi\sim1-10$. However, instrumental
  effects such as polarization leakage or frequency-dependent beams could spoil this coherence and
  lead to significant losses in the coverage of the $k_\parallel$-$k_\perp$ plane. We estimate
  that residuals with correlation lengths $\xi\lesssim0.1$ (as expected for polarised synchrotron
  emission, and corresponding to scales $k_\parallel\gtrsim0.01\,h\,{\rm Mpc}^{-1}$ at $z\sim1$)
  would significantly degrade the ability of 21cm maps to calibrate redshift distributions
  efficiently. We have furthermore shown that the calibration requirements of future photometric
  surveys can also be matched after accounting for the so-called ``foregrounds wedge''. Finally,
  it is worth noting that one of the main strength of the method is its reliance on
  cross-correlations between the spectroscopic and photometric samples, and that this
  cross-correlation should be very robust against systematic biases caused by foreground
  contamination.

  If we are to use IM to calibrate future redshift surveys, then we need to make sure that the
  observational set up satisfies minimum requirements and to do so, we have explored its
  dependence on experimental parameters. We have found that, for the noise levels of currently
  proposed experiments, we are primarily limited by cosmic variance and therefore there is no
  advantage in gaining depth at the cost of sky area: it is preferable to maximize the overlap
  between the HI maps and the galaxy imaging survey. Furthermore, if we are to accurately capture
  the longer wave-length modes, we need to resort to single dish observations, a promising but as
  yet untested mode of observation for MeerKAT and SKA. In the long term, ``$1/f$'' noise
  contributions will have to be controlled, and our analysis shows that the dishes must have a
  minimum size of $\sim15$ metres. If, on the other hand, we are to use interferometric
  observations (a method which is more tried and tested) then we need to ensure a minimum
  baseline of $\sim12$ metres to capture the large-scale angular modes. We have shown how
  MeerKAT, HIRAX and SKA fall well within these experimental parameter constraints.

  We must also note that our analysis has been conservative in terms of the ranges of scales
  that add up to the constraints on photo-$z$ parameters, only including angular scales in
  the regime where non-linear structure formation is believed to be well understood. It may,
  however, be possible to use even smaller scales for the purposes of photo-$z$ calibration
  \cite{2013arXiv1303.4722M}, in which case some of the conclusions drawn from this analysis
  could vary. In particular, the relative performance of HIRAX and SKA in terms of photo-$z$
  calibration could be significantly different, owing to the higher sensitivity of the SKA
  interferometer to small angular scales (see Figure \ref{fig:nk}).
  
  It is interesting to consider alternative approaches to sharpening photometric redshift
  measurements. Gravitational lensing of the CMB has recently been advocated as a promising approach,
  given its perfectly well-determined radial kernel (e.g. see \cite{2016MNRAS.461.4099B}). We
  can explore this possibility by considering correlations between LSST and a CMB lensing
  convergence $\kappa_{\rm CMB}$ map as in Section \ref{ssec:method.fisher}. We consider two
  different generations of CMB experiments: an ongoing ``Stage-3'' experiment, characterised
  by an rms noise level of $\sigma_T\sim8\,\mu{\rm K}\,{\rm arcmin}$, and a future ``Stage-4''
  experiment with $\sigma_T=1\,\mu{\rm K}\,{\rm arcmin}$. In both cases we assume very optimistic
  configurations, with a beam FWHM of 1 arcmin and using all angular scales $\ell\in[2,3000]$.
  We also fix all cosmological and bias parameters, and only consider varying $\Delta z$ and
  $\sigma_z$ for each LSST redshift bin. Figure \ref{fig:cmblens} shows the
  constraint on the photo-$z$ bias $\Delta z$ in the absence of CMB lensing data (green solid
  line), and using the cross-correlation with $\kappa_{\rm CMB}$ as measured by S3 (purple)
  and S4 (cyan) experiments. We clearly see that, even under these overly optimistic assumptions,
  adding CMB lensing information does not significantly help calibrate photo-$z$ uncertainties.
    
  Experiments to undertake intensity mapping of HI are an ongoing effort, and will give us an
  extremely promising new window on the Universe. We have argued that they will not only
  contribute in their own right to the further understanding of the large-scale structure of
  the Universe, but will also help improve the scientific returns from a plethora of up and
  coming optical surveys.

\section*{Acknowledgments}
  The authors would like to thank Elisabeth Krause, Mario Santos and An\v{z}e Slosar for useful
  comments and discussions. DA is supported by the Science and Technology Facilities Council
  and the Leverhume and Beecroft Trusts.
  PGF acnowledges support from Science Technology Facilities Council, the European Research
  Council and the Beecroft Trust. MJJ acknowledges support from The SA SKA project and the
  Oxford centre for astrophysical surveys which is funded through generous support from the
  Hintze Family Charitable Foundation. KM acknowledges support from the National Research
  Foundation of South Africa (Grant Number 98957). 
 
\bibliography{paper}

%merlin.mbs apsrev4-1.bst 2010-07-25 4.21a (PWD, AO, DPC) hacked
%Control: key (0)
%Control: author (8) initials jnrlst
%Control: editor formatted (1) identically to author
%Control: production of article title (-1) disabled
%Control: page (0) single
%Control: year (1) truncated
%Control: production of eprint (0) enabled
\begin{thebibliography}{64}%
\makeatletter
\providecommand \@ifxundefined [1]{%
 \@ifx{#1\undefined}
}%
\providecommand \@ifnum [1]{%
 \ifnum #1\expandafter \@firstoftwo
 \else \expandafter \@secondoftwo
 \fi
}%
\providecommand \@ifx [1]{%
 \ifx #1\expandafter \@firstoftwo
 \else \expandafter \@secondoftwo
 \fi
}%
\providecommand \natexlab [1]{#1}%
\providecommand \enquote  [1]{``#1''}%
\providecommand \bibnamefont  [1]{#1}%
\providecommand \bibfnamefont [1]{#1}%
\providecommand \citenamefont [1]{#1}%
\providecommand \href@noop [0]{\@secondoftwo}%
\providecommand \href [0]{\begingroup \@sanitize@url \@href}%
\providecommand \@href[1]{\@@startlink{#1}\@@href}%
\providecommand \@@href[1]{\endgroup#1\@@endlink}%
\providecommand \@sanitize@url [0]{\catcode `\\12\catcode `\$12\catcode
  `\&12\catcode `\#12\catcode `\^12\catcode `\_12\catcode `\%12\relax}%
\providecommand \@@startlink[1]{}%
\providecommand \@@endlink[0]{}%
\providecommand \url  [0]{\begingroup\@sanitize@url \@url }%
\providecommand \@url [1]{\endgroup\@href {#1}{\urlprefix }}%
\providecommand \urlprefix  [0]{URL }%
\providecommand \Eprint [0]{\href }%
\providecommand \doibase [0]{http://dx.doi.org/}%
\providecommand \selectlanguage [0]{\@gobble}%
\providecommand \bibinfo  [0]{\@secondoftwo}%
\providecommand \bibfield  [0]{\@secondoftwo}%
\providecommand \translation [1]{[#1]}%
\providecommand \BibitemOpen [0]{}%
\providecommand \bibitemStop [0]{}%
\providecommand \bibitemNoStop [0]{.\EOS\space}%
\providecommand \EOS [0]{\spacefactor3000\relax}%
\providecommand \BibitemShut  [1]{\csname bibitem#1\endcsname}%
\let\auto@bib@innerbib\@empty
%</preamble>
\bibitem [{\citenamefont {{Newman}}(2008)}]{2008ApJ...684...88N}%
  \BibitemOpen
  \bibfield  {author} {\bibinfo {author} {\bibfnamefont {J.~A.}\ \bibnamefont
  {{Newman}}},\ }\href {\doibase 10.1086/589982} {\bibfield  {journal}
  {\bibinfo  {journal} {\apj}\ }\textbf {\bibinfo {volume} {684}},\ \bibinfo
  {eid} {88-101} (\bibinfo {year} {2008})},\ \Eprint
  {http://arxiv.org/abs/0805.1409} {arXiv:0805.1409} \BibitemShut {NoStop}%
\bibitem [{\citenamefont {{Benjamin}}\ \emph {et~al.}(2010)\citenamefont
  {{Benjamin}}, \citenamefont {{van Waerbeke}}, \citenamefont {{M{\'e}nard}},\
  and\ \citenamefont {{Kilbinger}}}]{2010MNRAS.408.1168B}%
  \BibitemOpen
  \bibfield  {author} {\bibinfo {author} {\bibfnamefont {J.}~\bibnamefont
  {{Benjamin}}}, \bibinfo {author} {\bibfnamefont {L.}~\bibnamefont {{van
  Waerbeke}}}, \bibinfo {author} {\bibfnamefont {B.}~\bibnamefont
  {{M{\'e}nard}}}, \ and\ \bibinfo {author} {\bibfnamefont {M.}~\bibnamefont
  {{Kilbinger}}},\ }\href {\doibase 10.1111/j.1365-2966.2010.17191.x}
  {\bibfield  {journal} {\bibinfo  {journal} {\mnras}\ }\textbf {\bibinfo
  {volume} {408}},\ \bibinfo {pages} {1168} (\bibinfo {year} {2010})},\ \Eprint
  {http://arxiv.org/abs/1002.2266} {arXiv:1002.2266 [astro-ph.CO]} \BibitemShut
  {NoStop}%
\bibitem [{\citenamefont {{Matthews}}\ and\ \citenamefont
  {{Newman}}(2010)}]{2010ApJ...721..456M}%
  \BibitemOpen
  \bibfield  {author} {\bibinfo {author} {\bibfnamefont {D.~J.}\ \bibnamefont
  {{Matthews}}}\ and\ \bibinfo {author} {\bibfnamefont {J.~A.}\ \bibnamefont
  {{Newman}}},\ }\href {\doibase 10.1088/0004-637X/721/1/456} {\bibfield
  {journal} {\bibinfo  {journal} {\apj}\ }\textbf {\bibinfo {volume} {721}},\
  \bibinfo {pages} {456} (\bibinfo {year} {2010})},\ \Eprint
  {http://arxiv.org/abs/1003.0687} {arXiv:1003.0687 [astro-ph.CO]} \BibitemShut
  {NoStop}%
\bibitem [{\citenamefont {{Schmidt}}\ \emph {et~al.}(2013)\citenamefont
  {{Schmidt}}, \citenamefont {{M{\'e}nard}}, \citenamefont {{Scranton}},
  \citenamefont {{Morrison}},\ and\ \citenamefont
  {{McBride}}}]{2013MNRAS.431.3307S}%
  \BibitemOpen
  \bibfield  {author} {\bibinfo {author} {\bibfnamefont {S.~J.}\ \bibnamefont
  {{Schmidt}}}, \bibinfo {author} {\bibfnamefont {B.}~\bibnamefont
  {{M{\'e}nard}}}, \bibinfo {author} {\bibfnamefont {R.}~\bibnamefont
  {{Scranton}}}, \bibinfo {author} {\bibfnamefont {C.}~\bibnamefont
  {{Morrison}}}, \ and\ \bibinfo {author} {\bibfnamefont {C.~K.}\ \bibnamefont
  {{McBride}}},\ }\href {\doibase 10.1093/mnras/stt410} {\bibfield  {journal}
  {\bibinfo  {journal} {\mnras}\ }\textbf {\bibinfo {volume} {431}},\ \bibinfo
  {pages} {3307} (\bibinfo {year} {2013})},\ \Eprint
  {http://arxiv.org/abs/1303.0292} {arXiv:1303.0292} \BibitemShut {NoStop}%
\bibitem [{\citenamefont {{M{\'e}nard}}\ \emph {et~al.}(2013)\citenamefont
  {{M{\'e}nard}}, \citenamefont {{Scranton}}, \citenamefont {{Schmidt}},
  \citenamefont {{Morrison}}, \citenamefont {{Jeong}}, \citenamefont
  {{Budavari}},\ and\ \citenamefont {{Rahman}}}]{2013arXiv1303.4722M}%
  \BibitemOpen
  \bibfield  {author} {\bibinfo {author} {\bibfnamefont {B.}~\bibnamefont
  {{M{\'e}nard}}}, \bibinfo {author} {\bibfnamefont {R.}~\bibnamefont
  {{Scranton}}}, \bibinfo {author} {\bibfnamefont {S.}~\bibnamefont
  {{Schmidt}}}, \bibinfo {author} {\bibfnamefont {C.}~\bibnamefont
  {{Morrison}}}, \bibinfo {author} {\bibfnamefont {D.}~\bibnamefont {{Jeong}}},
  \bibinfo {author} {\bibfnamefont {T.}~\bibnamefont {{Budavari}}}, \ and\
  \bibinfo {author} {\bibfnamefont {M.}~\bibnamefont {{Rahman}}},\ }\href@noop
  {} {\bibfield  {journal} {\bibinfo  {journal} {ArXiv e-prints}\ } (\bibinfo
  {year} {2013})},\ \Eprint {http://arxiv.org/abs/1303.4722} {arXiv:1303.4722}
  \BibitemShut {NoStop}%
\bibitem [{\citenamefont {{van Daalen}}\ and\ \citenamefont
  {{White}}(2017)}]{2017arXiv170305326V}%
  \BibitemOpen
  \bibfield  {author} {\bibinfo {author} {\bibfnamefont {M.~P.}\ \bibnamefont
  {{van Daalen}}}\ and\ \bibinfo {author} {\bibfnamefont {M.}~\bibnamefont
  {{White}}},\ }\href@noop {} {\bibfield  {journal} {\bibinfo  {journal} {ArXiv
  e-prints}\ } (\bibinfo {year} {2017})},\ \Eprint
  {http://arxiv.org/abs/1703.05326} {arXiv:1703.05326} \BibitemShut {NoStop}%
\bibitem [{\citenamefont {{Rahman}}\ \emph
  {et~al.}(2016{\natexlab{a}})\citenamefont {{Rahman}}, \citenamefont
  {{Mendez}}, \citenamefont {{M{\'e}nard}}, \citenamefont {{Scranton}},
  \citenamefont {{Schmidt}}, \citenamefont {{Morrison}},\ and\ \citenamefont
  {{Budav{\'a}ri}}}]{2016MNRAS.460..163R}%
  \BibitemOpen
  \bibfield  {author} {\bibinfo {author} {\bibfnamefont {M.}~\bibnamefont
  {{Rahman}}}, \bibinfo {author} {\bibfnamefont {A.~J.}\ \bibnamefont
  {{Mendez}}}, \bibinfo {author} {\bibfnamefont {B.}~\bibnamefont
  {{M{\'e}nard}}}, \bibinfo {author} {\bibfnamefont {R.}~\bibnamefont
  {{Scranton}}}, \bibinfo {author} {\bibfnamefont {S.~J.}\ \bibnamefont
  {{Schmidt}}}, \bibinfo {author} {\bibfnamefont {C.~B.}\ \bibnamefont
  {{Morrison}}}, \ and\ \bibinfo {author} {\bibfnamefont {T.}~\bibnamefont
  {{Budav{\'a}ri}}},\ }\href {\doibase 10.1093/mnras/stw981} {\bibfield
  {journal} {\bibinfo  {journal} {\mnras}\ }\textbf {\bibinfo {volume} {460}},\
  \bibinfo {pages} {163} (\bibinfo {year} {2016}{\natexlab{a}})},\ \Eprint
  {http://arxiv.org/abs/1512.03057} {arXiv:1512.03057} \BibitemShut {NoStop}%
\bibitem [{\citenamefont {{Choi}}\ \emph {et~al.}(2016)\citenamefont {{Choi}},
  \citenamefont {{Heymans}}, \citenamefont {{Blake}}, \citenamefont
  {{Hildebrandt}}, \citenamefont {{Duncan}}, \citenamefont {{Erben}},
  \citenamefont {{Nakajima}}, \citenamefont {{Van Waerbeke}},\ and\
  \citenamefont {{Viola}}}]{2016MNRAS.463.3737C}%
  \BibitemOpen
  \bibfield  {author} {\bibinfo {author} {\bibfnamefont {A.}~\bibnamefont
  {{Choi}}}, \bibinfo {author} {\bibfnamefont {C.}~\bibnamefont {{Heymans}}},
  \bibinfo {author} {\bibfnamefont {C.}~\bibnamefont {{Blake}}}, \bibinfo
  {author} {\bibfnamefont {H.}~\bibnamefont {{Hildebrandt}}}, \bibinfo {author}
  {\bibfnamefont {C.~A.~J.}\ \bibnamefont {{Duncan}}}, \bibinfo {author}
  {\bibfnamefont {T.}~\bibnamefont {{Erben}}}, \bibinfo {author} {\bibfnamefont
  {R.}~\bibnamefont {{Nakajima}}}, \bibinfo {author} {\bibfnamefont
  {L.}~\bibnamefont {{Van Waerbeke}}}, \ and\ \bibinfo {author} {\bibfnamefont
  {M.}~\bibnamefont {{Viola}}},\ }\href {\doibase 10.1093/mnras/stw2241}
  {\bibfield  {journal} {\bibinfo  {journal} {\mnras}\ }\textbf {\bibinfo
  {volume} {463}},\ \bibinfo {pages} {3737} (\bibinfo {year} {2016})},\ \Eprint
  {http://arxiv.org/abs/1512.03626} {arXiv:1512.03626} \BibitemShut {NoStop}%
\bibitem [{\citenamefont {{Scottez}}\ \emph {et~al.}(2016)\citenamefont
  {{Scottez}}, \citenamefont {{Mellier}}, \citenamefont {{Granett}},
  \citenamefont {{Moutard}}, \citenamefont {{Kilbinger}} \emph
  {et~al.}}]{2016MNRAS.462.1683S}%
  \BibitemOpen
  \bibfield  {author} {\bibinfo {author} {\bibfnamefont {V.}~\bibnamefont
  {{Scottez}}}, \bibinfo {author} {\bibfnamefont {Y.}~\bibnamefont
  {{Mellier}}}, \bibinfo {author} {\bibfnamefont {B.~R.}\ \bibnamefont
  {{Granett}}}, \bibinfo {author} {\bibfnamefont {T.}~\bibnamefont
  {{Moutard}}}, \bibinfo {author} {\bibfnamefont {M.}~\bibnamefont
  {{Kilbinger}}},  \emph {et~al.},\ }\href {\doibase 10.1093/mnras/stw1500}
  {\bibfield  {journal} {\bibinfo  {journal} {\mnras}\ }\textbf {\bibinfo
  {volume} {462}},\ \bibinfo {pages} {1683} (\bibinfo {year} {2016})},\ \Eprint
  {http://arxiv.org/abs/1605.05501} {arXiv:1605.05501} \BibitemShut {NoStop}%
\bibitem [{\citenamefont {{Rahman}}\ \emph
  {et~al.}(2016{\natexlab{b}})\citenamefont {{Rahman}}, \citenamefont
  {{M{\'e}nard}},\ and\ \citenamefont {{Scranton}}}]{2016MNRAS.457.3912R}%
  \BibitemOpen
  \bibfield  {author} {\bibinfo {author} {\bibfnamefont {M.}~\bibnamefont
  {{Rahman}}}, \bibinfo {author} {\bibfnamefont {B.}~\bibnamefont
  {{M{\'e}nard}}}, \ and\ \bibinfo {author} {\bibfnamefont {R.}~\bibnamefont
  {{Scranton}}},\ }\href {\doibase 10.1093/mnras/stw256} {\bibfield  {journal}
  {\bibinfo  {journal} {\mnras}\ }\textbf {\bibinfo {volume} {457}},\ \bibinfo
  {pages} {3912} (\bibinfo {year} {2016}{\natexlab{b}})},\ \Eprint
  {http://arxiv.org/abs/1508.03046} {arXiv:1508.03046} \BibitemShut {NoStop}%
\bibitem [{\citenamefont {{Johnson}}\ \emph {et~al.}(2017)\citenamefont
  {{Johnson}}, \citenamefont {{Blake}}, \citenamefont {{Amon}}, \citenamefont
  {{Erben}}, \citenamefont {{Glazebrook}} \emph
  {et~al.}}]{2017MNRAS.465.4118J}%
  \BibitemOpen
  \bibfield  {author} {\bibinfo {author} {\bibfnamefont {A.}~\bibnamefont
  {{Johnson}}}, \bibinfo {author} {\bibfnamefont {C.}~\bibnamefont {{Blake}}},
  \bibinfo {author} {\bibfnamefont {A.}~\bibnamefont {{Amon}}}, \bibinfo
  {author} {\bibfnamefont {T.}~\bibnamefont {{Erben}}}, \bibinfo {author}
  {\bibfnamefont {K.}~\bibnamefont {{Glazebrook}}},  \emph {et~al.},\ }\href
  {\doibase 10.1093/mnras/stw3033} {\bibfield  {journal} {\bibinfo  {journal}
  {\mnras}\ }\textbf {\bibinfo {volume} {465}},\ \bibinfo {pages} {4118}
  (\bibinfo {year} {2017})},\ \Eprint {http://arxiv.org/abs/1611.07578}
  {arXiv:1611.07578} \BibitemShut {NoStop}%
\bibitem [{\citenamefont {Battye}\ \emph {et~al.}(2004)\citenamefont {Battye},
  \citenamefont {Davies},\ and\ \citenamefont {Weller}}]{Battye:2004re}%
  \BibitemOpen
  \bibfield  {author} {\bibinfo {author} {\bibfnamefont {R.~A.}\ \bibnamefont
  {Battye}}, \bibinfo {author} {\bibfnamefont {R.~D.}\ \bibnamefont {Davies}},
  \ and\ \bibinfo {author} {\bibfnamefont {J.}~\bibnamefont {Weller}},\ }\href
  {\doibase 10.1111/j.1365-2966.2004.08416.x} {\bibfield  {journal} {\bibinfo
  {journal} {Mon. Not. Roy. Astron. Soc.}\ }\textbf {\bibinfo {volume} {355}},\
  \bibinfo {pages} {1339} (\bibinfo {year} {2004})},\ \Eprint
  {http://arxiv.org/abs/astro-ph/0401340} {arXiv:astro-ph/0401340 [astro-ph]}
  \BibitemShut {NoStop}%
%%CITATION = ASTRO-PH/0401340;%%
\bibitem [{\citenamefont {{McQuinn}}\ \emph {et~al.}(2006)\citenamefont
  {{McQuinn}}, \citenamefont {{Zahn}}, \citenamefont {{Zaldarriaga}},
  \citenamefont {{Hernquist}},\ and\ \citenamefont
  {{Furlanetto}}}]{2006ApJ...653..815M}%
  \BibitemOpen
  \bibfield  {author} {\bibinfo {author} {\bibfnamefont {M.}~\bibnamefont
  {{McQuinn}}}, \bibinfo {author} {\bibfnamefont {O.}~\bibnamefont {{Zahn}}},
  \bibinfo {author} {\bibfnamefont {M.}~\bibnamefont {{Zaldarriaga}}}, \bibinfo
  {author} {\bibfnamefont {L.}~\bibnamefont {{Hernquist}}}, \ and\ \bibinfo
  {author} {\bibfnamefont {S.~R.}\ \bibnamefont {{Furlanetto}}},\ }\href
  {\doibase 10.1086/505167} {\bibfield  {journal} {\bibinfo  {journal} {\apj}\
  }\textbf {\bibinfo {volume} {653}},\ \bibinfo {pages} {815} (\bibinfo {year}
  {2006})},\ \Eprint {http://arxiv.org/abs/astro-ph/0512263} {astro-ph/0512263}
  \BibitemShut {NoStop}%
\bibitem [{\citenamefont {Chang}\ \emph {et~al.}(2008)\citenamefont {Chang},
  \citenamefont {Pen}, \citenamefont {Peterson},\ and\ \citenamefont
  {McDonald}}]{Chang:2007xk}%
  \BibitemOpen
  \bibfield  {author} {\bibinfo {author} {\bibfnamefont {T.-C.}\ \bibnamefont
  {Chang}}, \bibinfo {author} {\bibfnamefont {U.-L.}\ \bibnamefont {Pen}},
  \bibinfo {author} {\bibfnamefont {J.~B.}\ \bibnamefont {Peterson}}, \ and\
  \bibinfo {author} {\bibfnamefont {P.}~\bibnamefont {McDonald}},\ }\href
  {\doibase 10.1103/PhysRevLett.100.091303} {\bibfield  {journal} {\bibinfo
  {journal} {Phys. Rev. Lett.}\ }\textbf {\bibinfo {volume} {100}},\ \bibinfo
  {pages} {091303} (\bibinfo {year} {2008})},\ \Eprint
  {http://arxiv.org/abs/0709.3672} {arXiv:0709.3672 [astro-ph]} \BibitemShut
  {NoStop}%
%%CITATION = ARXIV:0709.3672;%%
\bibitem [{\citenamefont {{Wyithe}}\ and\ \citenamefont
  {{Loeb}}(2008)}]{2008MNRAS.383..606W}%
  \BibitemOpen
  \bibfield  {author} {\bibinfo {author} {\bibfnamefont {J.~S.~B.}\
  \bibnamefont {{Wyithe}}}\ and\ \bibinfo {author} {\bibfnamefont
  {A.}~\bibnamefont {{Loeb}}},\ }\href {\doibase
  10.1111/j.1365-2966.2007.12568.x} {\bibfield  {journal} {\bibinfo  {journal}
  {M.N.R.A.S.}\ }\textbf {\bibinfo {volume} {383}},\ \bibinfo {pages} {606}
  (\bibinfo {year} {2008})}\BibitemShut {NoStop}%
\bibitem [{\citenamefont {{Loeb}}\ and\ \citenamefont
  {{Wyithe}}(2008)}]{2008PhRvL.100p1301L}%
  \BibitemOpen
  \bibfield  {author} {\bibinfo {author} {\bibfnamefont {A.}~\bibnamefont
  {{Loeb}}}\ and\ \bibinfo {author} {\bibfnamefont {J.~S.~B.}\ \bibnamefont
  {{Wyithe}}},\ }\href {\doibase 10.1103/PhysRevLett.100.161301} {\bibfield
  {journal} {\bibinfo  {journal} {Physical Review Letters}\ }\textbf {\bibinfo
  {volume} {100}},\ \bibinfo {eid} {161301} (\bibinfo {year} {2008})},\ \Eprint
  {http://arxiv.org/abs/0801.1677} {arXiv:0801.1677} \BibitemShut {NoStop}%
\bibitem [{\citenamefont {{Peterson}}\ \emph {et~al.}(2009)\citenamefont
  {{Peterson}}, \citenamefont {{Aleksan}}, \citenamefont {{Ansari}} \emph
  {et~al.}}]{2009astro2010S.234P}%
  \BibitemOpen
  \bibfield  {author} {\bibinfo {author} {\bibfnamefont {J.~B.}\ \bibnamefont
  {{Peterson}}}, \bibinfo {author} {\bibfnamefont {R.}~\bibnamefont
  {{Aleksan}}}, \bibinfo {author} {\bibfnamefont {R.}~\bibnamefont {{Ansari}}},
   \emph {et~al.},\ }in\ \href@noop {} {\emph {\bibinfo {booktitle} {astro2010:
  The Astronomy and Astrophysics Decadal Survey}}},\ \bibinfo {series} {ArXiv
  Astrophysics e-prints}, Vol.\ \bibinfo {volume} {2010}\ (\bibinfo {year}
  {2009})\ p.\ \bibinfo {pages} {234},\ \Eprint
  {http://arxiv.org/abs/0902.3091} {arXiv:0902.3091 [astro-ph.IM]} \BibitemShut
  {NoStop}%
\bibitem [{\citenamefont {Bagla}\ \emph {et~al.}(2010)\citenamefont {Bagla},
  \citenamefont {Khandai},\ and\ \citenamefont {Datta}}]{Bagla:2009jy}%
  \BibitemOpen
  \bibfield  {author} {\bibinfo {author} {\bibfnamefont {J.}~\bibnamefont
  {Bagla}}, \bibinfo {author} {\bibfnamefont {N.}~\bibnamefont {Khandai}}, \
  and\ \bibinfo {author} {\bibfnamefont {K.~K.}\ \bibnamefont {Datta}},\ }\href
  {\doibase 10.1111/j.1365-2966.2010.16933.x} {\bibfield  {journal} {\bibinfo
  {journal} {Mon. Not. Roy. Astron. Soc.}\ }\textbf {\bibinfo {volume} {407}},\
  \bibinfo {pages} {567} (\bibinfo {year} {2010})},\ \Eprint
  {http://arxiv.org/abs/0908.3796} {arXiv:0908.3796 [astro-ph.CO]} \BibitemShut
  {NoStop}%
%%CITATION = ARXIV:0908.3796;%%
\bibitem [{\citenamefont {{Battye}}\ \emph {et~al.}(2013)\citenamefont
  {{Battye}}, \citenamefont {{Browne}}, \citenamefont {{Dickinson}},
  \citenamefont {{Heron}}, \citenamefont {{Maffei}},\ and\ \citenamefont
  {{Pourtsidou}}}]{2013MNRAS.434.1239B}%
  \BibitemOpen
  \bibfield  {author} {\bibinfo {author} {\bibfnamefont {R.~A.}\ \bibnamefont
  {{Battye}}}, \bibinfo {author} {\bibfnamefont {I.~W.~A.}\ \bibnamefont
  {{Browne}}}, \bibinfo {author} {\bibfnamefont {C.}~\bibnamefont
  {{Dickinson}}}, \bibinfo {author} {\bibfnamefont {G.}~\bibnamefont
  {{Heron}}}, \bibinfo {author} {\bibfnamefont {B.}~\bibnamefont {{Maffei}}}, \
  and\ \bibinfo {author} {\bibfnamefont {A.}~\bibnamefont {{Pourtsidou}}},\
  }\href {\doibase 10.1093/mnras/stt1082} {\bibfield  {journal} {\bibinfo
  {journal} {M.N.R.A.S.}\ }\textbf {\bibinfo {volume} {434}},\ \bibinfo {pages}
  {1239} (\bibinfo {year} {2013})},\ \Eprint {http://arxiv.org/abs/1209.0343}
  {arXiv:1209.0343 [astro-ph.CO]} \BibitemShut {NoStop}%
\bibitem [{\citenamefont {{Masui}}\ \emph {et~al.}(2013)\citenamefont
  {{Masui}}, \citenamefont {{Switzer}}, \citenamefont {{Banavar}},
  \citenamefont {{Bandura}}, \citenamefont {{Blake}} \emph
  {et~al.}}]{2013ApJ...763L..20M}%
  \BibitemOpen
  \bibfield  {author} {\bibinfo {author} {\bibfnamefont {K.~W.}\ \bibnamefont
  {{Masui}}}, \bibinfo {author} {\bibfnamefont {E.~R.}\ \bibnamefont
  {{Switzer}}}, \bibinfo {author} {\bibfnamefont {N.}~\bibnamefont
  {{Banavar}}}, \bibinfo {author} {\bibfnamefont {K.}~\bibnamefont
  {{Bandura}}}, \bibinfo {author} {\bibfnamefont {C.}~\bibnamefont {{Blake}}},
  \emph {et~al.},\ }\href {\doibase 10.1088/2041-8205/763/1/L20} {\bibfield
  {journal} {\bibinfo  {journal} {\apjl}\ }\textbf {\bibinfo {volume} {763}},\
  \bibinfo {eid} {L20} (\bibinfo {year} {2013})},\ \Eprint
  {http://arxiv.org/abs/1208.0331} {arXiv:1208.0331 [astro-ph.CO]} \BibitemShut
  {NoStop}%
\bibitem [{\citenamefont {{Switzer}}\ \emph {et~al.}(2013)\citenamefont
  {{Switzer}}, \citenamefont {{Masui}}, \citenamefont {{Bandura}},
  \citenamefont {{Calin}}, \citenamefont {{Chang}} \emph
  {et~al.}}]{2013MNRAS.434L..46S}%
  \BibitemOpen
  \bibfield  {author} {\bibinfo {author} {\bibfnamefont {E.~R.}\ \bibnamefont
  {{Switzer}}}, \bibinfo {author} {\bibfnamefont {K.~W.}\ \bibnamefont
  {{Masui}}}, \bibinfo {author} {\bibfnamefont {K.}~\bibnamefont {{Bandura}}},
  \bibinfo {author} {\bibfnamefont {L.-M.}\ \bibnamefont {{Calin}}}, \bibinfo
  {author} {\bibfnamefont {T.-C.}\ \bibnamefont {{Chang}}},  \emph {et~al.},\
  }\href {\doibase 10.1093/mnrasl/slt074} {\bibfield  {journal} {\bibinfo
  {journal} {\mnras}\ }\textbf {\bibinfo {volume} {434}},\ \bibinfo {pages}
  {L46} (\bibinfo {year} {2013})},\ \Eprint {http://arxiv.org/abs/1304.3712}
  {arXiv:1304.3712 [astro-ph.CO]} \BibitemShut {NoStop}%
\bibitem [{\citenamefont {{Bull}}\ \emph {et~al.}(2015)\citenamefont {{Bull}},
  \citenamefont {{Ferreira}}, \citenamefont {{Patel}},\ and\ \citenamefont
  {{Santos}}}]{2015ApJ...803...21B}%
  \BibitemOpen
  \bibfield  {author} {\bibinfo {author} {\bibfnamefont {P.}~\bibnamefont
  {{Bull}}}, \bibinfo {author} {\bibfnamefont {P.~G.}\ \bibnamefont
  {{Ferreira}}}, \bibinfo {author} {\bibfnamefont {P.}~\bibnamefont {{Patel}}},
  \ and\ \bibinfo {author} {\bibfnamefont {M.~G.}\ \bibnamefont {{Santos}}},\
  }\href {\doibase 10.1088/0004-637X/803/1/21} {\bibfield  {journal} {\bibinfo
  {journal} {\apj}\ }\textbf {\bibinfo {volume} {803}},\ \bibinfo {eid} {21}
  (\bibinfo {year} {2015})},\ \Eprint {http://arxiv.org/abs/1405.1452}
  {arXiv:1405.1452} \BibitemShut {NoStop}%
\bibitem [{\citenamefont {{Fonseca}}\ \emph {et~al.}(2017)\citenamefont
  {{Fonseca}}, \citenamefont {{Maartens}},\ and\ \citenamefont
  {{Santos}}}]{2017MNRAS.466.2780F}%
  \BibitemOpen
  \bibfield  {author} {\bibinfo {author} {\bibfnamefont {J.}~\bibnamefont
  {{Fonseca}}}, \bibinfo {author} {\bibfnamefont {R.}~\bibnamefont
  {{Maartens}}}, \ and\ \bibinfo {author} {\bibfnamefont {M.~G.}\ \bibnamefont
  {{Santos}}},\ }\href {\doibase 10.1093/mnras/stw3248} {\bibfield  {journal}
  {\bibinfo  {journal} {\mnras}\ }\textbf {\bibinfo {volume} {466}},\ \bibinfo
  {pages} {2780} (\bibinfo {year} {2017})},\ \Eprint
  {http://arxiv.org/abs/1611.01322} {arXiv:1611.01322} \BibitemShut {NoStop}%
\bibitem [{\citenamefont {{Newburgh}}\ \emph {et~al.}(2016)\citenamefont
  {{Newburgh}}, \citenamefont {{Bandura}}, \citenamefont {{Bucher}},
  \citenamefont {{Chang}}, \citenamefont {{Chiang}} \emph
  {et~al.}}]{2016SPIE.9906E..5XN}%
  \BibitemOpen
  \bibfield  {author} {\bibinfo {author} {\bibfnamefont {L.~B.}\ \bibnamefont
  {{Newburgh}}}, \bibinfo {author} {\bibfnamefont {K.}~\bibnamefont
  {{Bandura}}}, \bibinfo {author} {\bibfnamefont {M.~A.}\ \bibnamefont
  {{Bucher}}}, \bibinfo {author} {\bibfnamefont {T.-C.}\ \bibnamefont
  {{Chang}}}, \bibinfo {author} {\bibfnamefont {H.~C.}\ \bibnamefont
  {{Chiang}}},  \emph {et~al.},\ }in\ \href {\doibase 10.1117/12.2234286}
  {\emph {\bibinfo {booktitle} {Society of Photo-Optical Instrumentation
  Engineers (SPIE) Conference Series}}},\ \bibinfo {series} {\procspie}, Vol.\
  \bibinfo {volume} {9906}\ (\bibinfo {year} {2016})\ p.\ \bibinfo {pages}
  {99065X},\ \Eprint {http://arxiv.org/abs/1607.02059} {arXiv:1607.02059
  [astro-ph.IM]} \BibitemShut {NoStop}%
\bibitem [{\citenamefont {{Santos}}\ \emph {et~al.}(2015)\citenamefont
  {{Santos}}, \citenamefont {{Bull}}, \citenamefont {{Alonso}}, \citenamefont
  {{Camera}}, \citenamefont {{Ferreira}} \emph {et~al.}}]{2015aska.confE..19S}%
  \BibitemOpen
  \bibfield  {author} {\bibinfo {author} {\bibfnamefont {M.}~\bibnamefont
  {{Santos}}}, \bibinfo {author} {\bibfnamefont {P.}~\bibnamefont {{Bull}}},
  \bibinfo {author} {\bibfnamefont {D.}~\bibnamefont {{Alonso}}}, \bibinfo
  {author} {\bibfnamefont {S.}~\bibnamefont {{Camera}}}, \bibinfo {author}
  {\bibfnamefont {P.}~\bibnamefont {{Ferreira}}},  \emph {et~al.},\ }\href@noop
  {} {\bibfield  {journal} {\bibinfo  {journal} {Advancing Astrophysics with
  the Square Kilometre Array (AASKA14)}\ ,\ \bibinfo {eid} {19}} (\bibinfo
  {year} {2015})},\ \Eprint {http://arxiv.org/abs/1501.03989}
  {arXiv:1501.03989} \BibitemShut {NoStop}%
\bibitem [{\citenamefont {{McQuinn}}\ and\ \citenamefont
  {{White}}(2013)}]{2013MNRAS.433.2857M}%
  \BibitemOpen
  \bibfield  {author} {\bibinfo {author} {\bibfnamefont {M.}~\bibnamefont
  {{McQuinn}}}\ and\ \bibinfo {author} {\bibfnamefont {M.}~\bibnamefont
  {{White}}},\ }\href {\doibase 10.1093/mnras/stt914} {\bibfield  {journal}
  {\bibinfo  {journal} {\mnras}\ }\textbf {\bibinfo {volume} {433}},\ \bibinfo
  {pages} {2857} (\bibinfo {year} {2013})},\ \Eprint
  {http://arxiv.org/abs/1302.0857} {arXiv:1302.0857} \BibitemShut {NoStop}%
\bibitem [{\citenamefont {{Jasche}}\ and\ \citenamefont
  {{Wandelt}}(2012)}]{2012MNRAS.425.1042J}%
  \BibitemOpen
  \bibfield  {author} {\bibinfo {author} {\bibfnamefont {J.}~\bibnamefont
  {{Jasche}}}\ and\ \bibinfo {author} {\bibfnamefont {B.~D.}\ \bibnamefont
  {{Wandelt}}},\ }\href {\doibase 10.1111/j.1365-2966.2012.21423.x} {\bibfield
  {journal} {\bibinfo  {journal} {\mnras}\ }\textbf {\bibinfo {volume} {425}},\
  \bibinfo {pages} {1042} (\bibinfo {year} {2012})},\ \Eprint
  {http://arxiv.org/abs/1106.2757} {arXiv:1106.2757} \BibitemShut {NoStop}%
\bibitem [{\citenamefont {{Alonso}}\ and\ \citenamefont
  {{Ferreira}}(2015)}]{2015PhRvD..92f3525A}%
  \BibitemOpen
  \bibfield  {author} {\bibinfo {author} {\bibfnamefont {D.}~\bibnamefont
  {{Alonso}}}\ and\ \bibinfo {author} {\bibfnamefont {P.~G.}\ \bibnamefont
  {{Ferreira}}},\ }\href {\doibase 10.1103/PhysRevD.92.063525} {\bibfield
  {journal} {\bibinfo  {journal} {\prd}\ }\textbf {\bibinfo {volume} {92}},\
  \bibinfo {eid} {063525} (\bibinfo {year} {2015})},\ \Eprint
  {http://arxiv.org/abs/1507.03550} {arXiv:1507.03550} \BibitemShut {NoStop}%
\bibitem [{\citenamefont {{Gabasch}}\ \emph {et~al.}(2006)\citenamefont
  {{Gabasch}}, \citenamefont {{Hopp}}, \citenamefont {{Feulner}}, \citenamefont
  {{Bender}}, \citenamefont {{Seitz}} \emph {et~al.}}]{2006A&A...448..101G}%
  \BibitemOpen
  \bibfield  {author} {\bibinfo {author} {\bibfnamefont {A.}~\bibnamefont
  {{Gabasch}}}, \bibinfo {author} {\bibfnamefont {U.}~\bibnamefont {{Hopp}}},
  \bibinfo {author} {\bibfnamefont {G.}~\bibnamefont {{Feulner}}}, \bibinfo
  {author} {\bibfnamefont {R.}~\bibnamefont {{Bender}}}, \bibinfo {author}
  {\bibfnamefont {S.}~\bibnamefont {{Seitz}}},  \emph {et~al.},\ }\href
  {\doibase 10.1051/0004-6361:20053986} {\bibfield  {journal} {\bibinfo
  {journal} {\aap}\ }\textbf {\bibinfo {volume} {448}},\ \bibinfo {pages} {101}
  (\bibinfo {year} {2006})},\ \Eprint {http://arxiv.org/abs/astro-ph/0510339}
  {astro-ph/0510339} \BibitemShut {NoStop}%
\bibitem [{\citenamefont {{Blanton}}\ and\ \citenamefont
  {{Roweis}}(2007)}]{2007AJ....133..734B}%
  \BibitemOpen
  \bibfield  {author} {\bibinfo {author} {\bibfnamefont {M.~R.}\ \bibnamefont
  {{Blanton}}}\ and\ \bibinfo {author} {\bibfnamefont {S.}~\bibnamefont
  {{Roweis}}},\ }\href {\doibase 10.1086/510127} {\bibfield  {journal}
  {\bibinfo  {journal} {\aj}\ }\textbf {\bibinfo {volume} {133}},\ \bibinfo
  {pages} {734} (\bibinfo {year} {2007})},\ \Eprint
  {http://arxiv.org/abs/astro-ph/0606170} {astro-ph/0606170} \BibitemShut
  {NoStop}%
\bibitem [{\citenamefont {{LSST Collaboration}}\ \emph
  {et~al.}(2009)\citenamefont {{LSST Collaboration}} \emph
  {et~al.}}]{2009arXiv0912.0201L}%
  \BibitemOpen
  \bibfield  {author} {\bibinfo {author} {\bibnamefont {{LSST Collaboration}}}
  \emph {et~al.},\ }\href@noop {} {\bibfield  {journal} {\bibinfo  {journal}
  {arXiv e-prints}\ } (\bibinfo {year} {2009})},\ \Eprint
  {http://arxiv.org/abs/0912.0201} {arXiv:0912.0201 [astro-ph.IM]} \BibitemShut
  {NoStop}%
\bibitem [{\citenamefont {{Weinberg}}\ \emph {et~al.}(2004)\citenamefont
  {{Weinberg}}, \citenamefont {{Dav{\'e}}}, \citenamefont {{Katz}},\ and\
  \citenamefont {{Hernquist}}}]{2004ApJ...601....1W}%
  \BibitemOpen
  \bibfield  {author} {\bibinfo {author} {\bibfnamefont {D.~H.}\ \bibnamefont
  {{Weinberg}}}, \bibinfo {author} {\bibfnamefont {R.}~\bibnamefont
  {{Dav{\'e}}}}, \bibinfo {author} {\bibfnamefont {N.}~\bibnamefont {{Katz}}},
  \ and\ \bibinfo {author} {\bibfnamefont {L.}~\bibnamefont {{Hernquist}}},\
  }\href {\doibase 10.1086/380481} {\bibfield  {journal} {\bibinfo  {journal}
  {\apj}\ }\textbf {\bibinfo {volume} {601}},\ \bibinfo {pages} {1} (\bibinfo
  {year} {2004})},\ \Eprint {http://arxiv.org/abs/astro-ph/0212356}
  {astro-ph/0212356} \BibitemShut {NoStop}%
\bibitem [{\citenamefont {{Battye}}\ \emph {et~al.}(2012)\citenamefont
  {{Battye}}, \citenamefont {{Brown}}, \citenamefont {{Browne}}, \citenamefont
  {{Davis}}, \citenamefont {{Dewdney}}, \citenamefont {{Dickinson}},
  \citenamefont {{Heron}}, \citenamefont {{Maffei}}, \citenamefont
  {{Pourtsidou}},\ and\ \citenamefont {{Wilkinson}}}]{2012arXiv1209.1041B}%
  \BibitemOpen
  \bibfield  {author} {\bibinfo {author} {\bibfnamefont {R.~A.}\ \bibnamefont
  {{Battye}}}, \bibinfo {author} {\bibfnamefont {M.~L.}\ \bibnamefont
  {{Brown}}}, \bibinfo {author} {\bibfnamefont {I.~W.~A.}\ \bibnamefont
  {{Browne}}}, \bibinfo {author} {\bibfnamefont {R.~J.}\ \bibnamefont
  {{Davis}}}, \bibinfo {author} {\bibfnamefont {P.}~\bibnamefont {{Dewdney}}},
  \bibinfo {author} {\bibfnamefont {C.}~\bibnamefont {{Dickinson}}}, \bibinfo
  {author} {\bibfnamefont {G.}~\bibnamefont {{Heron}}}, \bibinfo {author}
  {\bibfnamefont {B.}~\bibnamefont {{Maffei}}}, \bibinfo {author}
  {\bibfnamefont {A.}~\bibnamefont {{Pourtsidou}}}, \ and\ \bibinfo {author}
  {\bibfnamefont {P.~N.}\ \bibnamefont {{Wilkinson}}},\ }\href@noop {}
  {\bibfield  {journal} {\bibinfo  {journal} {ArXiv e-prints}\ } (\bibinfo
  {year} {2012})},\ \Eprint {http://arxiv.org/abs/1209.1041} {arXiv:1209.1041
  [astro-ph.CO]} \BibitemShut {NoStop}%
\bibitem [{\citenamefont {{Newburgh}}\ \emph {et~al.}(2014)\citenamefont
  {{Newburgh}}, \citenamefont {{Addison}}, \citenamefont {{Amiri}},
  \citenamefont {{Bandura}}, \citenamefont {{Bond}} \emph
  {et~al.}}]{2014SPIE.9145E..4VN}%
  \BibitemOpen
  \bibfield  {author} {\bibinfo {author} {\bibfnamefont {L.~B.}\ \bibnamefont
  {{Newburgh}}}, \bibinfo {author} {\bibfnamefont {G.~E.}\ \bibnamefont
  {{Addison}}}, \bibinfo {author} {\bibfnamefont {M.}~\bibnamefont {{Amiri}}},
  \bibinfo {author} {\bibfnamefont {K.}~\bibnamefont {{Bandura}}}, \bibinfo
  {author} {\bibfnamefont {J.~R.}\ \bibnamefont {{Bond}}},  \emph {et~al.},\
  }in\ \href {\doibase 10.1117/12.2056962} {\emph {\bibinfo {booktitle}
  {Ground-based and Airborne Telescopes V}}},\ \bibinfo {series} {\procspie},
  Vol.\ \bibinfo {volume} {9145}\ (\bibinfo {year} {2014})\ p.\ \bibinfo
  {pages} {91454V},\ \Eprint {http://arxiv.org/abs/1406.2267} {arXiv:1406.2267
  [astro-ph.IM]} \BibitemShut {NoStop}%
\bibitem [{\citenamefont {{Bigot-Sazy}}\ \emph {et~al.}(2016)\citenamefont
  {{Bigot-Sazy}}, \citenamefont {{Ma}}, \citenamefont {{Battye}}, \citenamefont
  {{Browne}}, \citenamefont {{Chen}} \emph {et~al.}}]{2016ASPC..502...41B}%
  \BibitemOpen
  \bibfield  {author} {\bibinfo {author} {\bibfnamefont {M.-A.}\ \bibnamefont
  {{Bigot-Sazy}}}, \bibinfo {author} {\bibfnamefont {Y.-Z.}\ \bibnamefont
  {{Ma}}}, \bibinfo {author} {\bibfnamefont {R.~A.}\ \bibnamefont {{Battye}}},
  \bibinfo {author} {\bibfnamefont {I.~W.~A.}\ \bibnamefont {{Browne}}},
  \bibinfo {author} {\bibfnamefont {T.}~\bibnamefont {{Chen}}},  \emph
  {et~al.},\ }in\ \href@noop {} {\emph {\bibinfo {booktitle} {Frontiers in
  Radio Astronomy and FAST Early Sciences Symposium 2015}}},\ \bibinfo {series}
  {Astronomical Society of the Pacific Conference Series}, Vol.\ \bibinfo
  {volume} {502},\ \bibinfo {editor} {edited by\ \bibinfo {editor}
  {\bibfnamefont {L.}~\bibnamefont {{Qain}}}\ and\ \bibinfo {editor}
  {\bibfnamefont {D.}~\bibnamefont {{Li}}}}\ (\bibinfo {year} {2016})\
  p.~\bibinfo {pages} {41},\ \Eprint {http://arxiv.org/abs/1511.03006}
  {arXiv:1511.03006} \BibitemShut {NoStop}%
\bibitem [{\citenamefont {{Chen}}(2011)}]{2011SSPMA..41.1358C}%
  \BibitemOpen
  \bibfield  {author} {\bibinfo {author} {\bibfnamefont {X.}~\bibnamefont
  {{Chen}}},\ }\href {\doibase 10.1360/132011-972} {\bibfield  {journal}
  {\bibinfo  {journal} {Scientia Sinica Physica, Mechanica \& Astronomica}\
  }\textbf {\bibinfo {volume} {41}},\ \bibinfo {pages} {1358} (\bibinfo {year}
  {2011})}\BibitemShut {NoStop}%
\bibitem [{\citenamefont {{Santos}}\ \emph {et~al.}(2005)\citenamefont
  {{Santos}}, \citenamefont {{Cooray}},\ and\ \citenamefont
  {{Knox}}}]{2005ApJ...625..575S}%
  \BibitemOpen
  \bibfield  {author} {\bibinfo {author} {\bibfnamefont {M.~G.}\ \bibnamefont
  {{Santos}}}, \bibinfo {author} {\bibfnamefont {A.}~\bibnamefont {{Cooray}}},
  \ and\ \bibinfo {author} {\bibfnamefont {L.}~\bibnamefont {{Knox}}},\ }\href
  {\doibase 10.1086/429857} {\bibfield  {journal} {\bibinfo  {journal} {\apj}\
  }\textbf {\bibinfo {volume} {625}},\ \bibinfo {pages} {575} (\bibinfo {year}
  {2005})},\ \Eprint {http://arxiv.org/abs/astro-ph/0408515} {astro-ph/0408515}
  \BibitemShut {NoStop}%
\bibitem [{\citenamefont {{Wolz}}\ \emph {et~al.}(2014)\citenamefont {{Wolz}},
  \citenamefont {{Abdalla}}, \citenamefont {{Blake}}, \citenamefont {{Shaw}},
  \citenamefont {{Chapman}},\ and\ \citenamefont
  {{Rawlings}}}]{2014MNRAS.441.3271W}%
  \BibitemOpen
  \bibfield  {author} {\bibinfo {author} {\bibfnamefont {L.}~\bibnamefont
  {{Wolz}}}, \bibinfo {author} {\bibfnamefont {F.~B.}\ \bibnamefont
  {{Abdalla}}}, \bibinfo {author} {\bibfnamefont {C.}~\bibnamefont {{Blake}}},
  \bibinfo {author} {\bibfnamefont {J.~R.}\ \bibnamefont {{Shaw}}}, \bibinfo
  {author} {\bibfnamefont {E.}~\bibnamefont {{Chapman}}}, \ and\ \bibinfo
  {author} {\bibfnamefont {S.}~\bibnamefont {{Rawlings}}},\ }\href {\doibase
  10.1093/mnras/stu792} {\bibfield  {journal} {\bibinfo  {journal} {\mnras}\
  }\textbf {\bibinfo {volume} {441}},\ \bibinfo {pages} {3271} (\bibinfo {year}
  {2014})},\ \Eprint {http://arxiv.org/abs/1310.8144} {arXiv:1310.8144}
  \BibitemShut {NoStop}%
\bibitem [{\citenamefont {{Shaw}}\ \emph {et~al.}(2015)\citenamefont {{Shaw}},
  \citenamefont {{Sigurdson}}, \citenamefont {{Sitwell}}, \citenamefont
  {{Stebbins}},\ and\ \citenamefont {{Pen}}}]{2015PhRvD..91h3514S}%
  \BibitemOpen
  \bibfield  {author} {\bibinfo {author} {\bibfnamefont {J.~R.}\ \bibnamefont
  {{Shaw}}}, \bibinfo {author} {\bibfnamefont {K.}~\bibnamefont {{Sigurdson}}},
  \bibinfo {author} {\bibfnamefont {M.}~\bibnamefont {{Sitwell}}}, \bibinfo
  {author} {\bibfnamefont {A.}~\bibnamefont {{Stebbins}}}, \ and\ \bibinfo
  {author} {\bibfnamefont {U.-L.}\ \bibnamefont {{Pen}}},\ }\href {\doibase
  10.1103/PhysRevD.91.083514} {\bibfield  {journal} {\bibinfo  {journal}
  {\prd}\ }\textbf {\bibinfo {volume} {91}},\ \bibinfo {eid} {083514} (\bibinfo
  {year} {2015})},\ \Eprint {http://arxiv.org/abs/1401.2095} {arXiv:1401.2095}
  \BibitemShut {NoStop}%
\bibitem [{\citenamefont {{Alonso}}\ \emph {et~al.}(2015)\citenamefont
  {{Alonso}}, \citenamefont {{Bull}}, \citenamefont {{Ferreira}},\ and\
  \citenamefont {{Santos}}}]{2015MNRAS.447..400A}%
  \BibitemOpen
  \bibfield  {author} {\bibinfo {author} {\bibfnamefont {D.}~\bibnamefont
  {{Alonso}}}, \bibinfo {author} {\bibfnamefont {P.}~\bibnamefont {{Bull}}},
  \bibinfo {author} {\bibfnamefont {P.~G.}\ \bibnamefont {{Ferreira}}}, \ and\
  \bibinfo {author} {\bibfnamefont {M.~G.}\ \bibnamefont {{Santos}}},\ }\href
  {\doibase 10.1093/mnras/stu2474} {\bibfield  {journal} {\bibinfo  {journal}
  {\mnras}\ }\textbf {\bibinfo {volume} {447}},\ \bibinfo {pages} {400}
  (\bibinfo {year} {2015})},\ \Eprint {http://arxiv.org/abs/1409.8667}
  {arXiv:1409.8667} \BibitemShut {NoStop}%
\bibitem [{\citenamefont {{Planck Collaboration}}\ \emph
  {et~al.}(2016)\citenamefont {{Planck Collaboration}} \emph
  {et~al.}}]{2016A&A...594A..10P}%
  \BibitemOpen
  \bibfield  {author} {\bibinfo {author} {\bibnamefont {{Planck
  Collaboration}}} \emph {et~al.},\ }\href {\doibase
  10.1051/0004-6361/201525967} {\bibfield  {journal} {\bibinfo  {journal}
  {\aap}\ }\textbf {\bibinfo {volume} {594}},\ \bibinfo {eid} {A10} (\bibinfo
  {year} {2016})},\ \Eprint {http://arxiv.org/abs/1502.01588}
  {arXiv:1502.01588} \BibitemShut {NoStop}%
\bibitem [{\citenamefont {{Levi}}\ \emph {et~al.}(2013)\citenamefont {{Levi}},
  \citenamefont {{Bebek}}, \citenamefont {{Beers}}, \citenamefont {{Blum}},
  \citenamefont {{Cahn}} \emph {et~al.}}]{2013arXiv1308.0847L}%
  \BibitemOpen
  \bibfield  {author} {\bibinfo {author} {\bibfnamefont {M.}~\bibnamefont
  {{Levi}}}, \bibinfo {author} {\bibfnamefont {C.}~\bibnamefont {{Bebek}}},
  \bibinfo {author} {\bibfnamefont {T.}~\bibnamefont {{Beers}}}, \bibinfo
  {author} {\bibfnamefont {R.}~\bibnamefont {{Blum}}}, \bibinfo {author}
  {\bibfnamefont {R.}~\bibnamefont {{Cahn}}},  \emph {et~al.},\ }\href@noop {}
  {\bibfield  {journal} {\bibinfo  {journal} {ArXiv e-prints}\ } (\bibinfo
  {year} {2013})},\ \Eprint {http://arxiv.org/abs/1308.0847} {arXiv:1308.0847
  [astro-ph.CO]} \BibitemShut {NoStop}%
\bibitem [{\citenamefont {{Font-Ribera}}\ \emph {et~al.}(2014)\citenamefont
  {{Font-Ribera}}, \citenamefont {{McDonald}}, \citenamefont {{Mostek}},
  \citenamefont {{Reid}}, \citenamefont {{Seo}},\ and\ \citenamefont
  {{Slosar}}}]{2014JCAP...05..023F}%
  \BibitemOpen
  \bibfield  {author} {\bibinfo {author} {\bibfnamefont {A.}~\bibnamefont
  {{Font-Ribera}}}, \bibinfo {author} {\bibfnamefont {P.}~\bibnamefont
  {{McDonald}}}, \bibinfo {author} {\bibfnamefont {N.}~\bibnamefont
  {{Mostek}}}, \bibinfo {author} {\bibfnamefont {B.~A.}\ \bibnamefont
  {{Reid}}}, \bibinfo {author} {\bibfnamefont {H.-J.}\ \bibnamefont {{Seo}}}, \
  and\ \bibinfo {author} {\bibfnamefont {A.}~\bibnamefont {{Slosar}}},\ }\href
  {\doibase 10.1088/1475-7516/2014/05/023} {\bibfield  {journal} {\bibinfo
  {journal} {\jcap}\ }\textbf {\bibinfo {volume} {5}},\ \bibinfo {eid} {023}
  (\bibinfo {year} {2014})},\ \Eprint {http://arxiv.org/abs/1308.4164}
  {arXiv:1308.4164} \BibitemShut {NoStop}%
\bibitem [{\citenamefont {{Laureijs}}\ \emph {et~al.}(2011)\citenamefont
  {{Laureijs}}, \citenamefont {{Amiaux}}, \citenamefont {{Arduini}},
  \citenamefont {{Augu{\`e}res}}, \citenamefont {{Brinchmann}} \emph
  {et~al.}}]{2011arXiv1110.3193L}%
  \BibitemOpen
  \bibfield  {author} {\bibinfo {author} {\bibfnamefont {R.}~\bibnamefont
  {{Laureijs}}}, \bibinfo {author} {\bibfnamefont {J.}~\bibnamefont
  {{Amiaux}}}, \bibinfo {author} {\bibfnamefont {S.}~\bibnamefont {{Arduini}}},
  \bibinfo {author} {\bibfnamefont {J.~.}\ \bibnamefont {{Augu{\`e}res}}},
  \bibinfo {author} {\bibfnamefont {J.}~\bibnamefont {{Brinchmann}}},  \emph
  {et~al.},\ }\href@noop {} {\bibfield  {journal} {\bibinfo  {journal} {ArXiv
  e-prints}\ } (\bibinfo {year} {2011})},\ \Eprint
  {http://arxiv.org/abs/1110.3193} {arXiv:1110.3193 [astro-ph.CO]} \BibitemShut
  {NoStop}%
\bibitem [{\citenamefont {{Spergel}}\ \emph {et~al.}(2013)\citenamefont
  {{Spergel}}, \citenamefont {{Gehrels}}, \citenamefont {{Breckinridge}},
  \citenamefont {{Donahue}}, \citenamefont {{Dressler}} \emph
  {et~al.}}]{2013arXiv1305.5422S}%
  \BibitemOpen
  \bibfield  {author} {\bibinfo {author} {\bibfnamefont {D.}~\bibnamefont
  {{Spergel}}}, \bibinfo {author} {\bibfnamefont {N.}~\bibnamefont
  {{Gehrels}}}, \bibinfo {author} {\bibfnamefont {J.}~\bibnamefont
  {{Breckinridge}}}, \bibinfo {author} {\bibfnamefont {M.}~\bibnamefont
  {{Donahue}}}, \bibinfo {author} {\bibfnamefont {A.}~\bibnamefont
  {{Dressler}}},  \emph {et~al.},\ }\href@noop {} {\bibfield  {journal}
  {\bibinfo  {journal} {ArXiv e-prints}\ } (\bibinfo {year} {2013})},\ \Eprint
  {http://arxiv.org/abs/1305.5422} {arXiv:1305.5422 [astro-ph.IM]} \BibitemShut
  {NoStop}%
\bibitem [{\citenamefont {{Tegmark}}\ \emph {et~al.}(1998)\citenamefont
  {{Tegmark}}, \citenamefont {{Hamilton}}, \citenamefont {{Strauss}},
  \citenamefont {{Vogeley}},\ and\ \citenamefont
  {{Szalay}}}]{1998ApJ...499..555T}%
  \BibitemOpen
  \bibfield  {author} {\bibinfo {author} {\bibfnamefont {M.}~\bibnamefont
  {{Tegmark}}}, \bibinfo {author} {\bibfnamefont {A.~J.~S.}\ \bibnamefont
  {{Hamilton}}}, \bibinfo {author} {\bibfnamefont {M.~A.}\ \bibnamefont
  {{Strauss}}}, \bibinfo {author} {\bibfnamefont {M.~S.}\ \bibnamefont
  {{Vogeley}}}, \ and\ \bibinfo {author} {\bibfnamefont {A.~S.}\ \bibnamefont
  {{Szalay}}},\ }\href {\doibase 10.1086/305663} {\bibfield  {journal}
  {\bibinfo  {journal} {\apj}\ }\textbf {\bibinfo {volume} {499}},\ \bibinfo
  {pages} {555} (\bibinfo {year} {1998})},\ \Eprint
  {http://arxiv.org/abs/astro-ph/9708020} {astro-ph/9708020} \BibitemShut
  {NoStop}%
\bibitem [{\citenamefont {{Bond}}\ \emph {et~al.}(1998)\citenamefont {{Bond}},
  \citenamefont {{Jaffe}},\ and\ \citenamefont {{Knox}}}]{1998PhRvD..57.2117B}%
  \BibitemOpen
  \bibfield  {author} {\bibinfo {author} {\bibfnamefont {J.~R.}\ \bibnamefont
  {{Bond}}}, \bibinfo {author} {\bibfnamefont {A.~H.}\ \bibnamefont {{Jaffe}}},
  \ and\ \bibinfo {author} {\bibfnamefont {L.}~\bibnamefont {{Knox}}},\ }\href
  {\doibase 10.1103/PhysRevD.57.2117} {\bibfield  {journal} {\bibinfo
  {journal} {\prd}\ }\textbf {\bibinfo {volume} {57}},\ \bibinfo {pages} {2117}
  (\bibinfo {year} {1998})},\ \Eprint {http://arxiv.org/abs/astro-ph/9708203}
  {astro-ph/9708203} \BibitemShut {NoStop}%
\bibitem [{\citenamefont {{Newman}}\ \emph {et~al.}(2015)\citenamefont
  {{Newman}}, \citenamefont {{Abate}}, \citenamefont {{Abdalla}}, \citenamefont
  {{Allam}}, \citenamefont {{Allen}} \emph {et~al.}}]{2015APh....63...81N}%
  \BibitemOpen
  \bibfield  {author} {\bibinfo {author} {\bibfnamefont {J.~A.}\ \bibnamefont
  {{Newman}}}, \bibinfo {author} {\bibfnamefont {A.}~\bibnamefont {{Abate}}},
  \bibinfo {author} {\bibfnamefont {F.~B.}\ \bibnamefont {{Abdalla}}}, \bibinfo
  {author} {\bibfnamefont {S.}~\bibnamefont {{Allam}}}, \bibinfo {author}
  {\bibfnamefont {S.~W.}\ \bibnamefont {{Allen}}},  \emph {et~al.},\ }\href
  {\doibase 10.1016/j.astropartphys.2014.06.007} {\bibfield  {journal}
  {\bibinfo  {journal} {Astroparticle Physics}\ }\textbf {\bibinfo {volume}
  {63}},\ \bibinfo {pages} {81} (\bibinfo {year} {2015})},\ \Eprint
  {http://arxiv.org/abs/1309.5384} {arXiv:1309.5384} \BibitemShut {NoStop}%
\bibitem [{\citenamefont {{de Putter}}\ \emph {et~al.}(2014)\citenamefont {{de
  Putter}}, \citenamefont {{Dor{\'e}}},\ and\ \citenamefont
  {{Das}}}]{2014ApJ...780..185D}%
  \BibitemOpen
  \bibfield  {author} {\bibinfo {author} {\bibfnamefont {R.}~\bibnamefont {{de
  Putter}}}, \bibinfo {author} {\bibfnamefont {O.}~\bibnamefont {{Dor{\'e}}}},
  \ and\ \bibinfo {author} {\bibfnamefont {S.}~\bibnamefont {{Das}}},\ }\href
  {\doibase 10.1088/0004-637X/780/2/185} {\bibfield  {journal} {\bibinfo
  {journal} {\apj}\ }\textbf {\bibinfo {volume} {780}},\ \bibinfo {eid} {185}
  (\bibinfo {year} {2014})},\ \Eprint {http://arxiv.org/abs/1306.0534}
  {arXiv:1306.0534} \BibitemShut {NoStop}%
\bibitem [{\citenamefont {{The Dark Energy Survey
  Collaboration}}(2005)}]{2005astro.ph.10346T}%
  \BibitemOpen
  \bibfield  {author} {\bibinfo {author} {\bibnamefont {{The Dark Energy Survey
  Collaboration}}},\ }\href@noop {} {\bibfield  {journal} {\bibinfo  {journal}
  {ArXiv Astrophysics e-prints}\ } (\bibinfo {year} {2005})},\ \Eprint
  {http://arxiv.org/abs/astro-ph/0510346} {astro-ph/0510346} \BibitemShut
  {NoStop}%
\bibitem [{\citenamefont {{Camera}}\ \emph {et~al.}(2013)\citenamefont
  {{Camera}}, \citenamefont {{Santos}}, \citenamefont {{Ferreira}},\ and\
  \citenamefont {{Ferramacho}}}]{2013PhRvL.111q1302C}%
  \BibitemOpen
  \bibfield  {author} {\bibinfo {author} {\bibfnamefont {S.}~\bibnamefont
  {{Camera}}}, \bibinfo {author} {\bibfnamefont {M.~G.}\ \bibnamefont
  {{Santos}}}, \bibinfo {author} {\bibfnamefont {P.~G.}\ \bibnamefont
  {{Ferreira}}}, \ and\ \bibinfo {author} {\bibfnamefont {L.}~\bibnamefont
  {{Ferramacho}}},\ }\href {\doibase 10.1103/PhysRevLett.111.171302} {\bibfield
   {journal} {\bibinfo  {journal} {Physical Review Letters}\ }\textbf {\bibinfo
  {volume} {111}},\ \bibinfo {eid} {171302} (\bibinfo {year} {2013})},\ \Eprint
  {http://arxiv.org/abs/1305.6928} {arXiv:1305.6928 [astro-ph.CO]} \BibitemShut
  {NoStop}%
\bibitem [{\citenamefont {{Jeli{\'c}}}\ \emph {et~al.}(2010)\citenamefont
  {{Jeli{\'c}}}, \citenamefont {{Zaroubi}}, \citenamefont {{Labropoulos}},
  \citenamefont {{Bernardi}}, \citenamefont {{de Bruyn}},\ and\ \citenamefont
  {{Koopmans}}}]{2010MNRAS.409.1647J}%
  \BibitemOpen
  \bibfield  {author} {\bibinfo {author} {\bibfnamefont {V.}~\bibnamefont
  {{Jeli{\'c}}}}, \bibinfo {author} {\bibfnamefont {S.}~\bibnamefont
  {{Zaroubi}}}, \bibinfo {author} {\bibfnamefont {P.}~\bibnamefont
  {{Labropoulos}}}, \bibinfo {author} {\bibfnamefont {G.}~\bibnamefont
  {{Bernardi}}}, \bibinfo {author} {\bibfnamefont {A.~G.}\ \bibnamefont {{de
  Bruyn}}}, \ and\ \bibinfo {author} {\bibfnamefont {L.~V.~E.}\ \bibnamefont
  {{Koopmans}}},\ }\href {\doibase 10.1111/j.1365-2966.2010.17407.x} {\bibfield
   {journal} {\bibinfo  {journal} {\mnras}\ }\textbf {\bibinfo {volume}
  {409}},\ \bibinfo {pages} {1647} (\bibinfo {year} {2010})},\ \Eprint
  {http://arxiv.org/abs/1007.4135} {arXiv:1007.4135} \BibitemShut {NoStop}%
\bibitem [{\citenamefont {{Datta}}\ \emph {et~al.}(2010)\citenamefont
  {{Datta}}, \citenamefont {{Bowman}},\ and\ \citenamefont
  {{Carilli}}}]{2010ApJ...724..526D}%
  \BibitemOpen
  \bibfield  {author} {\bibinfo {author} {\bibfnamefont {A.}~\bibnamefont
  {{Datta}}}, \bibinfo {author} {\bibfnamefont {J.~D.}\ \bibnamefont
  {{Bowman}}}, \ and\ \bibinfo {author} {\bibfnamefont {C.~L.}\ \bibnamefont
  {{Carilli}}},\ }\href {\doibase 10.1088/0004-637X/724/1/526} {\bibfield
  {journal} {\bibinfo  {journal} {\apj}\ }\textbf {\bibinfo {volume} {724}},\
  \bibinfo {pages} {526} (\bibinfo {year} {2010})},\ \Eprint
  {http://arxiv.org/abs/1005.4071} {arXiv:1005.4071} \BibitemShut {NoStop}%
\bibitem [{\citenamefont {{Vedantham}}\ \emph {et~al.}(2012)\citenamefont
  {{Vedantham}}, \citenamefont {{Udaya Shankar}},\ and\ \citenamefont
  {{Subrahmanyan}}}]{2012ApJ...745..176V}%
  \BibitemOpen
  \bibfield  {author} {\bibinfo {author} {\bibfnamefont {H.}~\bibnamefont
  {{Vedantham}}}, \bibinfo {author} {\bibfnamefont {N.}~\bibnamefont {{Udaya
  Shankar}}}, \ and\ \bibinfo {author} {\bibfnamefont {R.}~\bibnamefont
  {{Subrahmanyan}}},\ }\href {\doibase 10.1088/0004-637X/745/2/176} {\bibfield
  {journal} {\bibinfo  {journal} {\apj}\ }\textbf {\bibinfo {volume} {745}},\
  \bibinfo {eid} {176} (\bibinfo {year} {2012})},\ \Eprint
  {http://arxiv.org/abs/1106.1297} {arXiv:1106.1297 [astro-ph.IM]} \BibitemShut
  {NoStop}%
\bibitem [{\citenamefont {{Liu}}\ \emph {et~al.}(2014)\citenamefont {{Liu}},
  \citenamefont {{Parsons}},\ and\ \citenamefont
  {{Trott}}}]{2014PhRvD..90b3019L}%
  \BibitemOpen
  \bibfield  {author} {\bibinfo {author} {\bibfnamefont {A.}~\bibnamefont
  {{Liu}}}, \bibinfo {author} {\bibfnamefont {A.~R.}\ \bibnamefont
  {{Parsons}}}, \ and\ \bibinfo {author} {\bibfnamefont {C.~M.}\ \bibnamefont
  {{Trott}}},\ }\href {\doibase 10.1103/PhysRevD.90.023019} {\bibfield
  {journal} {\bibinfo  {journal} {\prd}\ }\textbf {\bibinfo {volume} {90}},\
  \bibinfo {eid} {023019} (\bibinfo {year} {2014})},\ \Eprint
  {http://arxiv.org/abs/1404.4372} {arXiv:1404.4372} \BibitemShut {NoStop}%
\bibitem [{\citenamefont {{Seo}}\ and\ \citenamefont
  {{Hirata}}(2016)}]{2016MNRAS.456.3142S}%
  \BibitemOpen
  \bibfield  {author} {\bibinfo {author} {\bibfnamefont {H.-J.}\ \bibnamefont
  {{Seo}}}\ and\ \bibinfo {author} {\bibfnamefont {C.~M.}\ \bibnamefont
  {{Hirata}}},\ }\href {\doibase 10.1093/mnras/stv2806} {\bibfield  {journal}
  {\bibinfo  {journal} {\mnras}\ }\textbf {\bibinfo {volume} {456}},\ \bibinfo
  {pages} {3142} (\bibinfo {year} {2016})},\ \Eprint
  {http://arxiv.org/abs/1508.06503} {arXiv:1508.06503} \BibitemShut {NoStop}%
\bibitem [{\citenamefont {{Parsons}}\ \emph {et~al.}(2012)\citenamefont
  {{Parsons}}, \citenamefont {{Pober}}, \citenamefont {{Aguirre}},
  \citenamefont {{Carilli}}, \citenamefont {{Jacobs}},\ and\ \citenamefont
  {{Moore}}}]{2012ApJ...756..165P}%
  \BibitemOpen
  \bibfield  {author} {\bibinfo {author} {\bibfnamefont {A.~R.}\ \bibnamefont
  {{Parsons}}}, \bibinfo {author} {\bibfnamefont {J.~C.}\ \bibnamefont
  {{Pober}}}, \bibinfo {author} {\bibfnamefont {J.~E.}\ \bibnamefont
  {{Aguirre}}}, \bibinfo {author} {\bibfnamefont {C.~L.}\ \bibnamefont
  {{Carilli}}}, \bibinfo {author} {\bibfnamefont {D.~C.}\ \bibnamefont
  {{Jacobs}}}, \ and\ \bibinfo {author} {\bibfnamefont {D.~F.}\ \bibnamefont
  {{Moore}}},\ }\href {\doibase 10.1088/0004-637X/756/2/165} {\bibfield
  {journal} {\bibinfo  {journal} {\apj}\ }\textbf {\bibinfo {volume} {756}},\
  \bibinfo {eid} {165} (\bibinfo {year} {2012})},\ \Eprint
  {http://arxiv.org/abs/1204.4749} {arXiv:1204.4749 [astro-ph.IM]} \BibitemShut
  {NoStop}%
\bibitem [{\citenamefont {{Pober}}\ \emph {et~al.}(2014)\citenamefont
  {{Pober}}, \citenamefont {{Liu}}, \citenamefont {{Dillon}}, \citenamefont
  {{Aguirre}}, \citenamefont {{Bowman}}, \citenamefont {{Bradley}},
  \citenamefont {{Carilli}}, \citenamefont {{DeBoer}}, \citenamefont
  {{Hewitt}}, \citenamefont {{Jacobs}}, \citenamefont {{McQuinn}},
  \citenamefont {{Morales}}, \citenamefont {{Parsons}}, \citenamefont
  {{Tegmark}},\ and\ \citenamefont {{Werthimer}}}]{2014ApJ...782...66P}%
  \BibitemOpen
  \bibfield  {author} {\bibinfo {author} {\bibfnamefont {J.~C.}\ \bibnamefont
  {{Pober}}}, \bibinfo {author} {\bibfnamefont {A.}~\bibnamefont {{Liu}}},
  \bibinfo {author} {\bibfnamefont {J.~S.}\ \bibnamefont {{Dillon}}}, \bibinfo
  {author} {\bibfnamefont {J.~E.}\ \bibnamefont {{Aguirre}}}, \bibinfo {author}
  {\bibfnamefont {J.~D.}\ \bibnamefont {{Bowman}}}, \bibinfo {author}
  {\bibfnamefont {R.~F.}\ \bibnamefont {{Bradley}}}, \bibinfo {author}
  {\bibfnamefont {C.~L.}\ \bibnamefont {{Carilli}}}, \bibinfo {author}
  {\bibfnamefont {D.~R.}\ \bibnamefont {{DeBoer}}}, \bibinfo {author}
  {\bibfnamefont {J.~N.}\ \bibnamefont {{Hewitt}}}, \bibinfo {author}
  {\bibfnamefont {D.~C.}\ \bibnamefont {{Jacobs}}}, \bibinfo {author}
  {\bibfnamefont {M.}~\bibnamefont {{McQuinn}}}, \bibinfo {author}
  {\bibfnamefont {M.~F.}\ \bibnamefont {{Morales}}}, \bibinfo {author}
  {\bibfnamefont {A.~R.}\ \bibnamefont {{Parsons}}}, \bibinfo {author}
  {\bibfnamefont {M.}~\bibnamefont {{Tegmark}}}, \ and\ \bibinfo {author}
  {\bibfnamefont {D.~J.}\ \bibnamefont {{Werthimer}}},\ }\href {\doibase
  10.1088/0004-637X/782/2/66} {\bibfield  {journal} {\bibinfo  {journal}
  {\apj}\ }\textbf {\bibinfo {volume} {782}},\ \bibinfo {eid} {66} (\bibinfo
  {year} {2014})},\ \Eprint {http://arxiv.org/abs/1310.7031} {arXiv:1310.7031}
  \BibitemShut {NoStop}%
\bibitem [{\citenamefont {{Abazajian}}\ \emph {et~al.}(2016)\citenamefont
  {{Abazajian}}, \citenamefont {{Adshead}}, \citenamefont {{Ahmed}},
  \citenamefont {{Allen}}, \citenamefont {{Alonso}} \emph
  {et~al.}}]{2016arXiv161002743A}%
  \BibitemOpen
  \bibfield  {author} {\bibinfo {author} {\bibfnamefont {K.~N.}\ \bibnamefont
  {{Abazajian}}}, \bibinfo {author} {\bibfnamefont {P.}~\bibnamefont
  {{Adshead}}}, \bibinfo {author} {\bibfnamefont {Z.}~\bibnamefont {{Ahmed}}},
  \bibinfo {author} {\bibfnamefont {S.~W.}\ \bibnamefont {{Allen}}}, \bibinfo
  {author} {\bibfnamefont {D.}~\bibnamefont {{Alonso}}},  \emph {et~al.},\
  }\href@noop {} {\bibfield  {journal} {\bibinfo  {journal} {ArXiv e-prints}\ }
  (\bibinfo {year} {2016})},\ \Eprint {http://arxiv.org/abs/1610.02743}
  {arXiv:1610.02743} \BibitemShut {NoStop}%
\bibitem [{\citenamefont {{Calabrese}}\ \emph {et~al.}(2016)\citenamefont
  {{Calabrese}}, \citenamefont {{Alonso}},\ and\ \citenamefont
  {{Dunkley}}}]{2016arXiv161110269C}%
  \BibitemOpen
  \bibfield  {author} {\bibinfo {author} {\bibfnamefont {E.}~\bibnamefont
  {{Calabrese}}}, \bibinfo {author} {\bibfnamefont {D.}~\bibnamefont
  {{Alonso}}}, \ and\ \bibinfo {author} {\bibfnamefont {J.}~\bibnamefont
  {{Dunkley}}},\ }\href@noop {} {\bibfield  {journal} {\bibinfo  {journal}
  {ArXiv e-prints}\ } (\bibinfo {year} {2016})},\ \Eprint
  {http://arxiv.org/abs/1611.10269} {arXiv:1611.10269} \BibitemShut {NoStop}%
\bibitem [{\citenamefont {{Baxter}}\ \emph {et~al.}(2016)\citenamefont
  {{Baxter}}, \citenamefont {{Clampitt}}, \citenamefont {{Giannantonio}},
  \citenamefont {{Dodelson}}, \citenamefont {{Jain}} \emph
  {et~al.}}]{2016MNRAS.461.4099B}%
  \BibitemOpen
  \bibfield  {author} {\bibinfo {author} {\bibfnamefont {E.}~\bibnamefont
  {{Baxter}}}, \bibinfo {author} {\bibfnamefont {J.}~\bibnamefont
  {{Clampitt}}}, \bibinfo {author} {\bibfnamefont {T.}~\bibnamefont
  {{Giannantonio}}}, \bibinfo {author} {\bibfnamefont {S.}~\bibnamefont
  {{Dodelson}}}, \bibinfo {author} {\bibfnamefont {B.}~\bibnamefont {{Jain}}},
  \emph {et~al.},\ }\href {\doibase 10.1093/mnras/stw1584} {\bibfield
  {journal} {\bibinfo  {journal} {\mnras}\ }\textbf {\bibinfo {volume} {461}},\
  \bibinfo {pages} {4099} (\bibinfo {year} {2016})},\ \Eprint
  {http://arxiv.org/abs/1602.07384} {arXiv:1602.07384} \BibitemShut {NoStop}%
\bibitem [{\citenamefont {{Takahashi}}\ \emph {et~al.}(2012)\citenamefont
  {{Takahashi}}, \citenamefont {{Sato}}, \citenamefont {{Nishimichi}},
  \citenamefont {{Taruya}},\ and\ \citenamefont
  {{Oguri}}}]{2012ApJ...761..152T}%
  \BibitemOpen
  \bibfield  {author} {\bibinfo {author} {\bibfnamefont {R.}~\bibnamefont
  {{Takahashi}}}, \bibinfo {author} {\bibfnamefont {M.}~\bibnamefont {{Sato}}},
  \bibinfo {author} {\bibfnamefont {T.}~\bibnamefont {{Nishimichi}}}, \bibinfo
  {author} {\bibfnamefont {A.}~\bibnamefont {{Taruya}}}, \ and\ \bibinfo
  {author} {\bibfnamefont {M.}~\bibnamefont {{Oguri}}},\ }\href {\doibase
  10.1088/0004-637X/761/2/152} {\bibfield  {journal} {\bibinfo  {journal}
  {\apj}\ }\textbf {\bibinfo {volume} {761}},\ \bibinfo {eid} {152} (\bibinfo
  {year} {2012})},\ \Eprint {http://arxiv.org/abs/1208.2701} {arXiv:1208.2701}
  \BibitemShut {NoStop}%
\bibitem [{\citenamefont {{Blas}}\ \emph {et~al.}(2011)\citenamefont {{Blas}},
  \citenamefont {{Lesgourgues}},\ and\ \citenamefont
  {{Tram}}}]{2011JCAP...07..034B}%
  \BibitemOpen
  \bibfield  {author} {\bibinfo {author} {\bibfnamefont {D.}~\bibnamefont
  {{Blas}}}, \bibinfo {author} {\bibfnamefont {J.}~\bibnamefont
  {{Lesgourgues}}}, \ and\ \bibinfo {author} {\bibfnamefont {T.}~\bibnamefont
  {{Tram}}},\ }\href {\doibase 10.1088/1475-7516/2011/07/034} {\bibfield
  {journal} {\bibinfo  {journal} {\jcap}\ }\textbf {\bibinfo {volume} {7}},\
  \bibinfo {eid} {034} (\bibinfo {year} {2011})},\ \Eprint
  {http://arxiv.org/abs/1104.2933} {arXiv:1104.2933} \BibitemShut {NoStop}%
\bibitem [{\citenamefont {{Di Dio}}\ \emph {et~al.}(2013)\citenamefont {{Di
  Dio}}, \citenamefont {{Montanari}}, \citenamefont {{Lesgourgues}},\ and\
  \citenamefont {{Durrer}}}]{2013JCAP...11..044D}%
  \BibitemOpen
  \bibfield  {author} {\bibinfo {author} {\bibfnamefont {E.}~\bibnamefont {{Di
  Dio}}}, \bibinfo {author} {\bibfnamefont {F.}~\bibnamefont {{Montanari}}},
  \bibinfo {author} {\bibfnamefont {J.}~\bibnamefont {{Lesgourgues}}}, \ and\
  \bibinfo {author} {\bibfnamefont {R.}~\bibnamefont {{Durrer}}},\ }\href
  {\doibase 10.1088/1475-7516/2013/11/044} {\bibfield  {journal} {\bibinfo
  {journal} {\jcap}\ }\textbf {\bibinfo {volume} {11}},\ \bibinfo {eid} {044}
  (\bibinfo {year} {2013})},\ \Eprint {http://arxiv.org/abs/1307.1459}
  {arXiv:1307.1459} \BibitemShut {NoStop}%
\end{thebibliography}%

\appendix
\onecolumngrid
  \section{Individual clustering redshifts}\label{app:ind_phz}
    The idea of using clustering information to constrain the redshifts of individual objects
    of a given sample has been considered before in the literature, and shown to yield
    interesting results even in the absence of spectroscopic data \cite{2012MNRAS.425.1042J}.
    Here we outline the steps that should be taken to include intensity mapping information
    in this formalism.
  
    Our aim is to find the most general expression for the posterior distribution of the true
    redshifts of a set of galaxies for which we only have photometric data and an overlapping
    intensity mapping survey. We start by considering a data vector ${\bf d}$ consisting of:
    \begin{itemize}
      \item $\nv$: galaxy positions.
      \item $\mv$: galaxy magnitudes.
      \item $\dhi$: a map of the perturbation in the HI density across angles
            and redshift
    \end{itemize}
    For each galaxy we want to estimate a redshift $z_i$, so let ${\bf z}$ be a vector
    containing all those redshifts. We want to study the posterior probability
    $p({\bf z}|\dhi,\nv,\mv)$. Let us start by noting that, in the standard models of
    large-scale structure, both $\dhi$ and the galaxy distribution can be thought of
    as being biased and noisy representations of the underlying matter overdensity
    $\dmat$. Sampling the galaxy redshifts could then also give us information about
    $\dmat$, and therefore it's worth considering the joint distribution
    $p({\bf z},\dmat|\dhi,\nv,\mv)$.
    
    One can study this distribution by iteratively sampling the two conditional
    distributions:
    \begin{align}\nonumber
      \dmat^{i+1} \leftarrow p(\dmat|{\bf z}^i,\dhi,\mv,\nv),&\hspace{12pt}
      {\bf z}^{i+1} \leftarrow p({\bf z}|\dmat^{i+1},\dhi,\mv,\nv).\\
    \end{align}
    We outline these two steps below.
    
    \begin{itemize}
     \item {\bf Conditional density distribution.}
      We start by noting that, if the true redshifts ${\bf z}$ are known, then the
      photometric redshifts given by the magnitudes $\mv$ are of no use in constraining
      the overdensity field, and therefore
      \begin{align}
        p(\dmat|\dhi,{\bf z},\nv,\mv)&=p(\dmat|\dhi,{\bf z},\nv)\\
                                     &=p(\dmat|\dhi,\dgal),
      \end{align}
      where $\dgal({\bf z},\nv)$ is the galaxy overdensity uniquely defined by the 
      galaxy angular coordinates and redshifts. $p(\dmat|\dhi,\dgal)$ can then be decomposed
      using Bayes' theorem as:
      \begin{equation}
        p(\dmat|\dhi,\dgal)\propto p(\dgal|\dmat)\,p(\dhi|\dmat)\,p(\dmat),
      \end{equation}
      where, following the same philosophy as above, we have considered that
      $p(\dgal|\dmat,\dhi)=p(\dgal|\dmat)$, since $\dhi$ is just a noisy and biased
      realization of $\dmat$. All that remains is then to model the conditional distributions
      $p(\dgal|\dmat)$ and $p(\dhi|\dmat)$, which is by no means a cursory matter, but
      something that can certainly be accomplished in the regime of validity of structure
      formation models.

    \item {\bf Conditional redshift distribution.}
      Under the assumptions that galaxies are Poisson-distributed over the (biased) matter
      overdensity, and that the photometric redshift errors are independent of the environmental
      density, it is possible to show (e.g. \cite{2012MNRAS.425.1042J}) that the galaxy
      redshifts can be sampled individually with the distribution:
      \begin{equation}
        p(z|\delta,\hat{n},m)\propto
        \left[1+b_{\rm g}(\dmat)(z,\hat{n})\right]\,p(z|m),
      \end{equation}
      where $b_{\rm g}(\dmat)(z,\hat{n})$ is the galaxy overdensity along the angular direction of
      each galaxy, and $p(z|m)$ is the prior photo-$z$ distribution.
    \end{itemize}
    
  \section{Angular power spectra}\label{app:cls}
    This section describes the theoretical models used for the angular power spectra entering
    the computation of the Fisher matrix (Eq. \ref{eq:fisher}).    
    The cross-power spectrum between two tracers of the cosmic density field, $a$ and $b$,
    can be estimated as:
    \begin{equation}\label{eq:int_cl}
      C^{ab}_\ell=4\pi\int_0^\infty\frac{dk}{k}{\cal P}_\Phi(k)W^a_\ell(k)W^b_\ell(k),
    \end{equation}
    where ${\cal P}_\Phi(k)$ is the power spectrum of the primordial curvature perturbations
    and $W^a_\ell(k)$ is the window function for tracer $a$, containing information
    about the different contributions to the total anisotropy in that tracer and about
    its redshift distribution.
    
    In the case of galaxy clustering and intensity mapping, and neglecting contributions
    from magnification bias and large-scale relativistic effects, $W^a$ is given by:
    \begin{equation}
      W^a_\ell(k)=\int_0^\infty dz\,\phi_a(z)\left[b_a(z)T_\delta(k,z)j_\ell(k\chi(z))+
                   \frac{1+z}{H(z)}T_\theta(k,z)j''_\ell(k\chi(z))\right],
    \end{equation}
    where $H(z)$ and $\chi(z)$ are the expansion rate and radial comoving distance at
    redshift $z$ respectively, $\phi_a(z)$ is the source redshift distribution, and
    $T_\delta$ and $T_\theta$ are the transfer functions of the matter overdensity
    and velocity divergence fields. Note that, even though we include the effect of
    non-linearities using the non-linear transfer function for $\delta$ (through
    the prescription of \cite{2012ApJ...761..152T}), we only introduce the effect of
    redshift-space distortions at the linear level, and only consider a deterministic
    linear bias $b_a(z)$. This is, nevertheless, a more rigorous treatment than has
    been used in the literature, and the procedure used to mitigate the effect of
    non-linearities described in Section \ref{ssec:method.fisher} should minimize
    the corresponding impact on the forecasts presented here.
    
    For galaxy shear tracers of weak lensing, the expression for the window function
    is:
    \begin{equation}
      W^a_\ell(k)=-\frac{1}{2}\sqrt{\frac{(\ell+2)!}{(\ell-2)!}}\int_0^\infty
      \frac{dz}{H(z)}\int_z^\infty dz'\phi_a(z')\frac{\chi(z')-\chi(z)}{\chi(z')\chi(z)}
      T_{\phi+\psi}(k,z)\,j_\ell(k\chi(z)),
    \end{equation}
    where $T_{\phi+\psi}$ is the transfer function for the sum of the two metric
    potentials in the Newtonian gauge.
    
    The computation of Eq. \ref{eq:int_cl} was carried out using a modified version of
    the Boltzmann code CLASS \cite{2011JCAP...07..034B,2013JCAP...11..044D}.
    
  \section{Noise power spectrum for intensity mapping experiments}\label{app:noise_im}
    This section derives the expression for the noise power spectra of single-dish
    experiments and interferometers presented in Eq. \ref{eq:nl_im}. Similar derivations
    have been provided before in the literature (e.g. \cite{2015ApJ...803...21B}), but
    we include this calculation here for completeness.
    
    Throughout this section we will use a flat-sky approach, where angles on the sky are
    represented by a 2D Cartesian vector ${\bf x}$. In this approximation, the spherical
    harmonic transform of a field becomes a simple 2D Fourier transform:
    \begin{equation}
      f_{\ell m}\equiv\sum_{\ell m}f(\nv)Y_{\ell m}(\nv)\rightarrow
      f_{\bf l}\equiv\int\frac{(dx)^2}{2\pi}e^{i{\bf x}\cdot{\bf l}}f({\bf x}).
    \end{equation}
    We will also simplify the derivation by writing integrals as Riemann sums. For
    instance, the Fourier transform above will read:
    \begin{equation}\label{eq:fourier}
      f_{\bf l}=\sum_x\frac{(\Delta x)^2}{2\pi}e^{i{\bf x}\cdot{\bf l}}f({\bf x}).
    \end{equation}
    Note that, with this normalization, the definition for the power spectrum $P_f$
    of a stochastic field $f$ is
    \begin{equation}\label{eq:defpk}
     \langle f_{\bf l}f^*_{{\bf l}'}\rangle\equiv
     \frac{\delta_{{\bf l},{\bf l}'}}{(\Delta l)^2}\,P_f({\bf l}),
    \end{equation}
    where $\Delta l\equiv2\pi/\Delta x$.

    \subsection*{Single dish}
      The flux at angular position ${\bf x}$ measured by a single dish is the sky
      intensity $I$ integrated over the instrumental beam $B$:
      \begin{equation}
        S({\bf x})=N_B\sum_{x'}(\Delta x)^2I({\bf x}')\,B({\bf x}-{\bf x}'),
      \end{equation}
      where $N_B\equiv1/B(0)$. Inserting the definition \ref{eq:fourier} in the
      expression above, and using the orthogonality relation $\sum_x(\Delta x)^2
      \exp[i{\bf x}({\bf l}-{\bf l}')]=\delta_{{\bf l},{\bf l}'}(2\pi/\Delta l)^2$,
      one can relate the Fourier components of $S$ and $I$ as
      \begin{equation}
        I_{\bf l}=S_{\bf l}/[N_BB({\bf l})],\hspace{12pt}{\rm where}\hspace{12pt}
        B({\bf l})\equiv\sum_x\frac{(\Delta x)^2}{(2\pi)^2}e^{i{\bf x}\cdot{\bf l}}.
      \end{equation}
      
      The power spectrum for $I$ is then related to that of $S({\bf s})$ as
      $P_I({\bf l})=P_S({\bf l})/[N_BB({\bf l})]^2$. Assume now that $S$ is purely white
      noise with a per-pointing rms flux $\sigma_S$, such that its power spectrum
      is simply flat with an ampitude:
      \begin{equation}
        P_S=\sigma_S^2(\Delta x)^2=
        \left(\frac{2k_B\,T_{\rm sys}}{A_e\sqrt{\Delta\nu\,t_p}}\right)^2(\Delta x)^2=
        \left(\frac{2k_B\,T_{\rm sys}}{A_e}\right)^2
        \frac{\Omega_{\rm obs}}{\Delta\nu\,t_{\rm tot}},
      \end{equation}
      where $\Delta x$ is the angular separation between pointings, $T_{\rm sys}$ is the
      per-pointing rms temperature fluctuation, $A_e$ is the effective collecting area
      of the dish, $\Delta\nu$ is the channel frequency bandwidth, $t_p$ is the
      integration time per pointing, $\Omega_{\rm tot}$ is the total observed sky area
      and $t_{\rm tot}$ is the total integration time.
      
      Substituting this into the expression for $P_I$ and relating the intensity $I$ to
      a brightness temperature $T$ through the Rayleigh-Jeans law ($I=2k_BT/\lambda^2$),
      we obtain the temperature noise power spectrum:
      \begin{equation}
        P_T({\bf l})=\frac{T_{\rm sys}^24\pi f_{\rm sky}}{\eta^2N_{\rm dish}\Delta\nu\,t_{\rm tot}}
        B^{-2}({\bf l}),
      \end{equation}
      where $f_{\rm sky}$ is the observed sky fraction, we have considered the possibility
      of having $N_{\rm dish}$ independent dishes, and we have defined the quantity
      $\eta\equiv A_eN_B/\lambda^2$. Note that, for a circular aperture telescope,
      $N_B=4\lambda^2/(\pi d^2)$, where $d$ is the dish diameter, and therefore
      $\eta=A_e/[\pi(d/2)^2]$ is the ratio of the effective to total dish area, which we have
      labelled ``efficiency'' in Eq. \ref{eq:nl_im}.
      
    \subsection*{Interferometers}
      The visibility observed by a pair of antennas separated by a baseline
      ${\bf d}\equiv\lambda{\bf u}$ is
      \begin{equation}
        V({\bf u})\equiv\sum_x(\Delta x)^2T({\bf x})B({\bf x})e^{2\pi\,i\,{\bf u}\cdot{\bf x}}
        \hspace{6pt}\longrightarrow\hspace{6pt}
        T({\bf x})B({\bf x})=\sum_u(\Delta u)^2e^{-2\pi\,i\,{\bf u}\cdot{\bf x}}.
      \end{equation}
      Transforming this to Fourier space we find:
      \begin{align}
        T_{\bf l}=\sum_u\left(\sqrt{2\pi}\Delta u\right)^2V({\bf u})
        \sum_x\left(\frac{\Delta x}{2\pi}\right)^2
        \frac{e^{i{\bf x}\cdot({\bf l}-2\pi{\bf u})}}{B({\bf l})}=\frac{V({\bf l}/(2\pi))}{2\pi}
      \end{align}
      where, in the last step, we have used the small-angle approximation $B=1$.
      The variance of $T_{\bf l}$ is therefore given by
      \begin{equation}
        \langle|T_{\bf l}|^2\rangle=\frac{\langle|V({\bf l}/(2\pi))|^2\rangle}
        {(2\pi)^2n({\bf l}/(2\pi))(\Delta u)^2},
      \end{equation}
      where $n({\bf u})(\Delta u)^2$ is the number of baselines in a volume $(\Delta u)^2$ in
      ${\bf u}$-space.
      
    \begin{figure}
      \centering
      \includegraphics[width=0.49\textwidth]{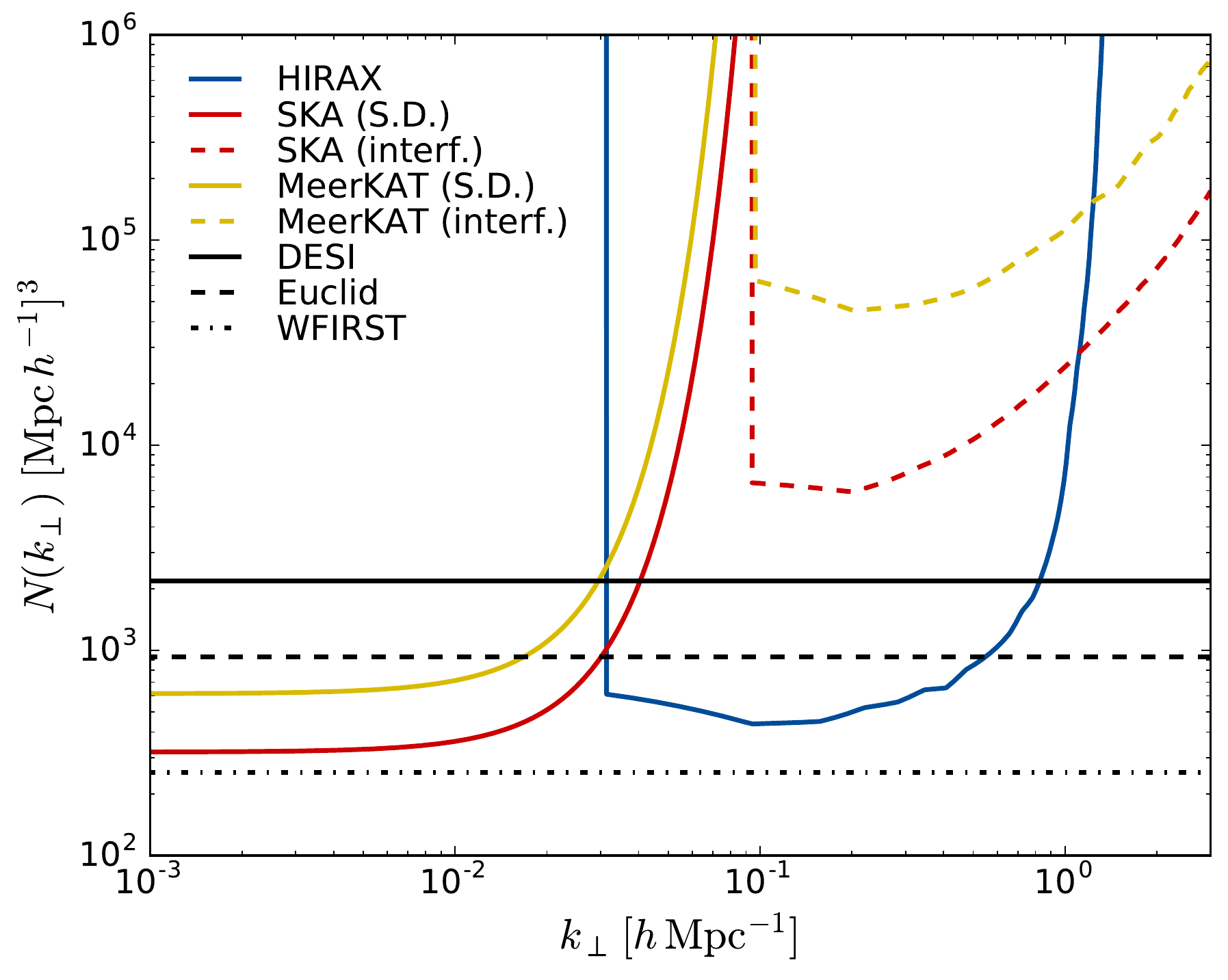}
      \caption{Noise power spectrum at $z=1.2$ as a function of transverse wavenumber $k_\perp$ for 
               HIRAX (blue), SKA (red) and MeerKAT (yellow). The curves for SKA and MeerKAT are
               separated into single-dish (solid) and interferometer (dashed). The shot-noise
               levels for the spectroscopic surveys DESI, Euclid and WFIRST at the same redshift
               are shown as black solid, dashed and dot-dashed respectively.}
      \label{fig:nk}
    \end{figure}
      In temperature units, the noise variance per visibility is given by
      $\langle|V({\bf u})|^2\rangle=[\lambda^2T_{\rm sys}/(A_e\sqrt{\Delta\nu\,t_p})]^2$.
      Relating baselines to Fourier coefficients as ${\bf u}={\bf l}/(2\pi)$ and recalling the
      definition of power spectrum (Eq. \ref{eq:defpk}), the noise power spectrum in temperature
      is then given by
      \begin{equation}
        P_T({\bf l})=\frac{\lambda^2T_{\rm sys}^2N_p}
        {A_e^2\,\Delta\nu\,t_{\rm tot}\,n({\bf u}={\bf l}/(2\pi))},
      \end{equation}
      where $N_p$ is the total number of pointings. Relating $n({\bf u})$ to the number density
      of physical baselines, and defining $N_p\Omega_p\equiv4\pi f_{\rm sky}$ we recover the
      expression for the noise power spectrum of interferometers in Eq. \ref{eq:nl_im}.
  
    \subsection*{Comparison with spectroscopic surveys}
      Converting the angular maps in different frequency channels into a three-dimensional map
      of the HI overdensity, we can relate the 3D noise power spectrum to its angular
      counterpart as:
      \begin{equation}
        P_{3D}({\bf k}_\parallel,{\bf k}_\perp)=\frac{c[(1+z)r(z)]^2}{\nu_{21}H(z)T^2_{\rm HI}(z)}
        P_T({\bf l}\equiv r{\bf k}_\perp),
      \end{equation}
      where $r$ is the comoving angular diameter distance to redshift $z$ and $T_{\rm HI}$ is
      the average 21cm temperature. This can then be directly compared with the shot-noise
      power spectrum of spectroscopic surveys, given by $P^{3D}=1/\bar{n}$, where $\bar{n}$
      is the 3D density of sources. The left panel of Figure \ref{fig:nk} shows the 3D noise
      power spectrum at $z\sim1.2$ as a function of the transverse wavenumber ${\bf k}_\perp$
      for the three IM experiments (HIRAX, SKA and MeerKAT) and the three next-generation
      spectroscopic surveys (DESI, Euclid and WFIRST) considered here.

\twocolumngrid

\end{document}